\documentclass[a4paper,11pt]{article}

\pdfoutput=1 

\usepackage{jcappub} 
\usepackage[numbers,sort&compress]{natbib}
\usepackage{graphicx}
\usepackage[british]{babel}
\usepackage[T1]{fontenc} 
\usepackage{amssymb,amsmath} 
\usepackage{makecell} 
\usepackage{comment}
\usepackage{pagecolor}

\title{Universality of Halo Shape and its Morphological Evolution across Cosmic Time}

\author[a,b]{Ayan Nanda}
\author[a,b]{Nishikanta Khandai,}
\author[c]{J. S. Bagla,}
\author[a,b,c]{Swati Gavas}

\def\knl{k_{\mathrm{nl}}}
\def\xnl{x_{\mathrm{nl}}}
\def\xnlsim{x_{\mathrm{nl}}^{\textrm{sim}}}
\def\xnlhs{x_{\mathrm{nl}}^{\textrm{hs}}}
\def\xnlmax{x_{\mathrm{nl}}^{\textrm{max}}}
\def\Nsnap{N_\textrm{snap}}

\def\Mfix{M_{\mathrm{fix}}}

\def\Mnl{M_{\mathrm{nl}}}
\def\Mvir{M_{\mathrm{vir}}}
\def\Rvir{R_{\mathrm{vir}}}
\def\Mhalo{M_{\textrm{h}}}
\def\Lbox{L_{\mathrm{box}}}

\def\Npart{N_{\mathrm{part}}}
\def\Lgrid{L_{\mathrm{grid}}}
\def\Ngrid{N_{\mathrm{grid}}}

\def\mpch{\textrm{Mpc/h}}

\def\kpch{\textrm{kpc/h}}
\def\beq{\begin{equation}}
\def\eeq{\end{equation}}
\def\lcdm{\Lambda\textrm{CDM}}
\def\msun{M_{\odot}}
\def\msunh{M_{\odot}/\textrm{h}}
\def\mdm{m_{\textrm{dm}}}
\def\neff{n_{\textrm{eff}}}

\newcommand{\software}[1]{{\small #1}}
\def\mpgadget{\software{MP-GADGET}\,\,}

\def\gadgetfour{\software{GADGET-4}\,\,}
\def\treepm{\software{TreePM}\,\,}
\def\rockstar{\software{ROCKSTAR}\,\,}
\def\subfind{\software{SUBFIND}\,\,}
\def\pylians{\software{Pylians}\,\,}

\def\colossus{\software{Colossus}\,\,}

\affiliation[a]{School of Physical Sciences, National Institute of Science Education and Research, Jatni 752050, India}
\affiliation[b]{Homi Bhabha National Institute, Training School Complex, Anushaktinagar, Mumbai 400094, India}
\affiliation[c]{
IISER Mohali, Knowledge city, Sector 81, SAS Nagar, Manauli PO 140306, Pujab, India}

\emailAdd{ayan.nanda@niser.ac.in}
\emailAdd{nkhandai@niser.ac.in}
\emailAdd{jasjeet@iisermohali.ac.in}
\emailAdd{swatigavas47@gmail.com}

\abstract{
We investigate the evolution of dark matter halo shapes in cosmological $N$-body simulations both in scale free Einstein-De Sitter (EdS) 
and $\lcdm$ cosmologies.
We compute the axis ratios ($q=b/a,s=c/a$) of well resolved central halos using the shape tensor.
These halos are identified using two different halo finding algorithms, \subfind and \rockstar.
We find that at fixed mass, halos become more spherical
with decreasing redshift.
The distribution $P(q,s)$ along with their median values ($q$ and $s$) 
shows self-similar behaviour as a function of mass scaled by the non-linear mass, $(M/\Mnl)$ 
across power-law spectral indices for scale free EdS models. 
However the median $q$ and $s$ show a tighter self-similar evolution as a function
of peak height $\nu=\delta_c/\sigma(M,z)$. We find
that the median $q(\nu)$ and $s(\nu)$  are consistent with an evolution along a universal curve described by
$y=\alpha-\delta\tanh \left[ \omega \left(\log_{10}(\nu) - \mu\right)\right]$
across the spectral indices ranging from $n=-1.0$ to $n=-2.2$. 
Our results hold for both \subfind and \rockstar, although there are some differences between them.
The universality of the evolution of median $q(\nu)$ and $s(\nu)$ also holds for the $\lcdm$ runs, although with a different behaviour 
at small $\nu$ compared to the scale free models.
The width of the distributions of $P(q)$ and $P(s)$ in
both, scale-free and $\Lambda$CDM, classes of simulations can be reduced 
further by classifying halos as oblate, triaxial and  prolate, each of
which also follows a universal behaviour. 
Although oblate halos are relatively rare at all redshifts, their fraction increases over time at the expense of the other two populations. 

}


\begin{document}
\maketitle
\flushbottom

\section{Introduction}
\label{sec:intro}

Within the standard cold dark matter (CDM) framework, dark matter halos originate from the collapse of high density peaks  in the matter distribution, arising from primordial density 
perturbations \cite{2007ApJS..170..377S,2009ApJS..180..330K} that grow through gravitational instability. This process proceeds hierarchically over cosmic time 
and forms a wealth of structures on a wide range of scales, i.e. smaller halos form earlier and they accrete matter, grow and even merge to form larger halos at later times, 
as demonstrated by numerical simulations \cite{1985ApJ...292..371D,1985ApJS...57..241E,1991ApJ...379...52W,1997Prama..49..161B,2005Natur.435..629S,2024MNRAS.527.4087J} 
and supported by observations\cite{2001MNRAS.328.1039C,2003AJ....126.2081A}. These halos provide deep gravitational potential wells for baryons to cool down, condense and form stars 
and galaxies at the centre \citep{1978MNRAS.183..341W}. Galaxies are therefore considered to be biased tracers of the underlying matter distribution. 
In this framework, it is generically expected that the properties of galaxies are correlated with the properties of dark matter halos in which they are 
embedded \cite{2017ARA&A..55...59N,2018MNRAS.473.2714S,2018ARA&A..56..435W,2025MNRAS.541.2304L}.

Over the past few decades, a number of studies~\cite{2009ApJ...696..620C,2010ApJ...710..903M,2013ApJ...770...57B,2018ARA&A..56..435W,2019MNRAS.488.3143B,
2026arXiv260210193T} have explored the relation between galaxy properties and halo properties and how they evolve with redshift. 
One of the strongest link found both in simulations and observations is a monotonic relation between galaxy stellar mass and halo mass. 
However the efficiency of galaxy formation increases with halo mass, peaks at $\Mhalo \sim 10^{12} M_{\odot}$ and drops thereafter. 
The most massive galaxies are redder in colour, have little to no star formation activity,  
have passively evolved over the past few billion years to host old population of stars
and have elliptical morphologies which are dispersion supported. 
These massive galaxies are hosted in the centre of massive cluster-sized dark matter halos. 

Although the radial density profile of dark matter halos is modelled assuming spherical symmetry and is well-described by an Navarro-Frenk-White (NFW) \cite{1996ApJ...462..563N} profile
which is motivated from N-body simulations, the global profile of halos i.e. its shape is non-spherical in cosmological 
simulations\cite{1991ApJ...378..496D,1991ApJ...368..325K,1992ApJ...399..405W,2002ApJ...574..538J,2006MNRAS.367.1781A,2012JCAP...05..030S}. Accounting for this deviation from sphericity is important when interpreting observations with the prediction 
of CDM models \cite{1988ApJ...327..507F,2007MNRAS.377...50H}. Studies have shown that the assumption of spherical symmetry affects a variety of halo properties, e.g. 
the halo mass function\cite{1974ApJ...187..425P} and the concentration parameter \cite{2002ApJ...574..538J,Sereno_2015}. In particular, fitting a spherical NFW profile to triaxial 
halos can systematically bias the inferred concentration parameter and mass estimates in lensing analysis, especially when the elongations 
along the line of sight are significant\cite{Corless_2007}. In addition, spherically averaged intrinsically ellipsoidal clusters introduce orientation-dependent biases and scatter 
in X-ray and SZ observables, even when the mean bias is small\cite{2012MNRAS.421.1399B}. Neglecting cluster triaxiality and gas non-isothermality introduces systematic biases in the 
determination of the Hubble constant from joint X-ray and SZ analyses\cite{2006ApJ...643..630W}. 
Furthermore, the axial symmetry in strong-lensing analyses of galaxy clusters can systematically bias the inferred inner density slope, particularly when cluster ellipticity 
and orientation are not properly accounted for\cite{2007MNRAS.381..171M}. Therefore, the shapes of dark matter halos can be used to constraint the nature of dark matter\cite{2024JCAP...10..027C,2025JCAP...01..086L,2026A&A...706A.340G}. 

The geometric assumptions of gravitational collapse directly determine the predicted abundance of halos in theoretical models. 
The Press-Schechter (PS) formalism \cite{1974ApJ...187..425P} employs a spherical collapse model with a constant overdensity threshold for collapse. 
It fails to accurately match the halo mass function (HMF) obtained from simulations by over predicting the number of halos below the characteristic mass ($M_*$) and underpredicting abundances
at larger masses. 
By accounting for triaxiality through the ellipsoidal collapse model, the Sheth-Tormen (ST) approach \cite{1999MNRAS.308..119S} modifies the constant collapse barrier
into a quadratic function of the mass variance. 
This shift from a fixed to a moving barrier incorporates the resistance to collapse caused by external tidal shear. 
Consequently, the ST mass function provides a superior fit to the  HMF from simulations compared the PS formalism, 
demonstrating that halo morphology is a fundamental driver of large-scale statistics. 
It is useful for predicting power spectrum through halo models since the abundance of halos directly influences the two-halo term in the matter power spectrum\cite{2002PhR...372....1C}.

Dark matter halos reside within diverse environments  of the cosmic web - nodes, filaments, sheets and voids and these environments play a crucial role in shaping the properties of halos, e.g shape, 
formation time, mass growth rate, mass accretion rate and concentration to name a few. The dependence of these secondary properties of halos on local environment affects small scale features like the local 
tidal environment around halos and large scale clustering strength\cite{2021MNRAS.503.2053R,2018MNRAS.476.3631P,2020MNRAS.499.4418R} consistent with observations\cite{2019MNRAS.483.4501A,2024MNRAS.527.3771A}. 
Since halos form hierarchically from initial density perturbations, their internal properties encode the information about the formation history, growth through mergers and anisotropic accretion. 
Halos that have assembled early and those that have formed recently experience different environments 
and merger histories and thereby influence halo morphology. 
Therefore it is  expected that the cosmic web environment regulates halo assembly and scale dependent clustering. 
The anisotropic accretion of matter, infall of satellites through cosmic filaments\cite{2015ApJ...813....6K}, 
the large scale tidal environments\cite{2024MNRAS.527.3771A} lead to the elongated structure of halos in a preferential 
direction \cite{2018MNRAS.481..414G}. 
The angular momentum vector of dark matter halos, aligned mostly with the minor axis, is small in magnitude. Halos are therefore not rotationally supported, 
rather they are supported by the anisotropic velocity dispersion which acts like an effective pressure \cite{2006MNRAS.367.1781A,2008gady.book.....B,1988ApJ...327..507F}. 

Over the last decade, hydrodynamical simulations have evolved rapidly in scale and complexity and are able to reproduce 
realistic galaxy populations. 
This progress naturally raises the question of how baryonic physics - 
gas cooling and heating, star formation, supermassive black hole growth and associated feedback processes are affected or affect the shapes of dark matter halos. 
The effect of gas cooling within halos make them more spherical than the halos formed in adiabatic simulations\cite{2004ApJ...611L..73K,2006PhRvD..74l3522G,2013MNRAS.429.3316B} whereas energy 
feedback from star formation and AGN expelling gas and suppress contraction, as a result of which it becomes more aspherical and the stellar distribution on galactic scale shows a median 
misalignment of 45-50 degree with respect to their host halo\cite{2015MNRAS.453..721V}. 

In this work, we investigate the evolution of the distribution of halo shapes for all halos and by morphological type (oblate, triaxial or prolate) for scale free models listed in Table~\ref{tab_scalefree}. 
This is done to test  self-similar evolution in scale-free models. We find that the distribution evolves along a characteristic universal curve across scale-free models. 
We then extend our study to a $\Lambda$CDM model (Table~\ref{tab_lcdm}) to explore 
if similar universal trends emerge. 
We present the results in terms of universal relations all halos and by morphological type.  We also investigate the mass function by halo morphology class
and the fractional change in counts of halos for different halo morphologies over cosmic time. Our results will be useful for extending and refining halo models. 

This paper is organized as follows: In section~\ref{sec:methods}, we briefly review the basic framework of Einstein-de Sitter universe, present scale-free simulations (Table~\ref{tab_scalefree}) and $\Lambda$CDM simulations (Table~\ref{tab_lcdm}), outline the two halo finders (\subfind \& \rockstar) used in our analysis. In section~\ref{sec:halo_shapes}, we present definitions, methods, resolution and convergence criteria used for halo shape measurements. Sections~\ref{sec:results1} and \ref{sec:result2} present our main results. Finally in section~\ref{sec:discussion_summary} we discuss the implication of our findings and summarize our conclusions.

\section{Methods}
\label{sec:methods}
In this section we describe our simulation suite and the halo catalogue generated by two halo finding algorithms.  

\subsection{Scale-free Simulations \& Self-Similarity}
Scale-free cosmologies, described by an initial power-law power spectrum $P(k) \propto k^n$ in an Einstein-De Sitter (EdS) background ($\Omega_\mathrm{m} = \Omega_{\mathrm{tot}} = 1 \Rightarrow \Omega_{\mathrm{k}} = 0$),
are useful models to explore the growth of linear perturbations which eventually collapse and form virialised non-linear structures in the universe. 
The only featuring scale in these models is the scale of non-linearity introduced by self-similar gravitational clustering. 
In linear theory, the growing mode for an EdS background evolves as $D_+(t) = a(t) \propto t^{2/3}$, 
where $a(t)$ is the scale factor. The non-linear (comoving) scale, $\xnl$, is defined as the scale at which the variance of linear fluctuations in mass, enclosed in a spherical volume of radius $\xnl$ is fixed by
\begin{equation}
\sigma^2(\xnl,a) = \int_0^{\infty} \frac{dk}{k} \frac{k^3 P(k,a)}{2\pi^2} \left|W(k,\xnl)\right|^2 = \beta \sim \mathcal{O}(1)
\label{eq_sigma2}
\end{equation}
$\beta$ is a number of order unity. The exact value of $\beta$ is not important, but once fixed equation~\ref{eq_sigma2} defines the nonlinear scale $\xnl = \xnl(a)$ 
and the corresponding nonlinear mass $\Mnl = \Mnl(a) = \frac{4\pi}{3}\xnl^3 \bar{\rho}_\textrm{m}$. We will use mass and length interchangeably in the
discussions that follow. Equation~\ref{eq_sigma2} is also used to normalize the power
spectrum to a particular value $\beta$ at a given redshift. 
The choice of normalization is arbitrary in scale-free cosmologies, and we set $\sigma(\xnl=8 \Lgrid,a=1) = 1$, 
where $\Lgrid$ denotes the grid length in the simulation. 
For $\Lambda$CDM cosmologies, $\sqrt{\beta} = \sigma_8 \equiv \sigma(\xnl=8\mpch,a=1)$ is fixed by observations. 
For scale free models the time evolution of $\xnl$ and $\Mnl$ can be calculated from linear theory as:
\beq
   \xnl \propto a^{\frac{2}{n+3}} \quad\quad \Mnl \propto a^{\frac{6}{n+3}}
   \label{eq_xnlofa}
\eeq

Since $\xnl = \xnl(a)$ is the only length scale  in scale-free cosmologies,
the time evolution of any statistical quantity of interest, say $Q(\left\{V_i\right\},a)$, exhibits a self-similar behaviour \cite{1977ApJS...34..425D, 
1980lssu.book.....P,1985ApJS...57..241E,1988MNRAS.235..715E}
when recast in terms of a set of dimensionless variables, $\left\{V_i/V_{\textrm{nl},i}\right\}$, where $V_i$ and $V_{\textrm{nl},i}$ are variables which are the same power-law functions of $x$ and $\xnl$ respectively.
\beq
Q(\left\{V_i\right\},a) \equiv \Tilde{Q}(\left\{V_i/V_{\textrm{nl},i}\right\})
\label{eq_self-similar}
\eeq

In scale free cosmologies fixing either  length or time fixes the other. 
One can therefore use this freedom to compare two snapshots, $i$ and $j$, with the same powerlaw index $n$:
\begin{equation}
     {\frac{L_{\textrm{box},i}^{new}}{L_{\textrm{box},j}^{new}} = \left(\frac{l_i\cdot N_{\textrm{part},i}^{1/3}}{l_j\cdot 
     N_{\textrm{part},j}^{1/3}}\right)\left(\frac{x_{nl,j}(a=1)}{x_{nl,i}(a=1)}\right)\left(\frac{a_i}{a_j}\right)^{-\frac{2}{n+3}}}
\end{equation}
Here the $\Lbox$ is the boxsize of the simulation, $\Npart$ is the total number of particles in the simulation, 
$l = (\Lbox/\Npart^{1/3})$ is the mean inter-particle separation. We note that the two snapshots under consideration 
can come from two distinct simulations or represent distinct epochs from the same simulation.

Over the past four decades scale-free simulations and the self-similar ansatz have been extensively used to understand aspects of gravitational instability and clustering; 
e.g. the scaling relation between clustering measures (power spectrum or correlation function) in the linear and non-linear regimes \cite{1991ApJ...374L...1H,1994MNRAS.271..781C,1995MNRAS.276L..25J,1996MNRAS.280L..19P,
1996ApJ...466..604P, 1998ApJ...509..517J,2005MNRAS.360..546R}, 
the deeply non-linear regime and the stable clustering hypothesis 
\cite{10.1093/mnras/253.2.295, 1996ApJ...466..604P, 1996ApJ...465...14C, 1997MNRAS.287..687J, 2003MNRAS.341.1311S,2005MNRAS.360..546R, 2009MNRAS.397.1275W,2014MNRAS.443.2126B,2017MNRAS.470.4099B}
the universality of halo abundances \cite{2023MNRAS.521.5960G}
finite volume effects on halo abundances  dark matter pair-velocities  \cite{2009MNRAS.395..918B}
and  discreteness effects \cite{2025JApA...46...33B}.
One can also use these models to test the accuracy of an $N$-body code by requiring that the observable
in equation~\ref{eq_self-similar} be self-similar. Indeed very large scale-free simulations have tested the accuracy
of the ABACUS code \cite{2021MNRAS.508..575G}, by requiring that statistical quantities like the halo abundance, dark matter clustering, accretion histories of halos, 
dark matter pair-velocities evolve self-similarly \cite{2021MNRAS.501.5051J, 2021MNRAS.501.5064L, 2021MNRAS.504.3550G, 2022MNRAS.509.2281G, 2022MNRAS.512.1829M, 2023MNRAS.525.1039M, 2024MNRAS.527.5603M, 2024MNRAS.532.1729S}.

We show in section~\ref{sec:results1} that the distribution of halo shapes is self-similar in scale-free simulations.

\subsection{N-body Simulations}
We use the \gadgetfour code \cite{2021MNRAS.506.2871S} with dark matter only particles for the scale-free and $\lcdm$ runs along with \mpgadget code\cite{yu_feng_2018_1451799} described in tables~\ref{tab_scalefree} and \ref{tab_lcdm} respectively.
The scale-free runs 
are the same as those used in \cite{2023MNRAS.521.5960G} to study the universality of the halo mass function (HMF), except for $n=-1.3$ which
has been added for this work.
\gadgetfour is run in \treepm \citep{2002JApA...23..185B} mode.
The number of grid points, $\Ngrid$, in one direction for the Particle-Mesh (PM) long range force computation is taken as $\Ngrid^3 = \Npart$ for all the runs.
This translates to the mean-interparticle separation to be 1 in units of grid scale, $\Lgrid = \Lbox/\Ngrid$. For the scale-free runs, length scales  (e.g. $\xnl$, $\Lbox$)
are in units of $\Lgrid$ and for the $\lcdm$ runs, they are in comoving $\textrm{Mpc/h}$. For the short range force, which uses the Barnes-Hut \cite{1986Natur.324..446B} oct-tree,
we have retained multipoles to order 3. This ensures that errors in force calculations are well below 1\%.
Due to finite volumes in simulations, power is missing on scales larger than $\Lbox$.
This causes significant errors in measuring various statistical quantities (like mass variance
$\sigma(M)$, correlation function $\xi(x) $, power spectrum $\Delta^2(k) = \frac{k^3 P(k)}{2\pi^2}$,
HMF ($\phi(M) = dn/d\log{M}$) \cite{2005MNRAS.358.1076B, 2006MNRAS.370..993B, 2006MNRAS.370..691P,2009MNRAS.395..918B},
since modes beyond the box size are ignored while generating the initial condition.
The error in the mass variance increases as we go to more negative indices. Keeping this in mind, we have chosen the final output  in terms of $\xnl$
such that the error in mass variance at $\xnlmax$ is below $1\%$.
We refer the reader to \cite{2025JApA...46...33B} for details of finite volume
considerations and other accuracy parameters for the simulations.

The initial condition is generated using second order Lagrangian perturbation theory (2LPT).
All power-law power spectra are normalized such that $\sigma_{lin}(x_{nl}=8;a=1)=1$. 
The starting time epoch $z_{i}$ is chosen such that the  mass variance at $\Lgrid$ is of  order  $10^{-2}$ at $z=z_i$.
For the $\lcdm$ runs we have taken the cosmological parameters of Planck data \cite{2014A&A...571A..16P}.
The cosmological parameters used are
$\Omega_m=0.307115, \Omega_{\Lambda}=0.692885, \Omega_b= 0.048206, \sigma_8=0.8228, h=0.6777, n_s=0.96$. 
$\Omega_m, \Omega_\Lambda, \Omega_b$ are the dimensionless matter, cosmological constant and baryon densities. $h$ is the Hubble's constant in units of 100 km/s and $n_s$ is
the primordial spectral index.
The initial power spectrum is generated using the \colossus package\cite{2018ApJS..239...35D}.
Details of the $\lcdm$ runs are given in table~\ref{tab_lcdm}. 

\begin{table}[h]
  \centering
  \begin{tabular}{|c |c |c |c |c |c |c |c |} 
    \hline
    $C_1$ & $C_2$ & $C_3$ & $C_4$ & $C_5$ & $C_6$ & $C_7$ & $C_8$ \\
    $n$ & $z_{i}$ & $\Lbox$ & $\Npart^{1/3}$ & $\Nsnap$ & $\xnlsim$ & $\xnlhs$ & $\xnlmax$ \\
    \hline
    $-1.0$ & $160$ & $1024$ & $1024$ & $7$  & $1.3-49.4$  & $10.2-49.1$  & $54.8$ \\
    \hline
    $-1.3$ & $116$ & $1024$ & $1024$ & $6$  & $1.3-33.4$  & $9.6-29.2$  & $33.4$ \\
    \hline
    $-1.5$ & $96$ & $1024$ & $1024$ & $5$  & $1.3-49.4$  & $7.9-22.5$  & $27.8$ \\
    \hline
    $-1.8$ & $70$ & $1024$ & $1024$ & $4$  & $1.3-49.4$  & $6.1-13.3$  & $14.5$ \\
    \hline
    $-2.0$ & $57$ & $1536$ & $1536$ & $5$  & $0.2-10.2$  & $3.6-10.2$  & $11.4$ \\
    \hline
    $-2.2$ & $46$ & $1536$ & $1536$ & $6$  & $0.2-10.2$  & $2.1-3.6$  & $4.4$ \\
    \hline
  \end{tabular}
  \caption{\textbf{Details of scale-free runs}: The columns $C_1 - C_8$ denote the following.
    $C_1$: fixed spectral index, $n$, of the power-law model.
    $C_2$: Starting redshift, $z_i$, of the simulation.
    $C_3$: Length of the simulation box, $\Lbox$.
    $C_4$: Cube root of the number of particles, $\Npart$, used in simulation
    $C_5$: Number of snapshots, $\Nsnap$, used for the analysis.
    $C_6$: Range of non-linear scales, ($\xnlsim$), covered in the simulation.
    $C_7$: Range of $\xnlhs$ used to compute halo shapes.
    $C_8$: Maximum allowed value of $\xnl$ to keep errors in $\sigma$ below  1\%, finite volume effects.}
  \label{tab_scalefree}
\end{table}
  
One  approximate way to relate scale free runs to the $\lcdm$ ones is via the effective spectral index defined as below:
\begin{equation}
n_{\mathrm{eff}}= -2 \left.\frac{d \ln \sigma(R)}{d \ln R}\right|_{R=R_L} - 3
\end{equation}
where $R_L$ is the comoving lagrangian radius corresponding to the given halo mass $M$. 
$\neff$\cite{1995MNRAS.276L..25J,1996MNRAS.280L..19P,1996ApJ...466..604P,2009MNRAS.397.1275W,2023MNRAS.521.5960G}  can be thought as a proxy for time or a scale (which is turning non-linear at that time). 
In this work we will also consider this mapping in section~\ref{sec:results1}.
In a smooth power spectrum without the features of baryonic acoustic oscillation imprinted on it, $\neff \in [\sim -1.8,\sim -2.2]$ ranges from  cluster to galactic scales.

\begin{table}[h]
    \centering
    \begin{tabular}{|c|c|c|c|c|c|c|c|}
    \hline
    $C_1$ & $C_2$ & $C_3$ & $C_4$ & $C_5$ & $C_6$ & $C_7$ & $C_8$ \\
    Code & $z_i$ & $L_{\text{box}}$  & $N_{\text{part}}$  & $m_{dm} $ & $\epsilon $  & $z_{\text{range}}$ & $N_{\text{realizations}}$ \\ 
     &  & [Mpc/$h$] & &  $[M_{\odot}/h]$ & $ [\text{kpc}/h]$  & & \\
    \hline
    \mpgadget & $159$  & $75$ & $1008^3$ & $3.60 \times 10^7$ & $4.63$ & $6-0$ & $1$  \\ 
    \hline
    \mpgadget & $159$  & $100$ & $1008^3$ & $8.54 \times 10^7$ & $3.31$ & $5-0$ & $1$  \\
    \hline
    \mpgadget & $159$  & $125$ & $1008^3$ & $1.67 \times 10^8$ & $4.13$ & $5-0$ & $1$  \\ 
    \hline
    \mpgadget & $159$  & $150$ & $1008^3$ & $2.89 \times 10^8$ & $4.96$ & $4-0$ & $1$  \\ 
    \hline
    \mpgadget & $159$  & $250$ & $1344^3$ & $5.63 \times 10^8$ & $6.20$ & $4-0$ & $1$  \\ 
    \hline
    \gadgetfour & $159$ & $128$  & $1024^3$ & $1.68\times 10^8$ & $4.00$  & $3-0$ & $1$ \\
    \hline
    \gadgetfour & $159$ & $512$  & $1024^3$ & $1.07\times 10^{10}$ & $17.00$  & $2.1-0$ & $1$ \\
    \hline
    \gadgetfour & $159$ & $1024$ & $1024^3$ & $8.52 \times 10^{10}$ & $35.00$  & $0.9-0$ & $2$ \\
    \hline
    \end{tabular}
    \caption{\textbf{Details of $\lcdm$ runs}: The columns $C_1 - C_8$ denote the following.
      $C_1$: Simulation suite.
      $C_2$: Starting redshift, $z_i$, of the simulation.
      $C_3$: Comoving length of the  simulation box, $\Lbox$.
      $C_4$: Number of particles, $\Npart$, used in the simulation.
      $C_5$: Mass of the single dark matter particle, $\mdm$, in units of $\msunh$.
      $C_6$: The gravitational softening length in unit of $\kpch$.
      $C_7$: The redshift range of the simulation used for the analysis. 
      $C_8$: The number of realizations for each simulation.}
    \label{tab_lcdm}
\end{table}

\begin{figure}[h]
    \centering
    \includegraphics[width=0.95\linewidth]{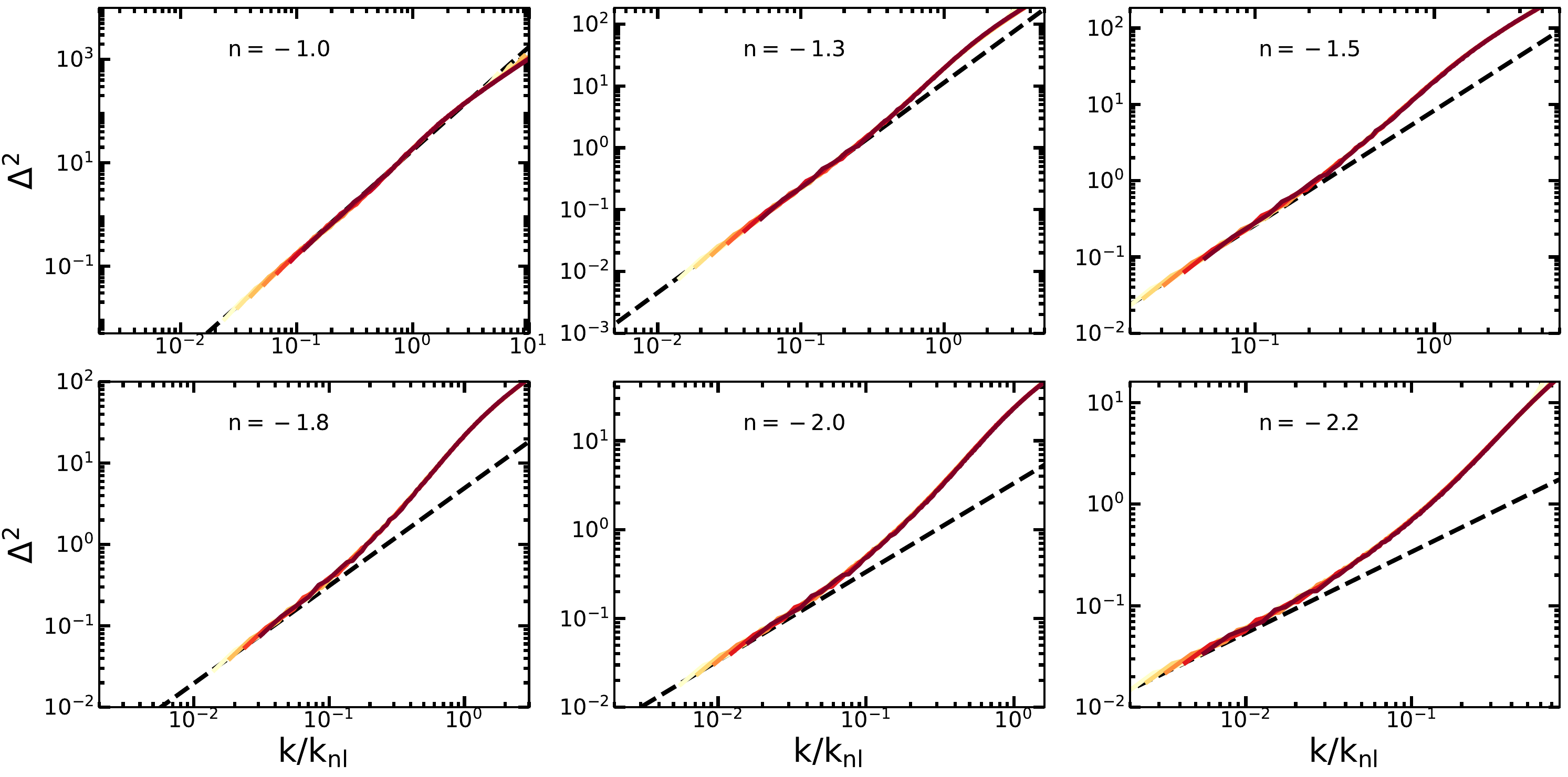}
    \caption{The dimensionless power spectrum($\Delta^2$) as a function of $k/\knl$ for the six scale-free models listed in table~\ref{tab_scalefree}.
      Lighter colours represent earlier epochs. The black dashed line is the linear spectrum, $P(k) \propto k^n$ used in the simulation for the initial condition.}
    \label{fig_delsq}
\end{figure}

\subsection{Self-similar Evolution of Power Spectrum}
In figure~\ref{fig_delsq} we have plotted the dimensionless power spectrum, $\Delta^2(k)$, as a function of scaled wave number ($k/\knl$)
for different epochs for each of the six indices listed in table~\ref{tab_scalefree}.  The lighter colours represent earlier times in the run or larger values of $\knl$.
The dashed line is the linear power spectrum used for generating the initial conditions. The self-similar evolution of the power spectrum is seen across
all models and in all regimes, i.e. the linear to the quasi-linear to the non-linear regimes.
The slope asymptotes to $\sim 2 $ in the non-linear regime consistent with \cite{1997MNRAS.286.1023B, 2023MNRAS.521.5960G}.

\subsection{Halo and subhalo finders: \subfind and \rockstar}
\begin{figure}[!htb] 
  \centering
  \includegraphics[width=0.95 \linewidth]{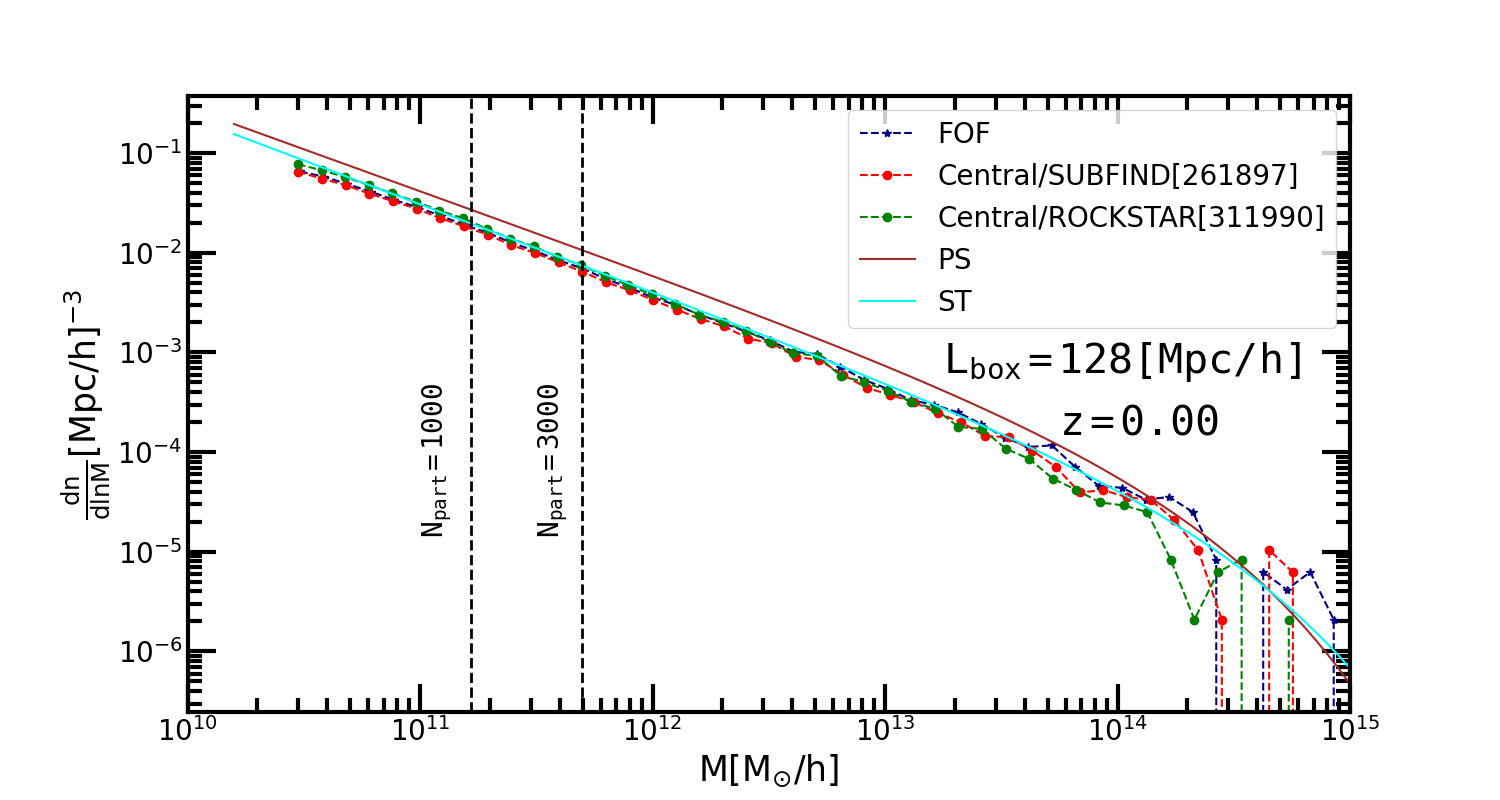}
  \caption{The figure shows the abundance of halo mass generated from \subfind(solid dashed red) and \rockstar(solid dashed dotted red) algorithm.
    The brown and cyan solid lines are the analytical Press-Schechter and Sheth-Tormen mass function respectively.
    The vertical lines corresponds to the $1000 \,\,\textrm{\&}\,\, 3000$ particles (left-right).
    32 particles is the minimum FOF resolution of halos.
    The 1000-3000 particle count is the minimum requirement to accurately identify
    shapes. Numbers in the square bracket represent the number of central halos in each of the halo finder with $\Npart\geq100$.}
    \label{fig_hmf_l128}
\end{figure}

We use two substructure finders, \subfind \cite{2001MNRAS.328..726S} and  \rockstar \cite{2013ApJ...762..109B} for this work.
This is to explore if our results are qualitatively similar or are sensitive to the substructure finder.
Both algorithms begin by first identifying a catalogue of groups of particles using the percolation -- Friends-of-Friends (FOF) algorithm \cite{1985ApJ...292..371D} with a
linking length of 0.2 times the mean interparticle separation. The substructure finder then works on each FOF group to identify bound
substructures within this group. While running \subfind and \rockstar we have kept the minimum number of particles in the group (subgroup) to be 32(20).
In \subfind, local density peaks are first identified within the parent FOF group. The region around the density peak is
expanded to associate more particles with the density peak until a saddle point is reached in the isodensity contour. A binding criterion
is then used to remove any unbound particle. The largest subgroup  of particles is referred to as the central halo and the smaller subgroups as satellites.
The \rockstar  algorithm starts with the same FOF method. It then builds a hierarchy of FOF subgroups in phase space (position and velocity)
by progressively and adaptively reducing the linking length such that $70\%$ of the particles in each group is captured in the next hierarchy of subgroups.
This process continues until seed halos with a minimum of 20 particles are left in the final level of the subgroup tree.
The seed halos are placed at the deepest level in the subgroup tree. Once the particles are assigned to subgroups at each level,
the unbound particles are removed from them. Here we only consider halos in level 1 of the substructure tree.
As in \subfind we define the largest subgroup in the first level of the tree as the central halo and the rest as satellites.
For this work we have decided to only consider central halos for two reasons:
\begin{itemize}
  \item The central halo being the most massive subhalo of the FOF group 
    is relatively less disturbed unless it is undergoing a major (i.e $\sim$ mass) merger. Such mergers are rare in the mass ranges we are working with
    and do not affect our results (which are statistical in nature) were we to remove such objects. We will show in the next section that the resolution 
    requirement is $\gtrsim 1000$ particles, making similar mass mergers even rarer. The central halo dominates the HMF as can be seen
    in figure~\ref{fig_hmf_l128} where we plot the FOF HMF (blue-dashed), the \rockstar central HMF(green-dashed) and the \subfind central HMF (red-dashed)
    for the $\Lbox=128 \textrm{Mpc/h}$-$\lcdm$ run at $z=0$. Comparison is made
    with the PS (brown-solid) and ST (cyan-solid) HMFs. Both central HMFs are close to the FOF and ST HMF. The differences occur at larger masses due to poor statistics for this
    particular box. The PS HMF is known
    to overproduce halos at lower masses and underproduce at higher masses compared to the ST HMF and those obtained from simulations. The satellite HMF, which
    we have not plotted (due to reasons given in the next point) represent a small fraction of the total FOF and central HMF. 

  \item We find that the central halos in both \rockstar and \subfind are well identified and matched in both algorithms. The differences appear in the matching
    of satellites which are not as well resolved as the central halo. 
\end{itemize}

\begin{figure}[h]
  \centering
  \begin{minipage}{0.45\linewidth}
    \centering
    \includegraphics[width=\linewidth]{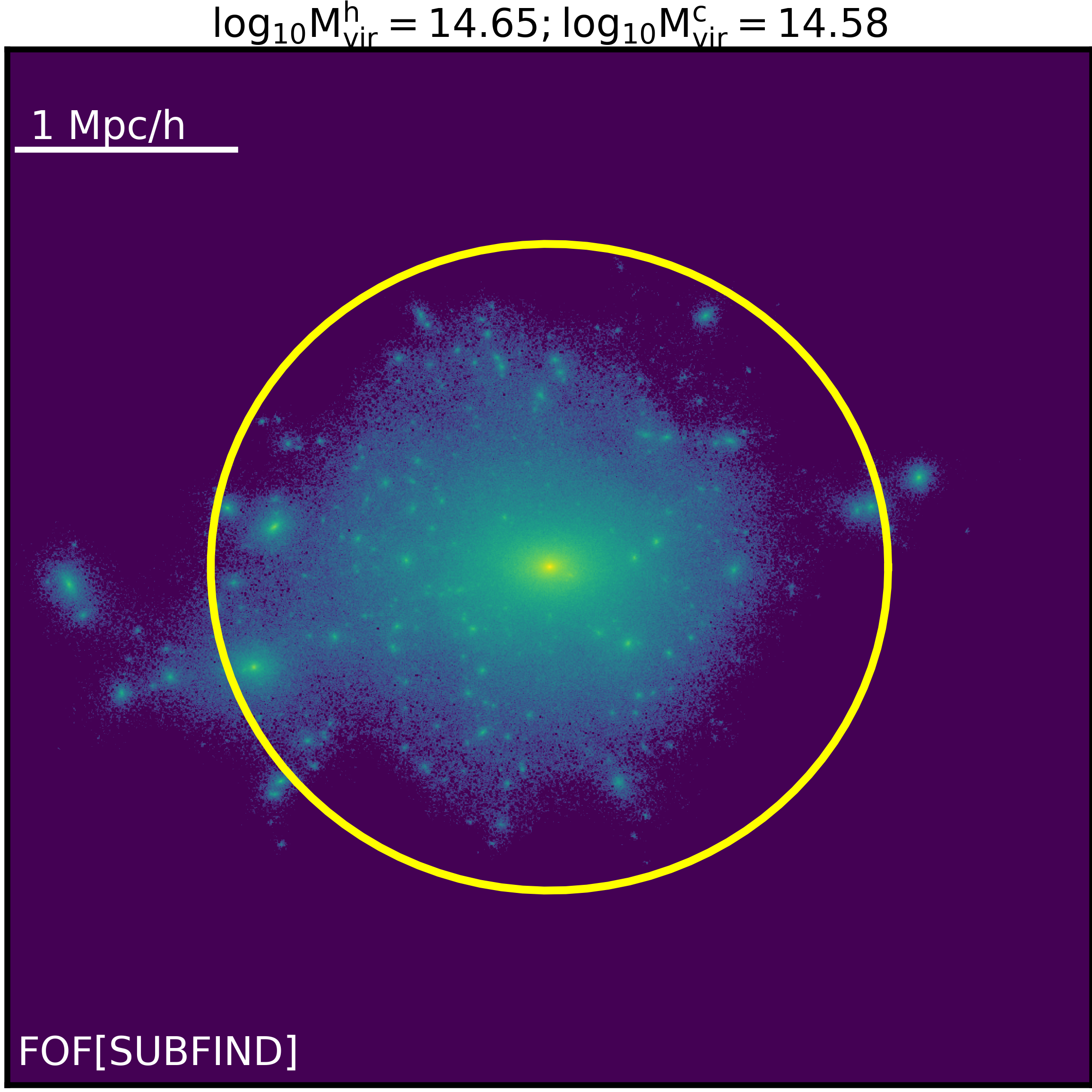} \\
    \includegraphics[width=\linewidth]{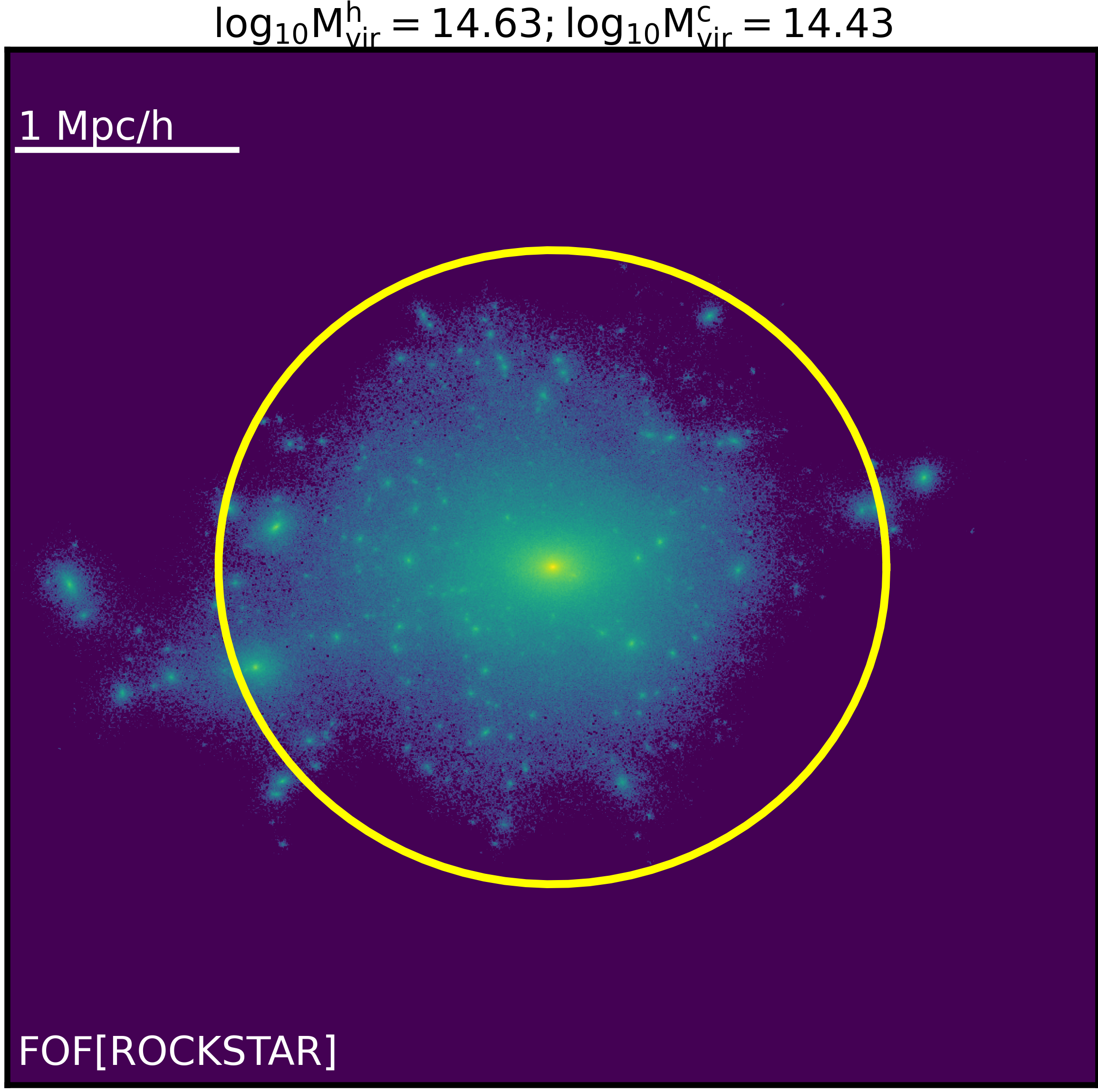}
  \end{minipage}
  \begin{minipage}{0.45\linewidth}
    \centering
    \includegraphics[width=\linewidth]{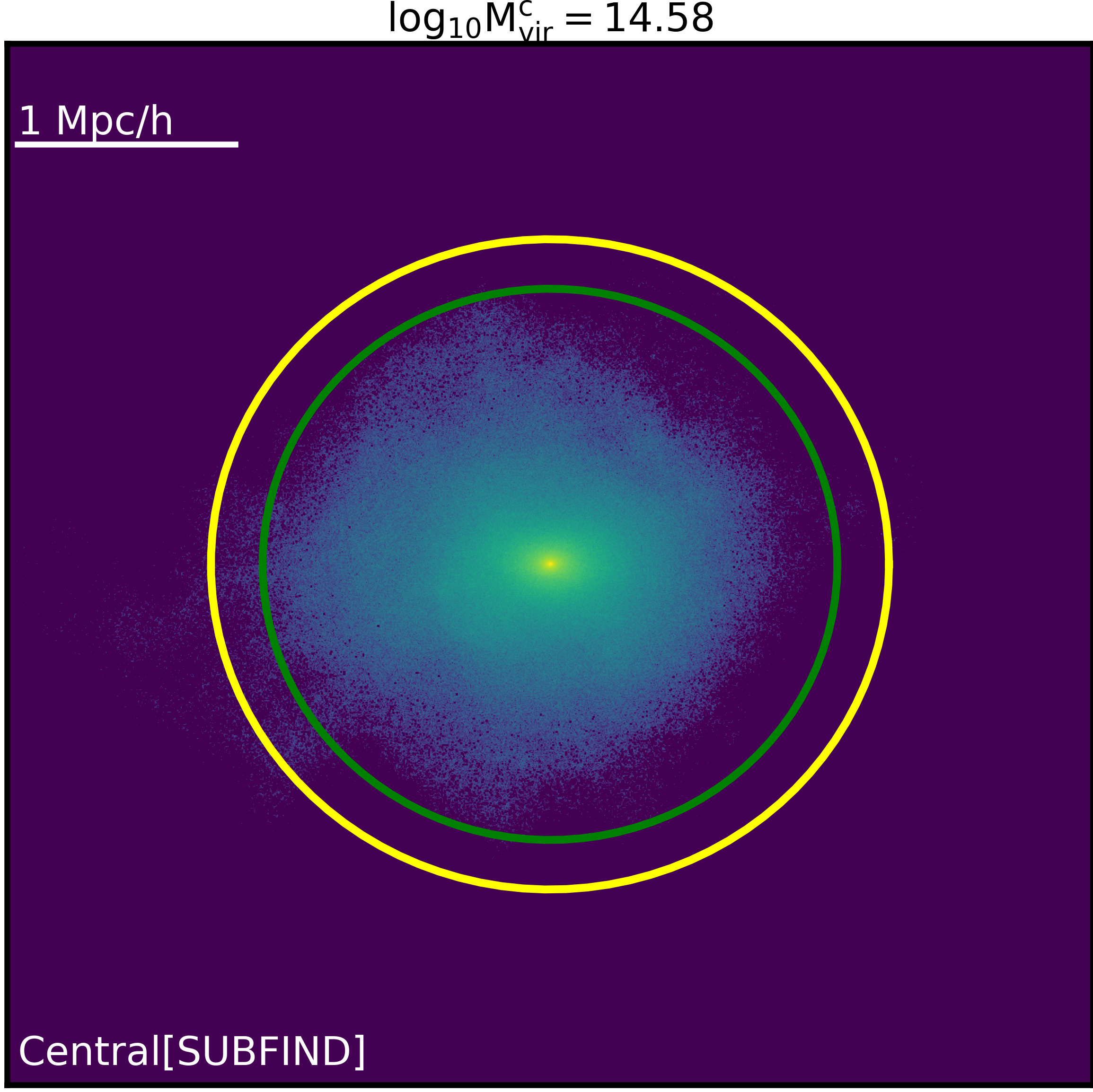} \\
    \includegraphics[width=\linewidth]{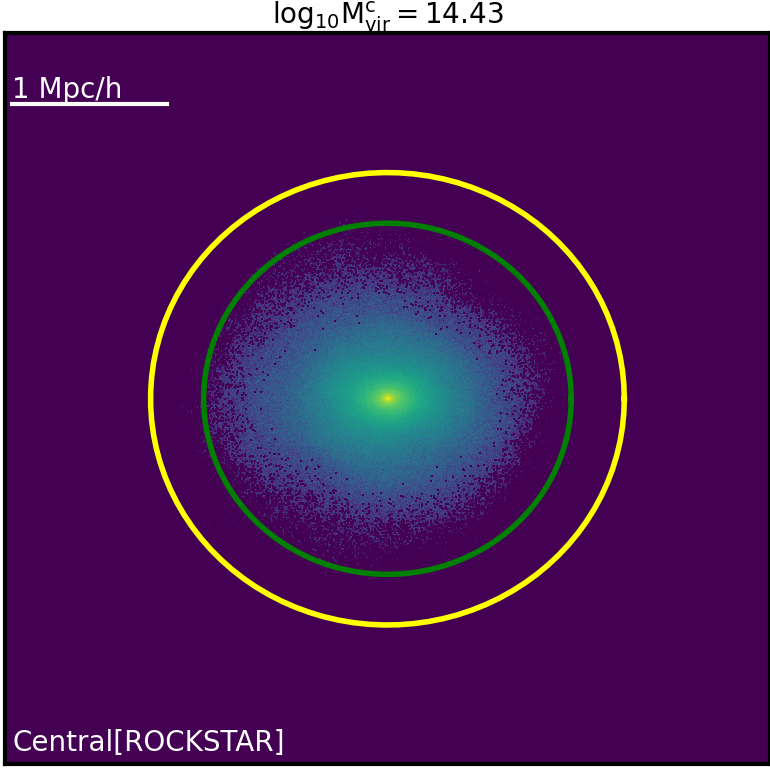}
  \end{minipage}
  \caption{The top (bottom) panels show particle distributions in a massive cluster-sized halo using the \subfind (\rockstar) algorithms.
    Particle distributions for the FOF and central halos are plotted in the left and right panels respectively.
    The yellow and green circles are the virial radii of the FOF and central halos respectively. The mass of the FOF,
    $\Mvir^{\textrm{h}}$, and central, $\Mvir^{\textrm{c}}$, halos are indicated in the corresponding panels.
    The pixels are colour-coded by projected density, with lighter colours representing denser regions. The projected density is more spherical compared
    to the next example in figure~\ref{fig_SFRS_ex2}.}
  \label{fig_SFRS_ex1}
\end{figure}

\begin{figure}[h]
    \centering
    \begin{minipage}{0.45\linewidth}
        \centering
        \includegraphics[width=\linewidth]{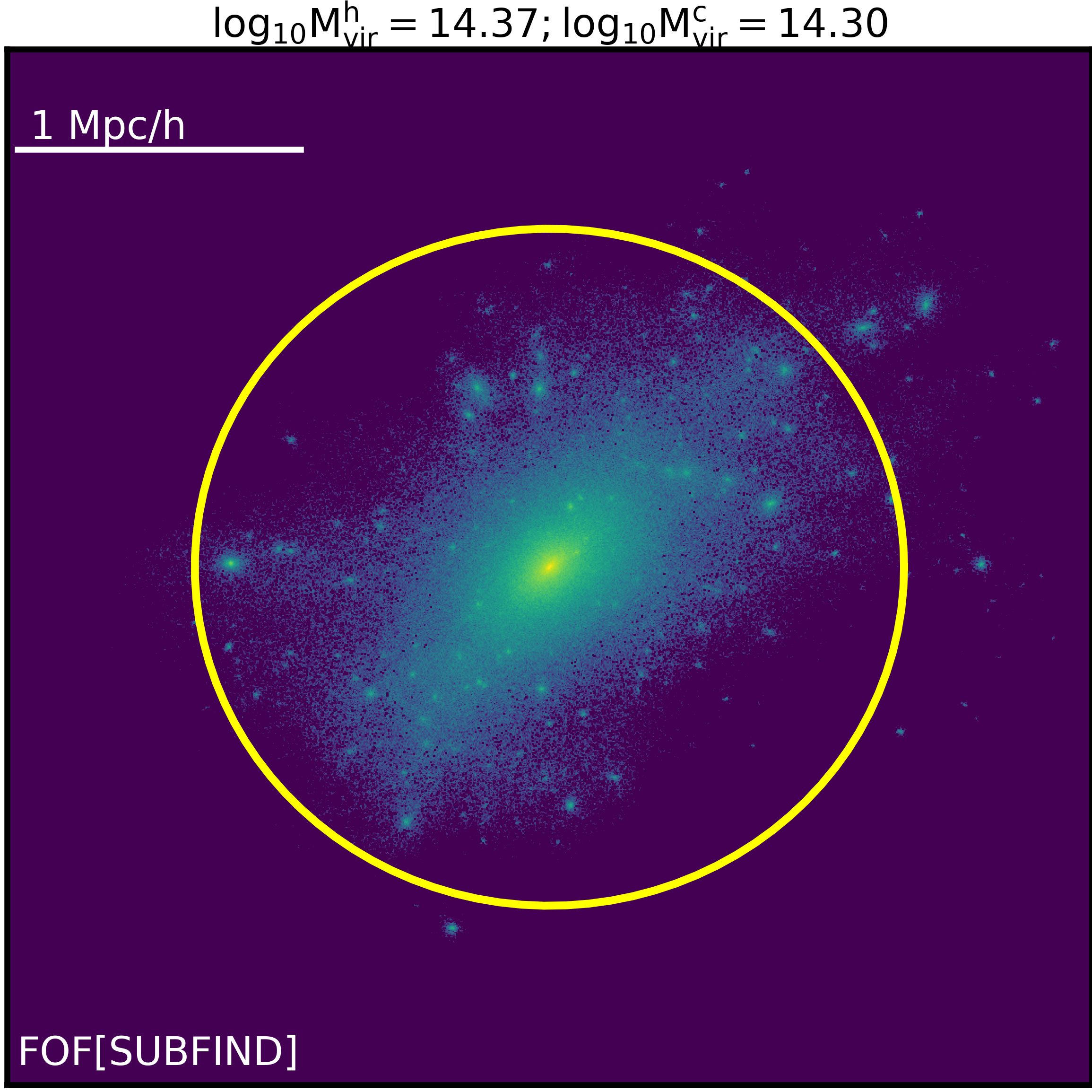} \\
        \includegraphics[width=\linewidth]{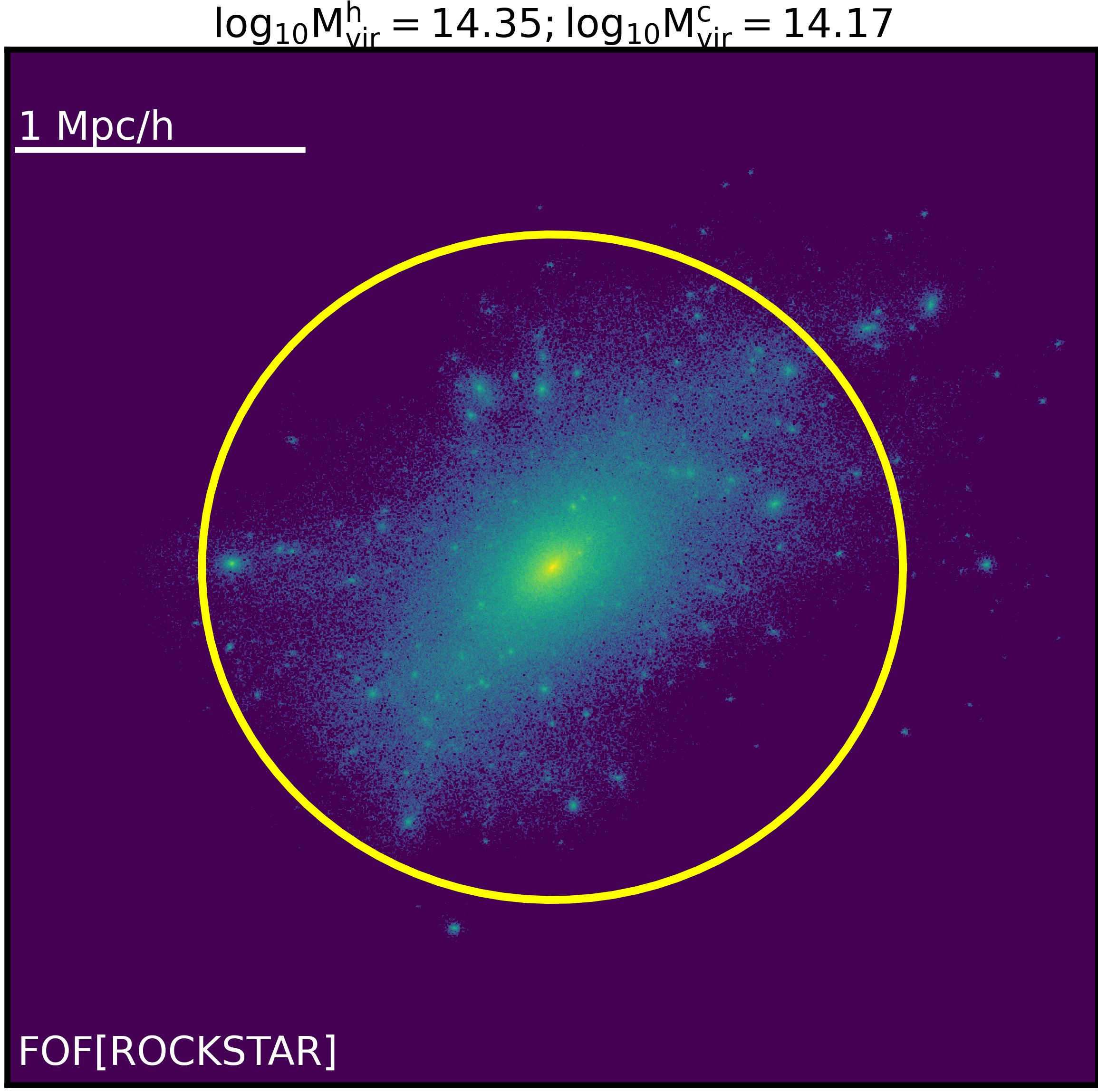}
    \end{minipage}
    \begin{minipage}{0.45\linewidth}
        \centering
        \includegraphics[width=\linewidth]{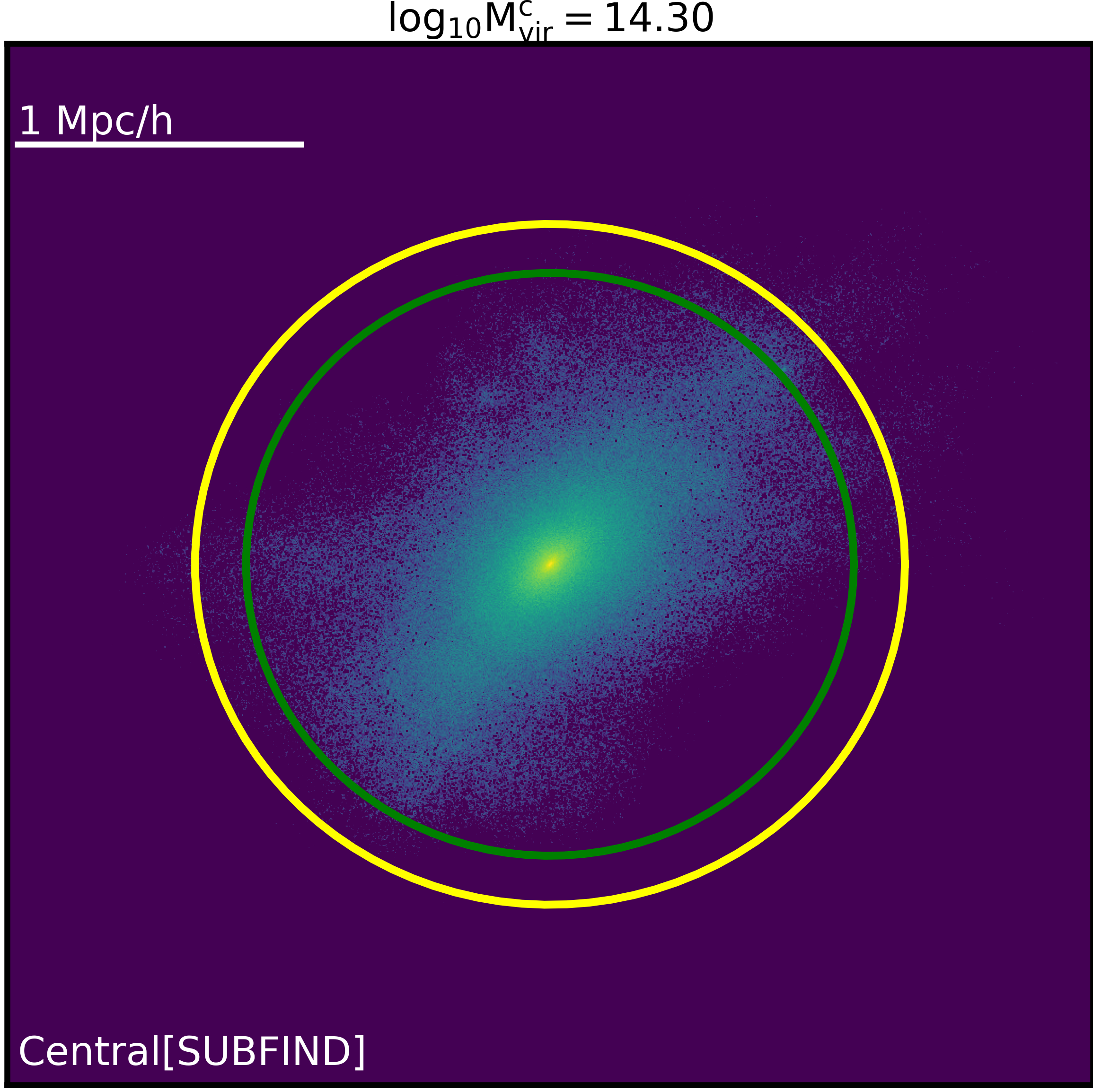} \\
        \includegraphics[width=\linewidth]{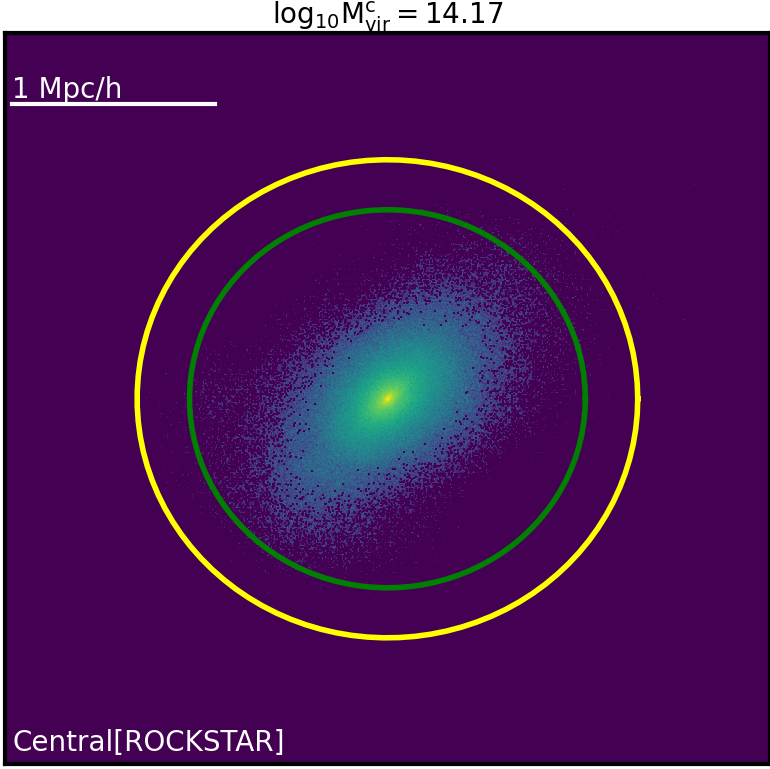}
        \end{minipage}
    \caption{Another example of a cluster sized halo but is more elliptical in compared to figure~\ref{fig_SFRS_ex1} in projection.
      The labels, colours, markers are the same as in figure~\ref{fig_SFRS_ex1}}  
    \label{fig_SFRS_ex2}
\end{figure}

In figures~\ref{fig_SFRS_ex1} and \ref{fig_SFRS_ex2} we show two examples of the matter distribution in
cluster sized halos $\Mvir \sim 10^{14}\msun/h$. The left panel is for the FOF halo and the right panel is for the central halo.
The top and bottom panels are for the \subfind and \rockstar algorithms. The yellow circle represents the virial radius of the FOF
halo and the green circle is the virial radius of the central halo. 
The virial mass, $\Mvir$, is defined as the mass within a sphere of radius $\Rvir$  whose density is $\Delta$ times the background density, $\bar{\rho}$, with
$\Delta = 18\pi^2 \approx 200 $ derived from the spherical collapse model using the formula\cite{1998ApJ...495...80B}. 
\begin{equation}
   \Mvir = \frac{4}{3} \pi \Rvir^3 \bar{\rho} \Delta
    \label{eq_mvir_def}
\end{equation}
The pixels are colour-coded by projected density with lighter
colours representing denser regions. Figure~\ref{fig_SFRS_ex1} has a more spherical distribution in projection
as compared to figure~\ref{fig_SFRS_ex2}. The left panels look the same for both \subfind and \rockstar (as it should be) but the differences
appear in the right panel, i.e. the central halo. At these masses \subfind central halos are more extended (and therefore more massive) compared
to \rockstar central halos consistent with the abundances seen in figure~\ref{fig_hmf_l128}.
The diffuse dark matter particles around satellites are retained in \subfind, but these do not appear in \rockstar.
The green circles, i.e. the virial radius of the central halo will be used to compute their shapes as we discuss in the
next section.

\section{Halo Shapes}
\label{sec:halo_shapes}
In this section we discuss the definitions and methods existing in the literature for computing halo shape.
We investigate  the requirement of minimum particle number in halo as well as the error tolerance criterion for computing the  shape of the halo accurately.

\subsection{Definitions \& Methods}
Different definitions exist to describe the triaxial-ellipsoidal matter distribution in halos.
The definitions \citep{2011ApJS..197...30Z} differ in the weights, $w_n$, that are assigned when defining the $3\times 3$ shape tensor, $S_{ij}$:
\begin{equation}
  S_{ij} = \sum_{n=1}^{N} \frac{x_{i,n}x_{j,n}}{w_n}
  \label{eq_shape_tensor}
\end{equation}
The summation in $n$, runs over all the $N$ particles associated with the halo, with $x_{i,n}$ denoting the $i^{\mathrm{th}}$ component of the coordinate vector
$\mathbf{x}$ of the $n^{\mathrm{th}}$ particle and $w_n$ is the weight associated with the $n^{\mathrm{th}}$ particle.
$(i,j)$ take values $(1,2,3)$ to denote the $(x,y,z)$ coordinates of particle $n$ with respect to the halo centre.
For this work, we consider the halo centre to be the location of the potential minimum in the halo, i.e. the central halo.

Diagonalising the shape tensor,  $S_{ij}$ we obtain the eigenvector ($\hat{e}_a,\hat{e}_b,\hat{e}_c$) which 
represent the directions of the three major axes which gives us the orientation of
the halo and the square root of the corresponding eigenvalues 
($\sqrt{\lambda_a} (\equiv a) \geq \sqrt{\lambda_b} (\equiv b) \geq \sqrt{\lambda_c} (\equiv c) $) 
correspond to the length of the axes.
Alternately a triaxial ellipsoid is characterized by the axis ratios $\left(q=\frac{b}{a},s=\frac{c}{a}\right)$.
The halo approaches a spherical shape in the limit $q \rightarrow 1, s \rightarrow 1$.  
The weight $w_n=1$ corresponds to the
unweighted shape tensor and the weights $w_n=(r_n)^2$ or $w_n=(r_{n,ell})^2$ to the weighted (or reduced) tensor.
We refer to these three definitions of the shape tensor as $D_1$, $D_2$, \textrm{and} $D_3$ respectively. 
Here $(r_n)^2=(x_n)^2 + (y_n)^2 + (z_n)^2$ is called spherical radius
and $(r_{n,ell})^2= (x_n)^2 + (y_n)^2/q^2 + (z_n)^2/s^2 $ is called the elliptical radius,
in the eigenvector coordinate system  $(\hat{e}_a,\hat{e}_b,\hat{e}_c)$, with the origin at the halo centre.

There are two methods in computing the shape (determining the axis ratios $(q,s)$)
and orientation, (determining the eigenvectors of $S_{ij}$, $(\hat{e}_a,\hat{e}_b,\hat{e}_c)$), of the halo.
Both methods \citep{2006MNRAS.367.1781A} involve iteratively converging to the axis ratios and orientation of the halo
given the shape tensor for a given tolerance.

In the first approach, the major axis $a$ is kept fixed to $\Rvir$ and other axes
are scaled $\left(b=\sqrt{\frac{\lambda_2}{\lambda_1}}a,c=\sqrt{\frac{\lambda_3}{\lambda_1}}a\right)$ at every iteration. The iteration begins
by considering all particles of the halo under consideration (central halo in this work) within $\Rvir$ (the virial radius of the central halo).
We refer to this set as the master set of particles. The eigenvalues and eigenvectors of $S_{ij}$ are then obtained using this master set.
The axes $(a,b,c)$ are rescaled as described above and the volume and boundary of the ellipsoid are determined using the rescaled axes
and the eigenvectors $(\hat{e}_a,\hat{e}_b,\hat{e}_c)$. For every subsequent 
iteration, only particles from the master set that lie within the ellipsoid are used in recomputing
the axes and orientation. The eigenvalues are up to a constant and rescaling them does not affect the results nor their
ratios $(q,s)$. The iteration is stopped when a tolerance in both axis ratios is achieved. This tolerance also guarantees
a tolerance in the eigenvectors and hence the orientation and boundary of the halo deforms from a sphere
and converges to an ellipsoid. We refer to this approach as the major-axis fixed, $A_{\textrm{fix}}$, method.

In the second method \cite{2012JCAP...05..030S} we rescale all the axes by  $\frac{\Rvir}{(abc)^{1/3}}$
at every iteration. This guarantees that the volume of the ellipsoid is kept fixed to the volume
of the sphere of virial radius $\Rvir$ at every step. As in the first method, the master set of particles within the ellipsoid is
used at every step to determine the shape tensor and the iteration is stopped when a tolerance in determining
the axis ratios is achieved. We refer to this method as the volume-fixed, $V_{\textrm{fix}}$, method.

\begin{figure}[h]
  \centering
  \includegraphics[width=0.95\linewidth]{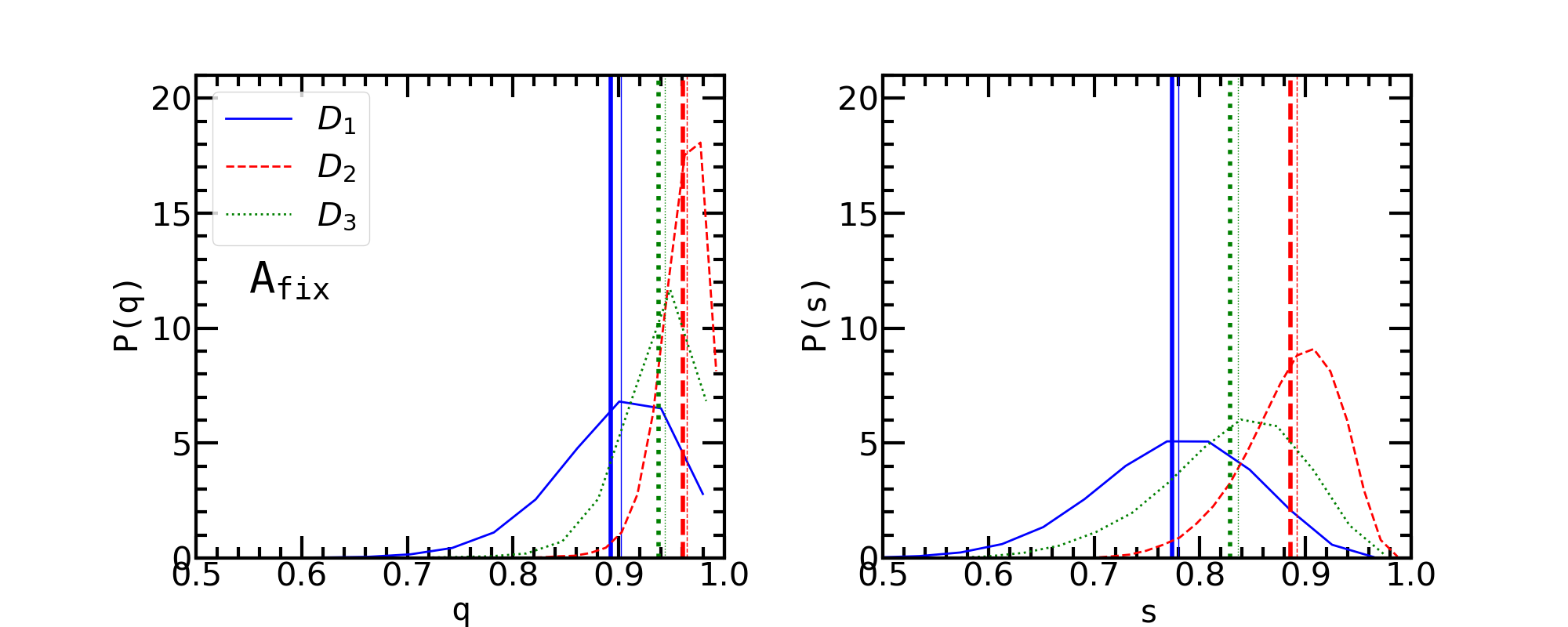}\\
  \includegraphics[width=0.95\linewidth]{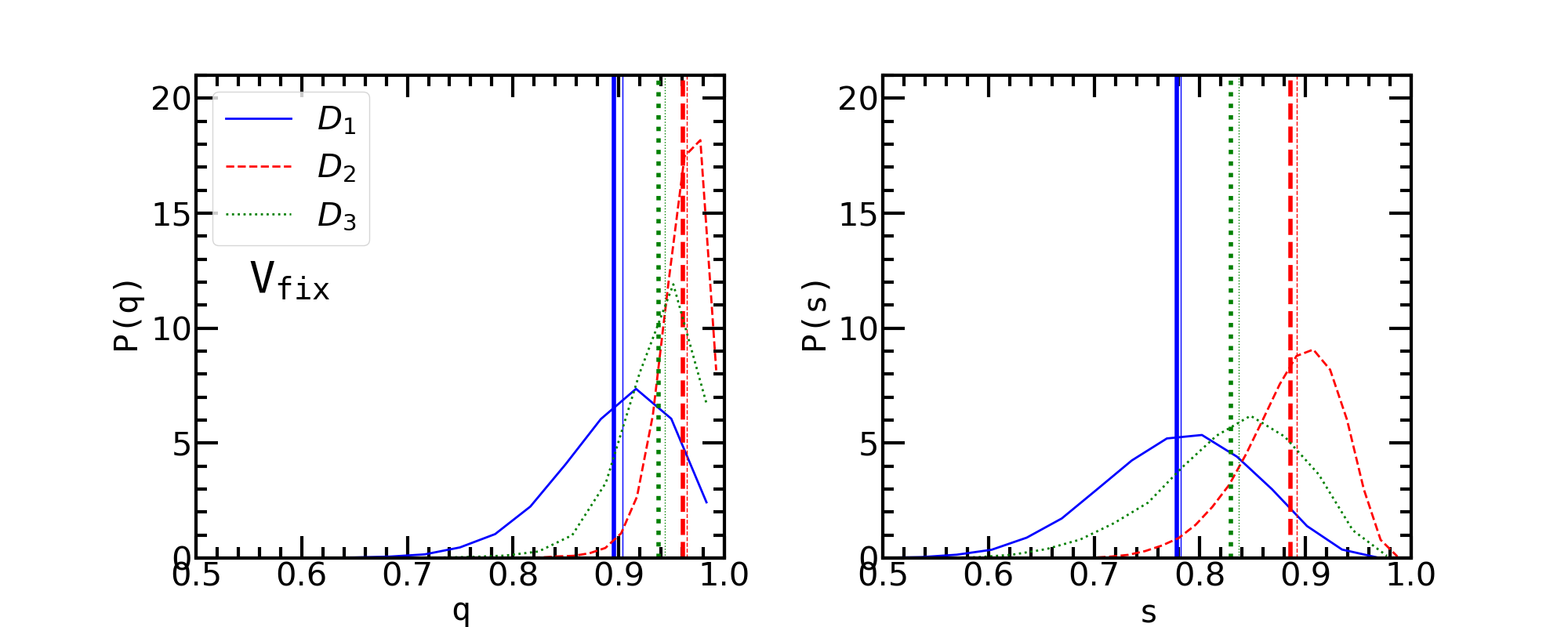}
  \caption{The figure shows the distribution $P(q)$ and $P(s)$ of axis ratios $q$ and $s$ in the left and right panels for the $D_1$ (unweighted), $D_2$ ($r^2$-weighted) and
    $D_3$ ($r^2_{\textrm{ell}}$-weighted) definitions of the shape tensor represented in the solid, dashed and dotted lines respectively.
    This is done for the scale-free $n=-1.0$ model at $x_{nl}=49.4$
    for halos in the mass range of $1000-10000$ particles. There are about 60,000 halos in this mass range.
    The top row is for the $A_{\textrm{fix}}$ (major-axis fixed) method and the bottom row is for the $V_{\textrm{fix}}$ (volume fixed) method.
    The vertical lines are the mean (thick) and median (thin) of the respective distributions.}
  \label{fig_shape_definition}
\end{figure}

In Figure \ref{fig_shape_definition} we look at the distribution $P(q)$ (left) and $P(s)$ (right) for the three definitions, $D_1$ (solid),
$D_2$ (dashed) and $D_3$ (dotted), of the shape tensor, $S_{ij}$. The top row is for the  $A_{\textrm{fix}}$ (major-axis fixed) method and the bottom row
is for the  $V_{\textrm{fix}}$ (volume fixed) method.  The vertical lines are the mean (thick) and median (thin) of the respective distributions. This is done
for the scale free model $n=-1.0$ at $\xnl = 49.4$ for halos in the mass range $10^3 - 10^4$ particles. There are
about 60,000 halos in this mass range.

Comparing the methods (top and bottom rows) we find that they give near identical results. The unweighted definition are less skewed as compared
to the weighted definitions and the locations of the peak, median and mean of the unweighted definition occur at lower values of $q$ and $s$
compared to the weighted definitions. The spread in
the unweighted definition is however larger compared to the weighted definitions. Comparing the weights in the unweighted and weighted distributions
in equation~\ref{eq_shape_tensor} the shape in the weighted case is dominated by particles closer to the halo centre when compared to the unweighted case.
Since we are computing shapes with particles in the enclosed volume, the weighted definition is  more appropriate compared to the unweighted case
\cite{2006MNRAS.367.1781A,2011ApJS..197...30Z, 2019MNRAS.484..476C}.

Spherically averaged density profiles of halos  can be approximated to a two-parameter 
Navarro-Frenk-White (NFW) \cite{1996ApJ...462..563N, 1997ApJ...490..493N} profile given by
\beq
\rho(r) = \frac{\rho_0}{\frac{r}{r_s} \left( 1 + \frac{r}{r_s}\right)^2},
\label{eq_nfw}
\eeq
where $r_s$ is the scale radius and $\rho_0$ describes the amplitude of the profile.
One can use this profile to estimate the half mass radius.
The fractional mass of the halo as a function of fractional radius $x=r/\Rvir$ is given by:
\beq
\frac{M(<r)}{\Mvir} = \frac{\int_0^r \rho(r) 4\pi r^2 dr}{\int_0^{\Rvir} \rho(r)4\pi r^2 dr} = \frac{\left[ \ln(1+xc) -\frac{xc}{1+xc}\right]}{\left[ \ln(1+c) -\frac{c}{1+c}\right]},
\label{eq_mfrac_nfw}
\eeq
where the concentration parameter $c$ is defined as $\Rvir = cr_s$. For a $\Lambda$CDM model at $z=0$ halo concentrations range from $c \sim 10-4$ for halos in the
mass range $\Mhalo \sim 10^{11} - 10^{14} \msun/h$ \citep{2016MNRAS.457.4340K}. Plugging these numbers in equation~\ref{eq_mfrac_nfw} we find
that the half-mass radius ranges from $0.36 \Rvir$ to $0.45\Rvir$ in the mass range  $\Mhalo \sim 10^{11} - 10^{14} \msun/h$. This suggests that
when considering the shape of halos with particles enclosed in a volume choosing the weighted definition is more appropriate since the majority of the particles
are closer to the centre. On the other hand the unweighted definition is more appropriate if we
were to measure the shape in radial bins \cite{2011ApJS..197...30Z,2019MNRAS.484..476C}. Given these considerations we will consider
the $D_3$  definition of $S_{ij}$ and the  $\Mfix$ method to determine the shape of the halo.
Since the distributions $P(q)$ and $P(s)$ are skewed we will use the median and percentile spread to describe these distributions.
Our  results do not change qualitatively if we use the mean and rms. We have also checked
that neither the definition nor method change the qualitative behaviour of results: i.e a universal curve describes the evolution of  $P(q)$ and $P(s)$.
A number of studies have used one of the three definitions and the two methods outlined above
\citep{2006MNRAS.367.1781A, 2011ApJS..197...30Z,2012JCAP...05..030S,2019MNRAS.484..476C,2014MNRAS.443.3208D, 2014MNRAS.441..470T, 2015MNRAS.448.3522T, 2015MNRAS.453..469T,
  2021MNRAS.503.2053R, 2022MNRAS.516.5849R}. 

\subsection{Resolution \& Convergence Criteria}
In this section we investigate the resolution criteria in accurately determining the halo shape. Instead of determining the shape on an object to object basis or
on analytical (or idealised) halo profiles \citep{2006MNRAS.367.1781A,2011ApJS..197...30Z}, we will look for statistical convergence based on the simulations itself,
i.e. the convergence on the distributions
$P(q)$ and $P(s)$. We also explore the tolerance required for these distributions to converge.
For this section we again use the halo catalogue of the scale-free run $n=-1.0$ at $\xnl = 49.4$.
In order to obtain the distribution of shape parameters accurately we
consider the mass range of halos which have $\sim$ thousand to $\sim$ ten thousand particles instead of larger mass halos, which are better resolved but are much fewer in number.

\begin{figure*}[t]
    \centering
    \includegraphics[width=0.95 \linewidth]{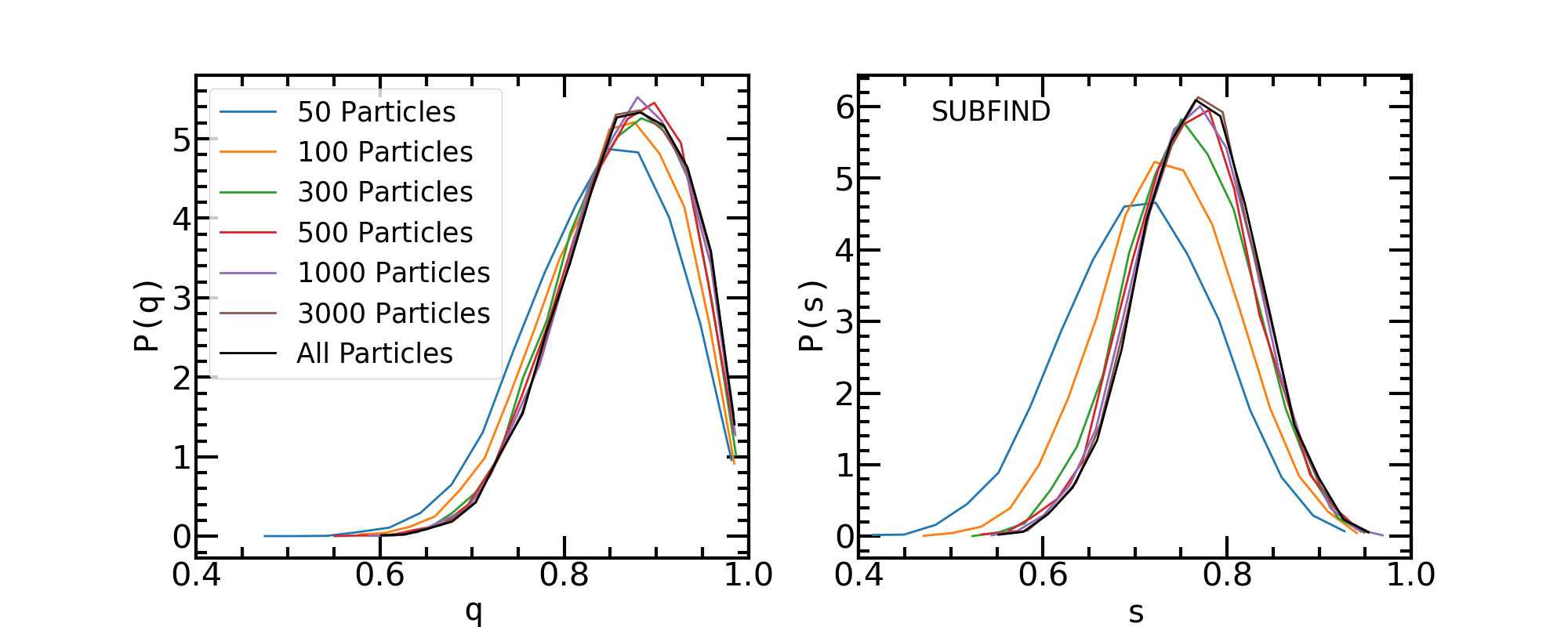}
    \includegraphics[width=0.95 \linewidth]{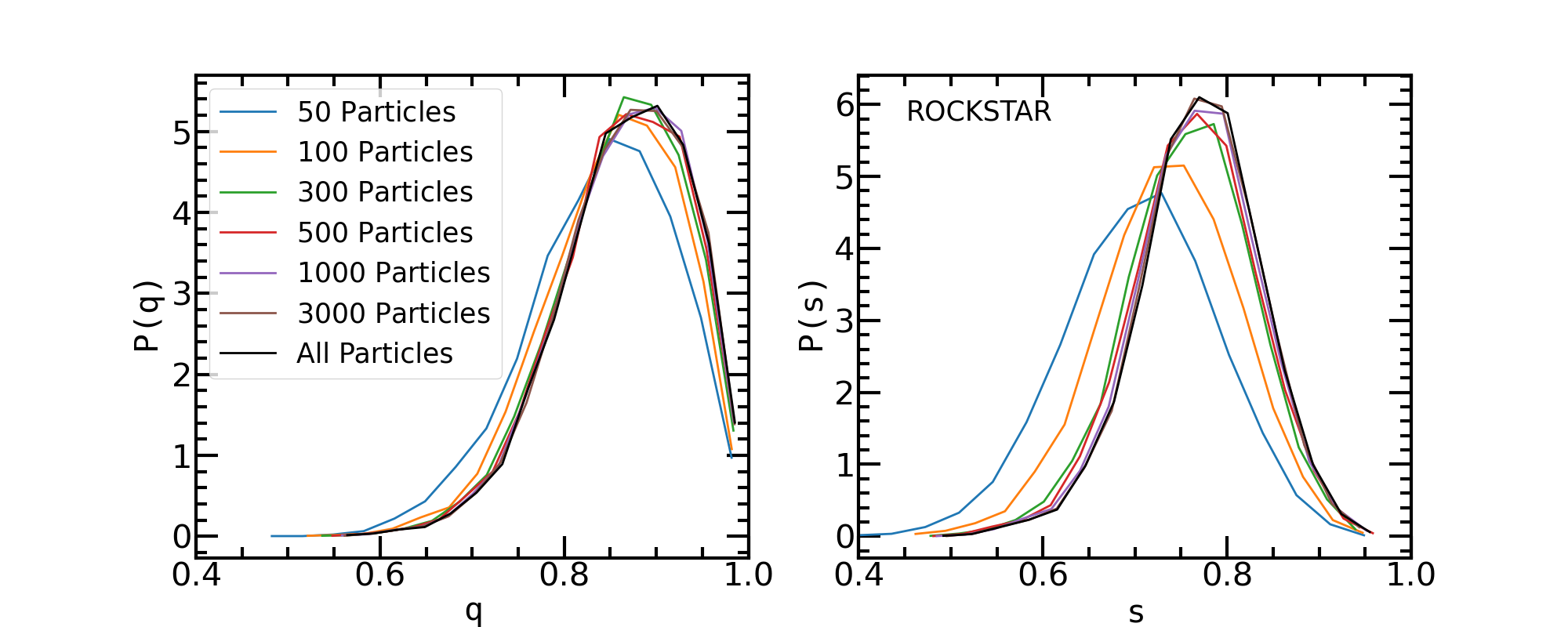}
    \caption{Normalized distribution of the axis ratios, $P(q)$(left) and $P(s)$(right) for the scale-free $n=-1.0$ run at $\xnl=49.4$ for halos with $5000-10000$ particles.
      Shape measurements are done with the the number of down-sampled particles in halos  $N_{\textrm{ran}} = (50,100,300,500,1000,3000)$ . This
      is shown as the blue, orange, green, red, purple, brown curves respectively. The black curve is computed by considering all particles. The top and bottom rows
      represent the results for \subfind and \rockstar.}
    \label{fig_convergence_random}
\end{figure*}

The first test is based on randomly sampling particles in these well resolved halos (5000-10000 particles) and looking at how the distribution $P(q)$ and $P(s)$
converge as the number of down-sampled particles increase.  
We consider shape measurements when the number of down-sampled particles in halos are $N_{\textrm{ran}} = (50,100,300,500,1000,3000)$. 
Comparison is made with shape measurements using all  particles in halos. Our results are shown in figure~\ref{fig_convergence_random}.
The left and right columns are for $P(q)$ and $P(s)$ and the top and bottom rows are for \subfind and \rockstar.
The different colours represent the curves for  $N_{\textrm{ran}}$. We look for convergence by comparing with the shape measurements
considering all particles (black curve). We find that there is good convergence for  $N_{\textrm{ran}} \geq 500$ consistent with  \cite{2014MNRAS.441..470T} however
a more conservative value $N_{\textrm{ran}} \geq 3000$ gives excellent convergence.

\begin{figure}[!htb]
    \centering
    \includegraphics[width=0.827\linewidth]{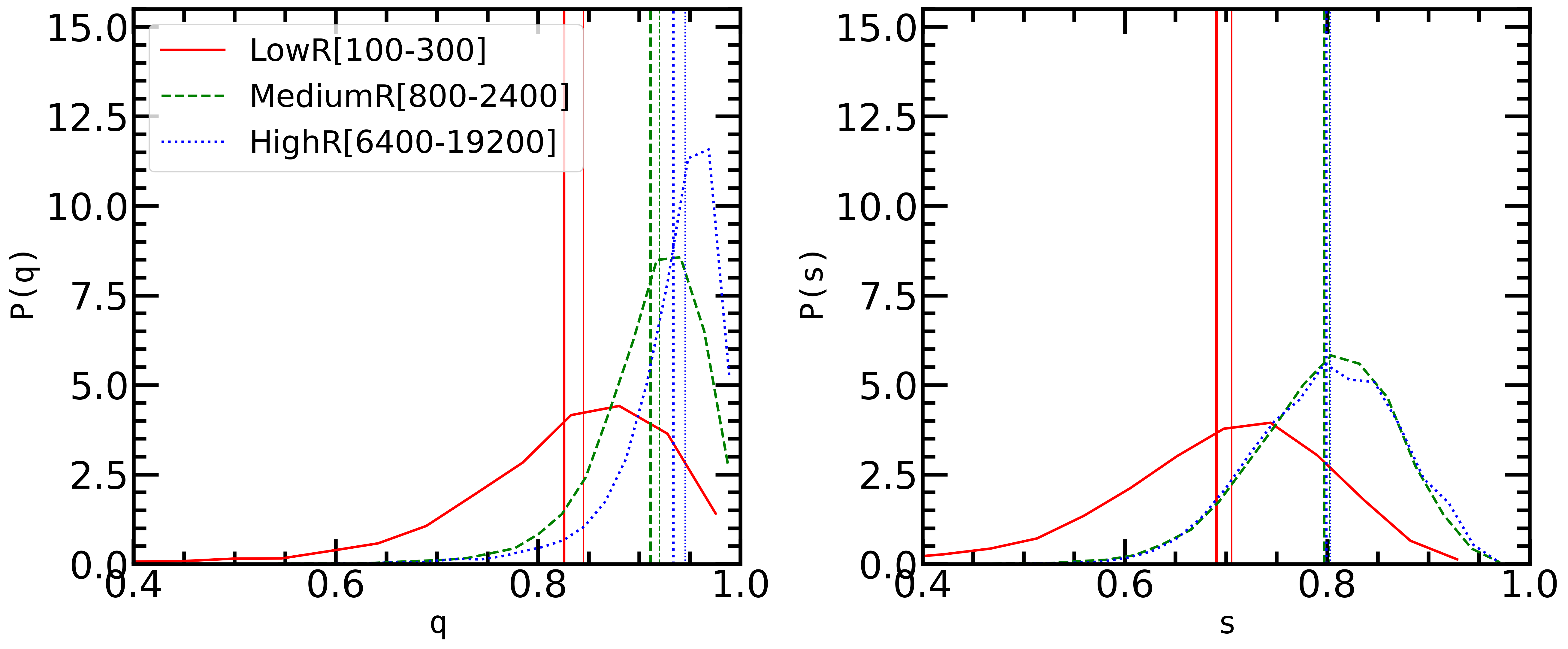}
    \caption{The figure shows the distributions  $P(q)$ (left) and $P(s)$ (right) using \subfind, from the low (solid) , medium (medium) and high (dotted)
      resolution simulations for $n=-1.0$ at the final output, $\xnl=49.4$. These were run with $256^3,512^2,1024^3$ particles respectively. The  particle (or mass) range
      used to compute the distributions are indicated in the plot. The thick and thin vertical lines represent the mean and median of the distributions. }
    \label{fig_res_lmh}
\end{figure}

In the second test we look at a low and medium resolution run of the same scale-free $n=-1.0$ simulation at $\xnl=49.4$. We do this by running the low and medium resolution simulations with
$\Npart = 256^3$ and $\Npart = 512^3$ particles. This time we consider halo masses in the high resolution run with $6400-19200$ particles. This particle (or mass) range
is equivalent to $800-2400$ particles in the medium resolution run and $100-300$ particles in the low-resolution run.
In Figure \ref{fig_res_lmh}, we have plotted the distribution of axis ratios for the low (solid), medium (dashed) and high (dotted) resolution runs.
The thick and thin vertical lines represent the mean and median of the distributions. 
The distribution has reasonable convergence when comparing the medium and high resolution simulations. The medium and high resolution runs give shapes
that are more spherical compared to the low resolution run. 
This analysis suggests that we need a few thousand particles to get a reasonable convergence of the axis ratios.

\begin{figure}[tbp]
    \centering
    \includegraphics[width=\linewidth]{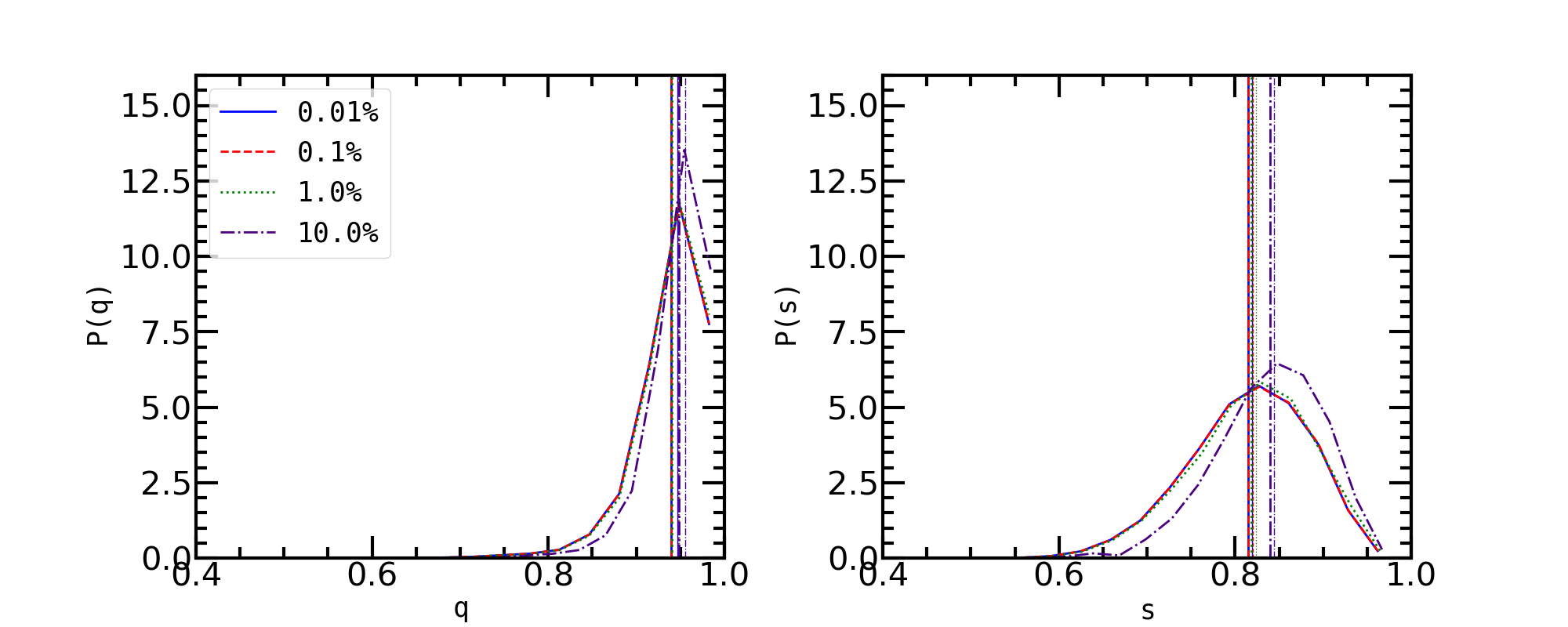}
    \caption{Axis ratio  distributions  $P(q)$(left) and  $P(s)$ (right) in the mass range of $3000-10000$ particles at the final output $\xnl=49.4$ for scale-free model $n=-1.0.$.
      The solid, dashed, dotted and dot-dashed lines represent the results for a tolerance of 0.01\%, 0.1\%, 1\%, 10\% respectively.  The thick and thin vertical lines are mean and median values.}
    \label{fig_tolerance}
\end{figure}

$q$ and $s$ values are limited to the $[0,1]$  range, with typical median values ranging from 0.6 to 0.95 as can be seen in the particular
examples of this section in figures~\ref{fig_shape_definition},\ref{fig_convergence_random} and \ref{fig_res_lmh}.
The spread $\sigma_q$ and $\sigma_p$ in their distribution ranges from 0.05 to 0.1 which is about 10\% of the median values as will be shown in figure~\ref{fig_scalefree_universal_all}
and \ref{fig_scalefree_universal_otp}..
We therefore expect that a  tolerance of less than 1\% will not affect our estimate of the distributions of $q$ and $s$.
In figure~\ref{fig_tolerance} we look at the convergence of the distributions $P(q)$ and $P(s)$ for a tolerance of 0.01\%, 0.1\%, 1\%, 10\% in both $q$ and $s$ measurements.
This is done for the scale-free run for $n=-1.0$ at $\xnl = 49.4$.
This is shown as the solid, dashed, dotted and dot-dashed curves respectively. The thick and thin vertical lines are mean and median values.
As expected the distributions converge for a tolerance $\leq 1\%$. Even for a 10\% tolerance, there is reasonable agreement with respect to the more stringent tolerance
values.

For the rest of the paper, we will use a minimum 3000 particles in a halo and a 1\% tolerance in $q$ and $s$ for our analysis. These numbers are broadly consistent with
\cite{2006MNRAS.367.1781A,2012JCAP...05..030S,2014MNRAS.441..470T,2021MNRAS.503.2053R}
although a higher number is required for radial shape computations \cite{2002ApJ...574..538J, 2011ApJS..197...30Z, 2019MNRAS.484..476C}.

\section{Results I: Scale-Free Models}
\label{sec:results1}
In this section, we study the evolution of the halo shape parameters ($q,s$) as a function of the rescaled mass $\left(\frac{M}{\Mnl}\right)$ and the equivalent
variable, the peak height $\left(\nu=\frac{\delta_c}{\sigma(M,z)}\right)$, across all scale-free models ranging from $n=-1.0$ to $n=-2.2$
listed in table~\ref{tab_scalefree}. Halos can be further classified as oblate(O), triaxial(T) or prolate(P) based on their their triaxiality parameter \cite{1991ApJ...383..112F} $\tau$:
\begin{equation}
    \tau = \frac{1-q^2}{1-s^2}
    \label{eq_triaxiality_def}
\end{equation}
An ellipsoid is said to be oblate ($a\approx b > c $) if \,\,\,$0<\tau<1/3$,
triaxial($a>b>c$) if \,\,\,$1/3 <\tau< 2/3$ and prolate ($ a>b\approx c$) if \,\,\,$2/3<\tau<1$.
We use the nomenclature OTP, to describe the triaxiality classification of halos which can be one of: oblate, triaxial or prolate.
We present results for the evolution of shape parameters for all halos and separately for the OTP halos. 
The time variable is $\Mnl$ which is equivalent to $\xnl$ (equation~\ref{eq_xnlofa})
and has been chosen from the range listed in table~\ref{tab_scalefree}.
For a fixed mass $M$, both $(M/\Mnl)$ and $\nu$
decrease as the scale factor $a$ (or time) increases.
In the context of the spherical collapse model, a value of   $\left(\frac{M}{M_{nl}}\right) \simeq 1$
or  $\left(\nu=\frac{\delta_c}{\sigma(M,z)}\right) \simeq 1.7$ represents the largest mass scale
which is turning around and collapsing at a particular epoch.

\subsection{Self-similar Evolution of Axis ratios}
\begin{figure*}[h]
    \minipage{0.48\textwidth}
    \includegraphics[width=\linewidth]{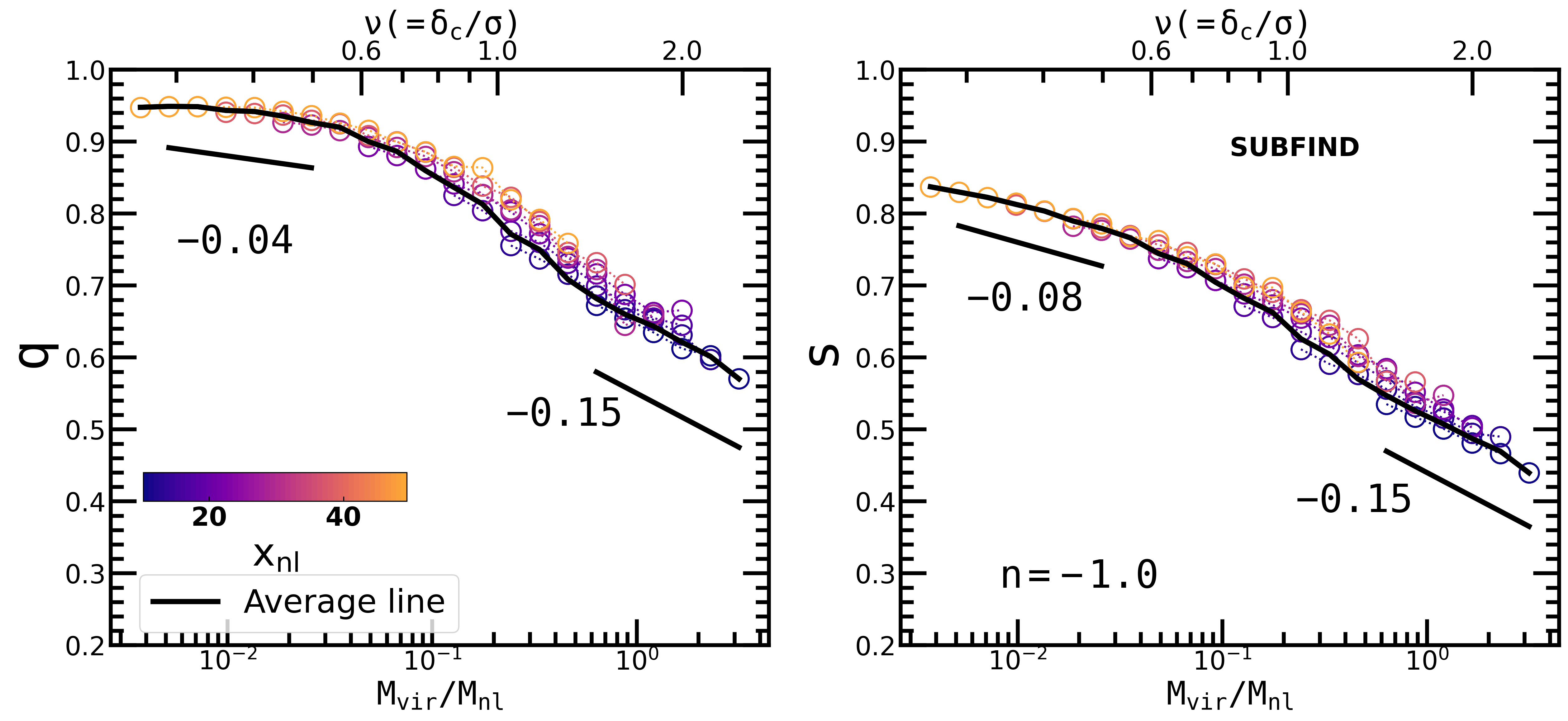}
    \includegraphics[width=\linewidth]{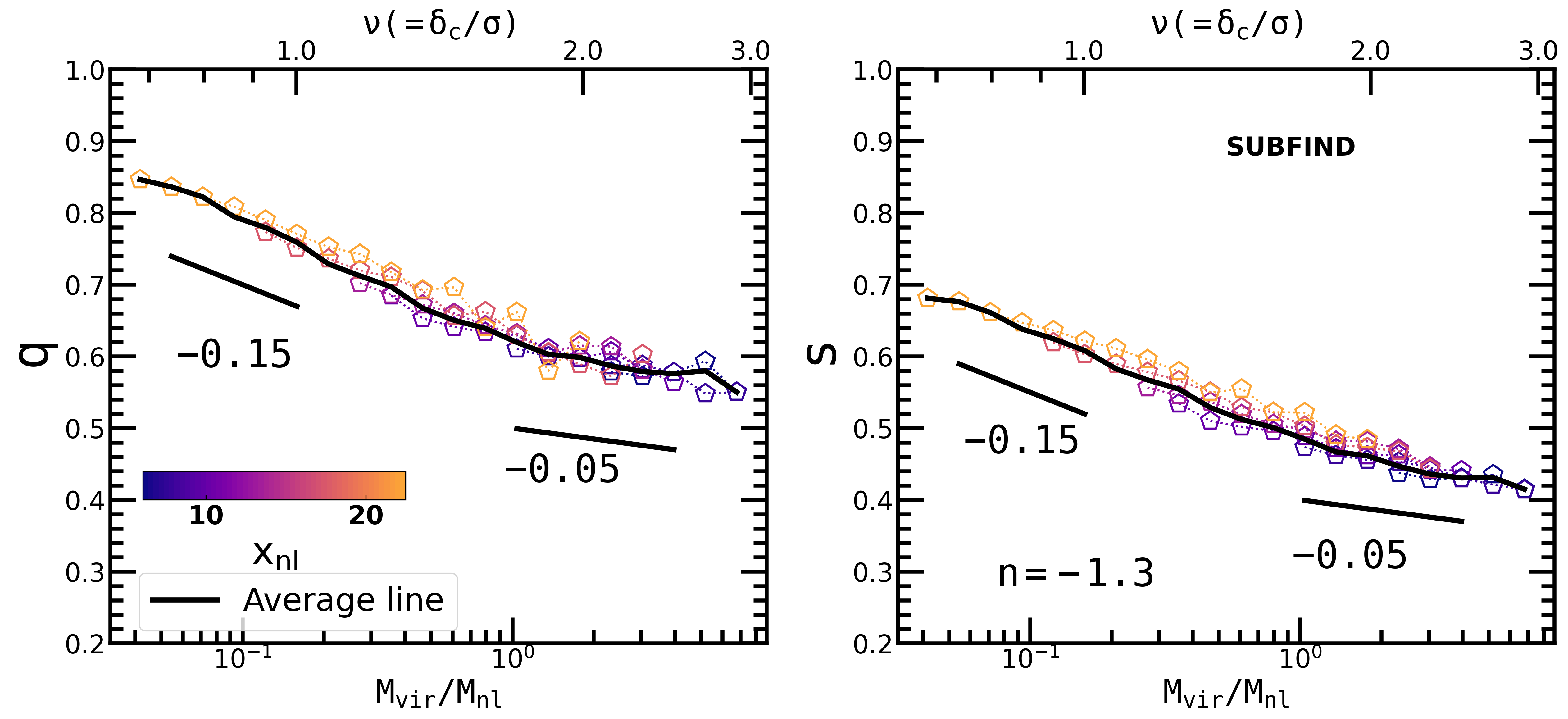}
    \includegraphics[width=\linewidth]{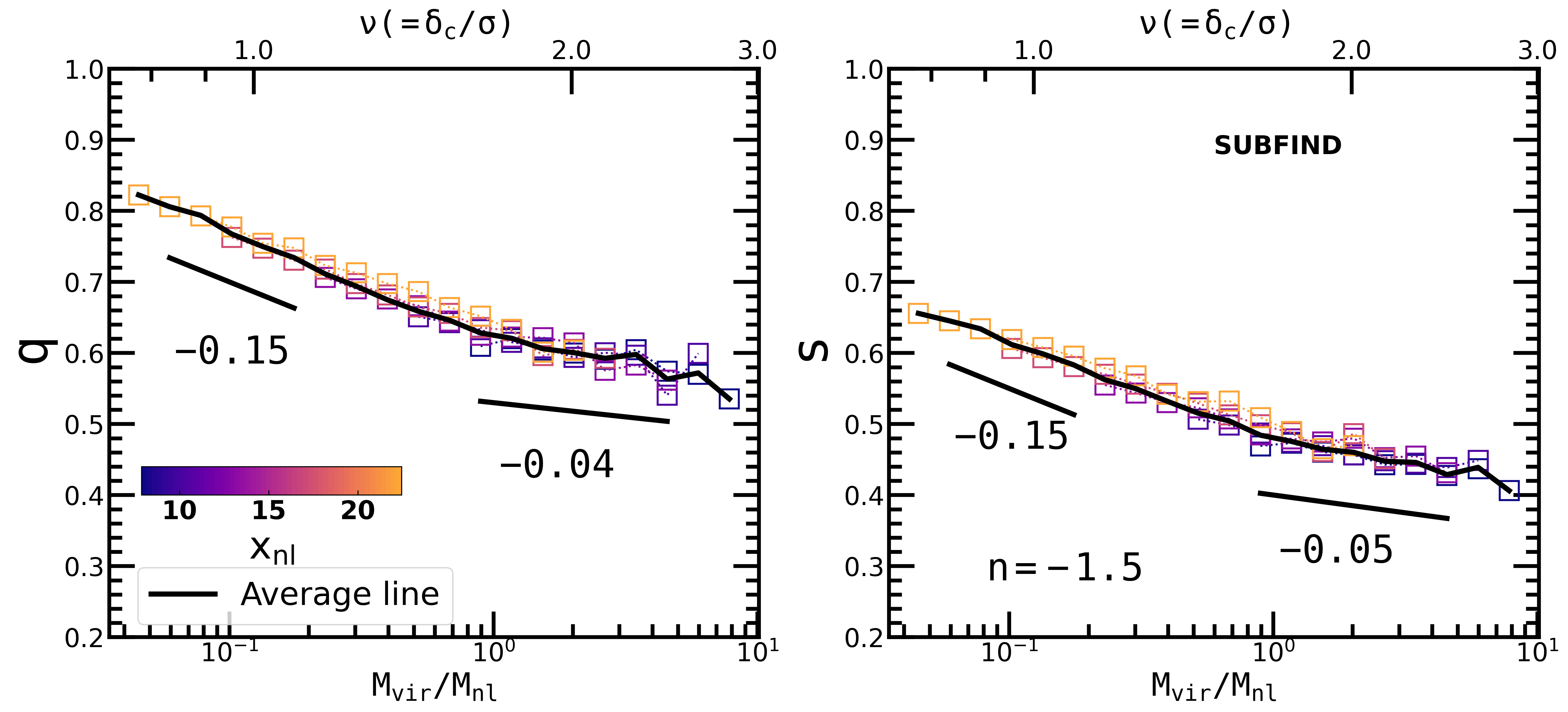}
    \includegraphics[width=\linewidth]{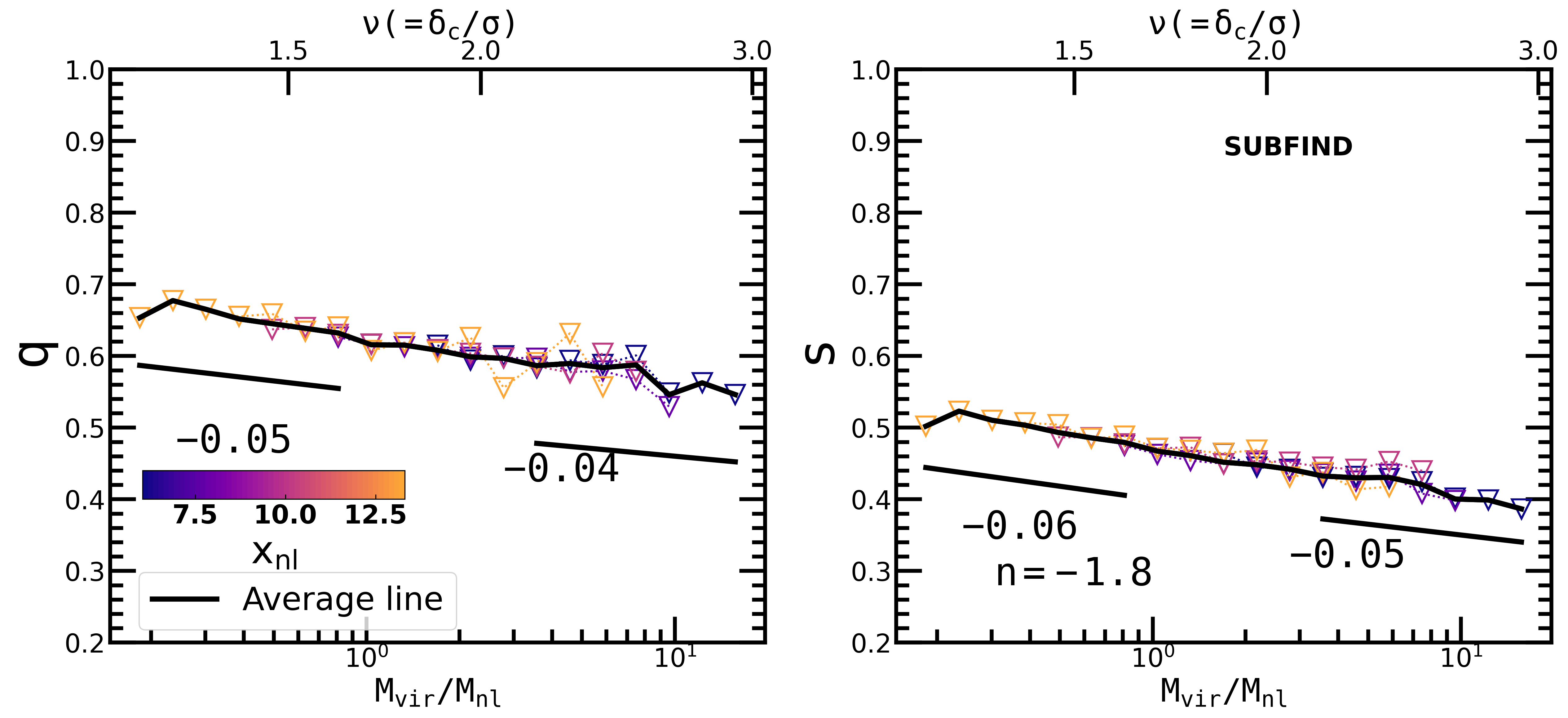}
    \includegraphics[width=\linewidth]{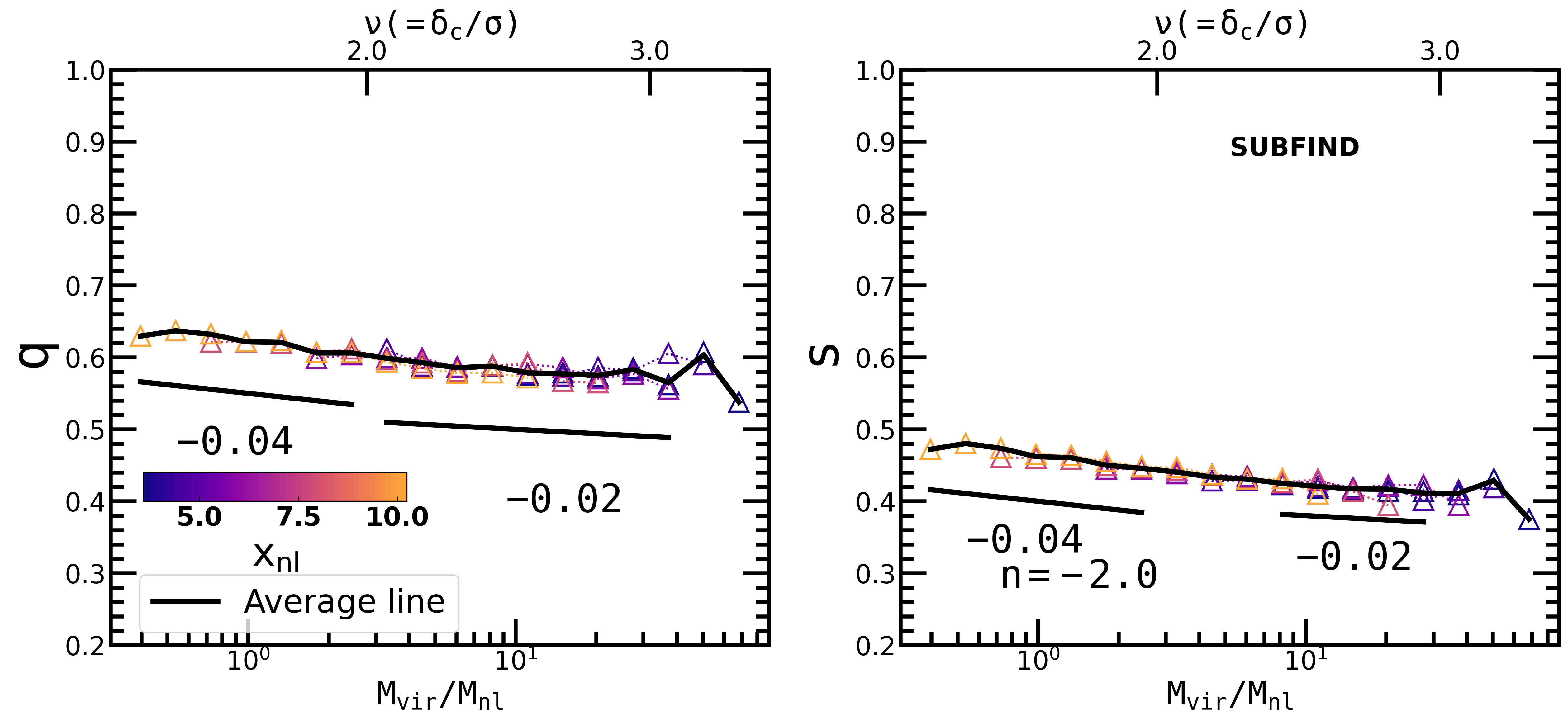}
    \includegraphics[width=\linewidth]{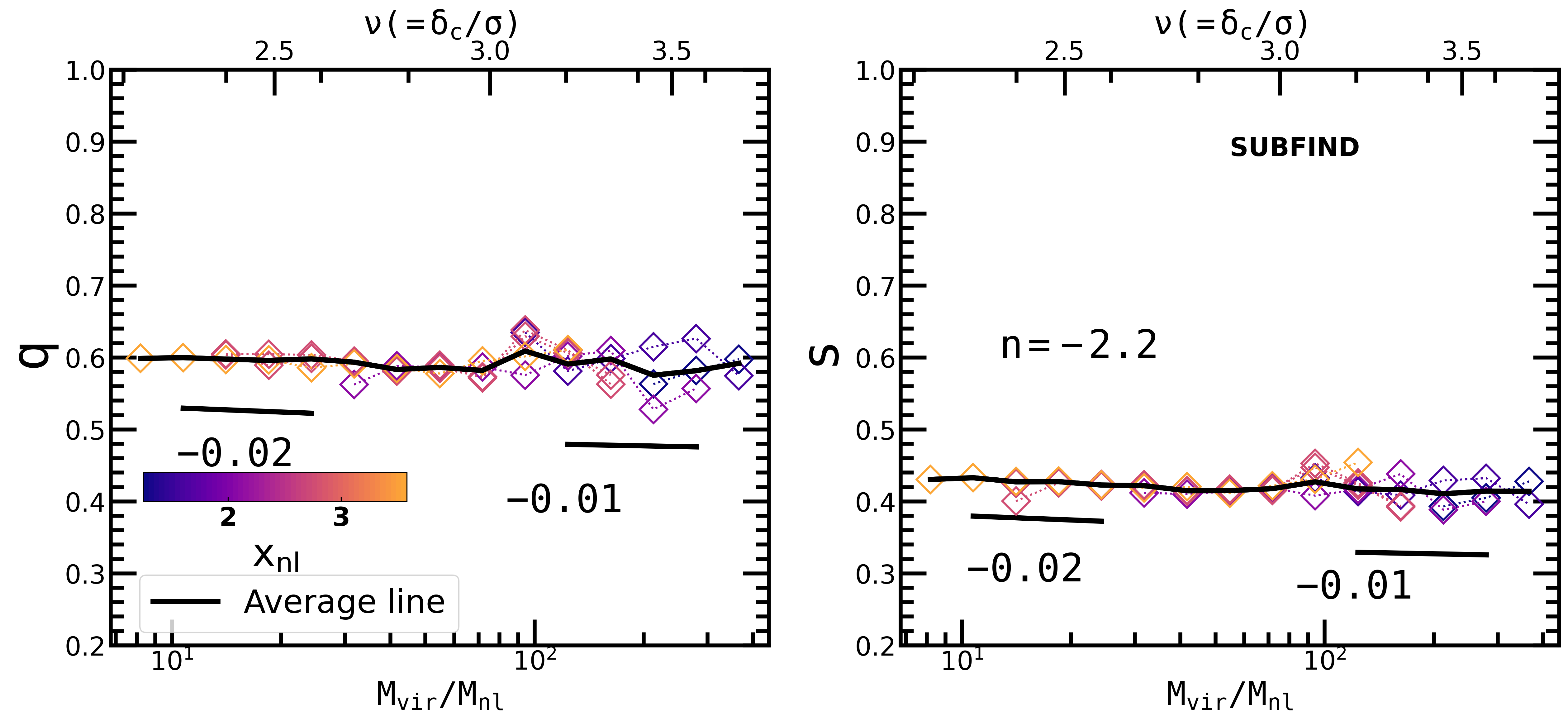}
    \endminipage\hfill
    \minipage{0.48\textwidth}
    \includegraphics[width=\linewidth]{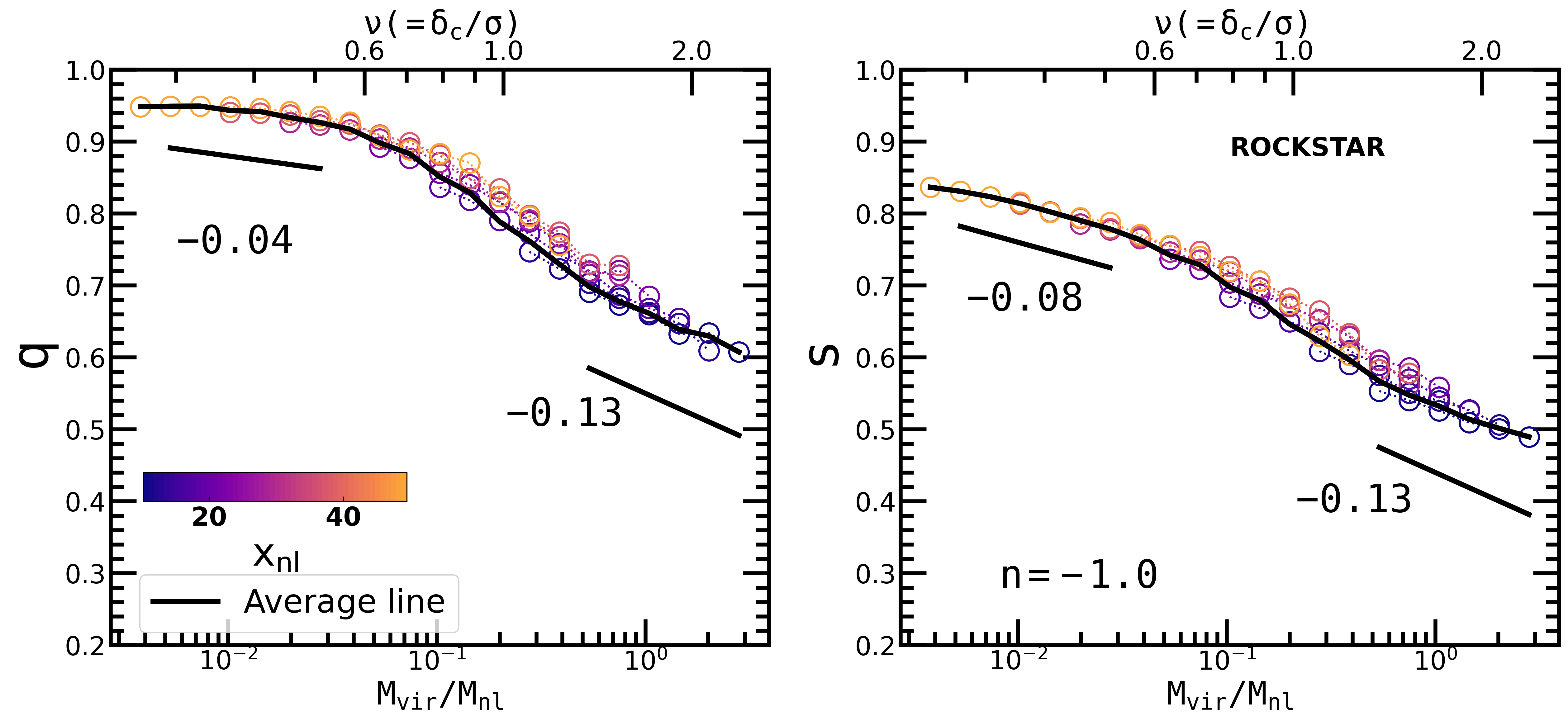}
    \includegraphics[width=\linewidth]{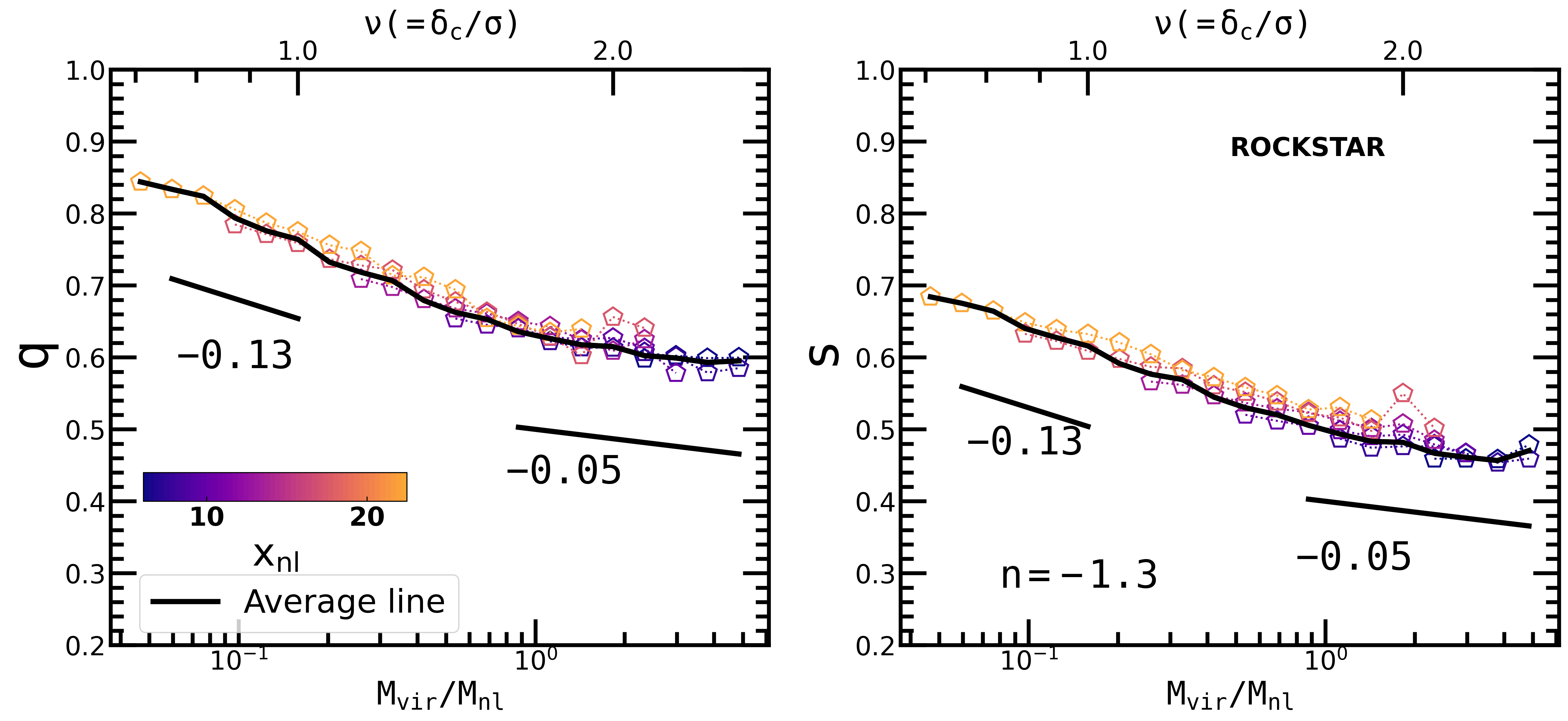}
    \includegraphics[width=\linewidth]{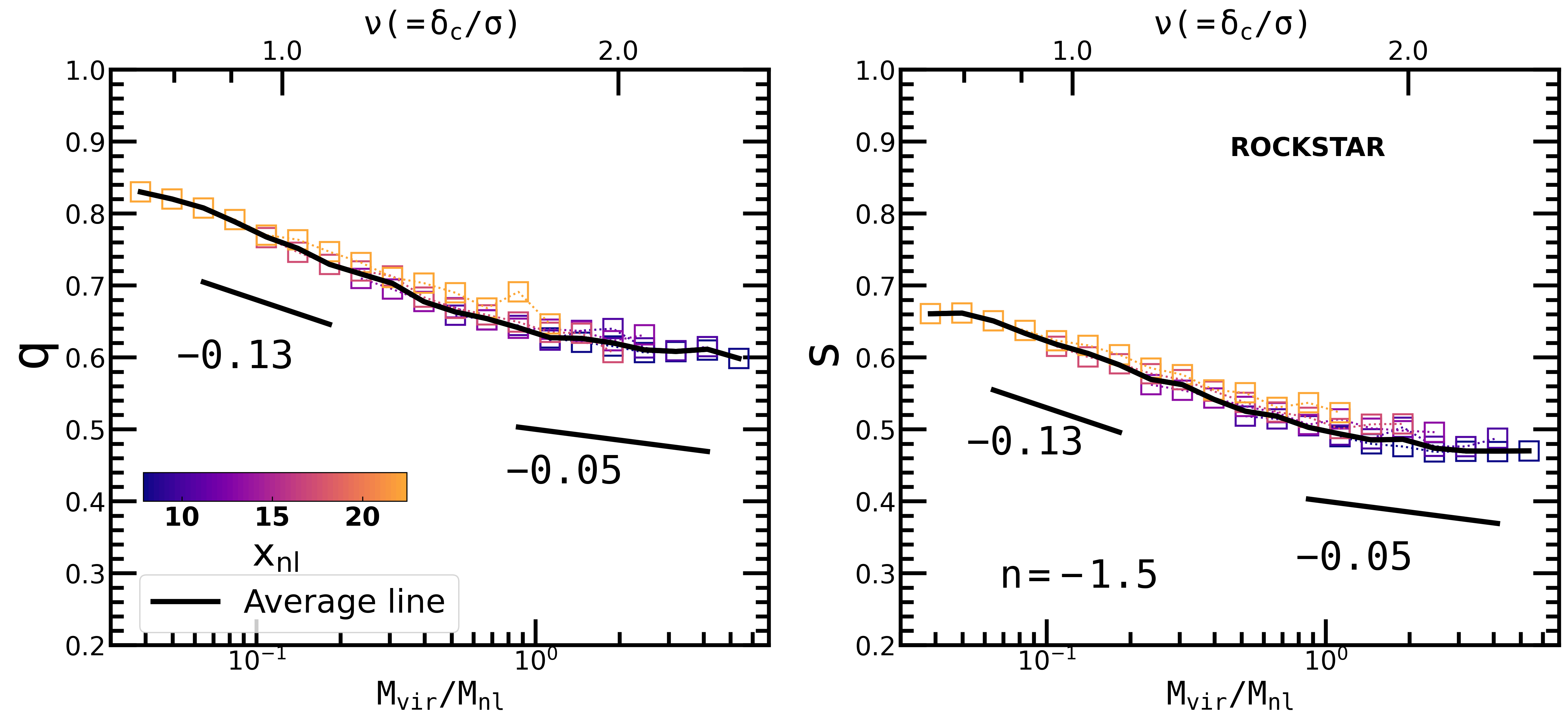}
    \includegraphics[width=\linewidth]{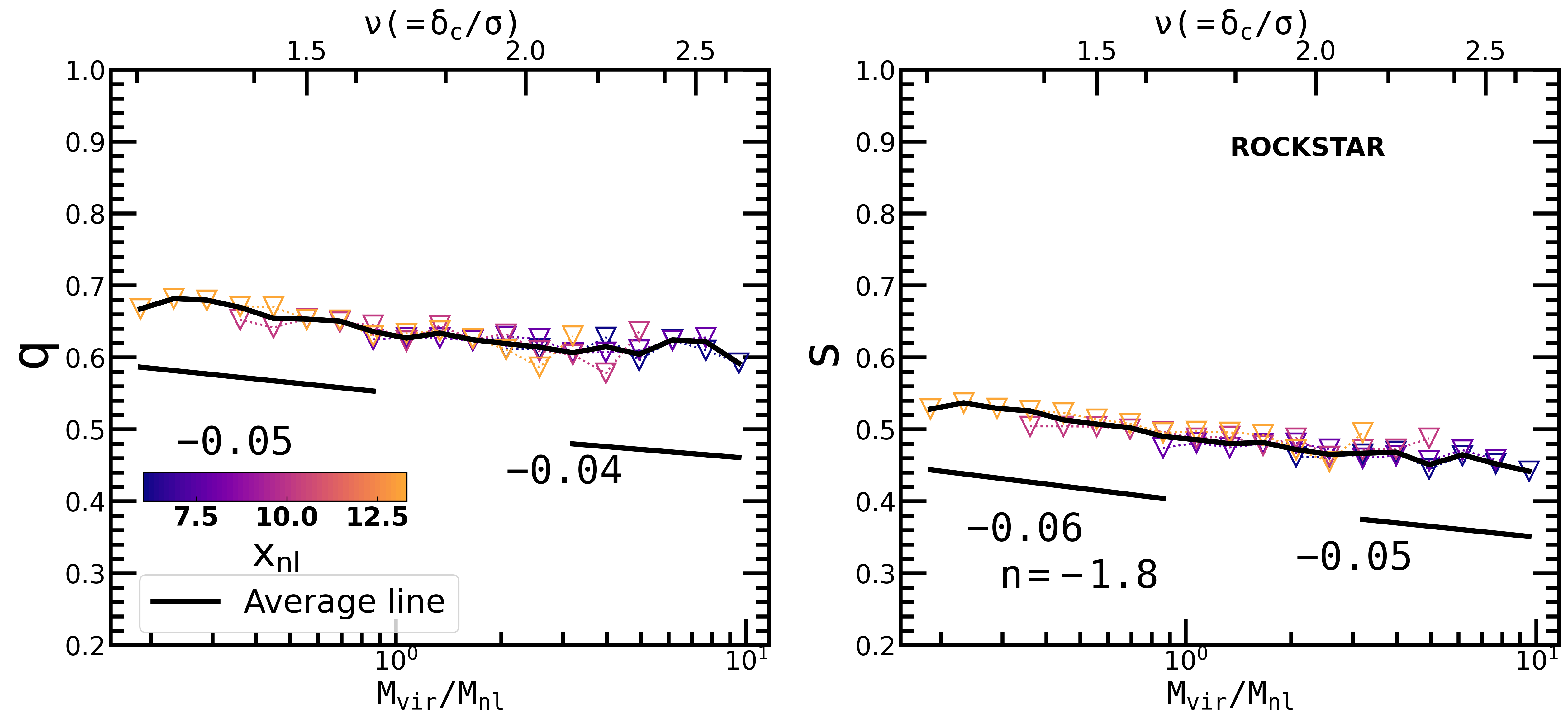}
    \includegraphics[width=\linewidth]{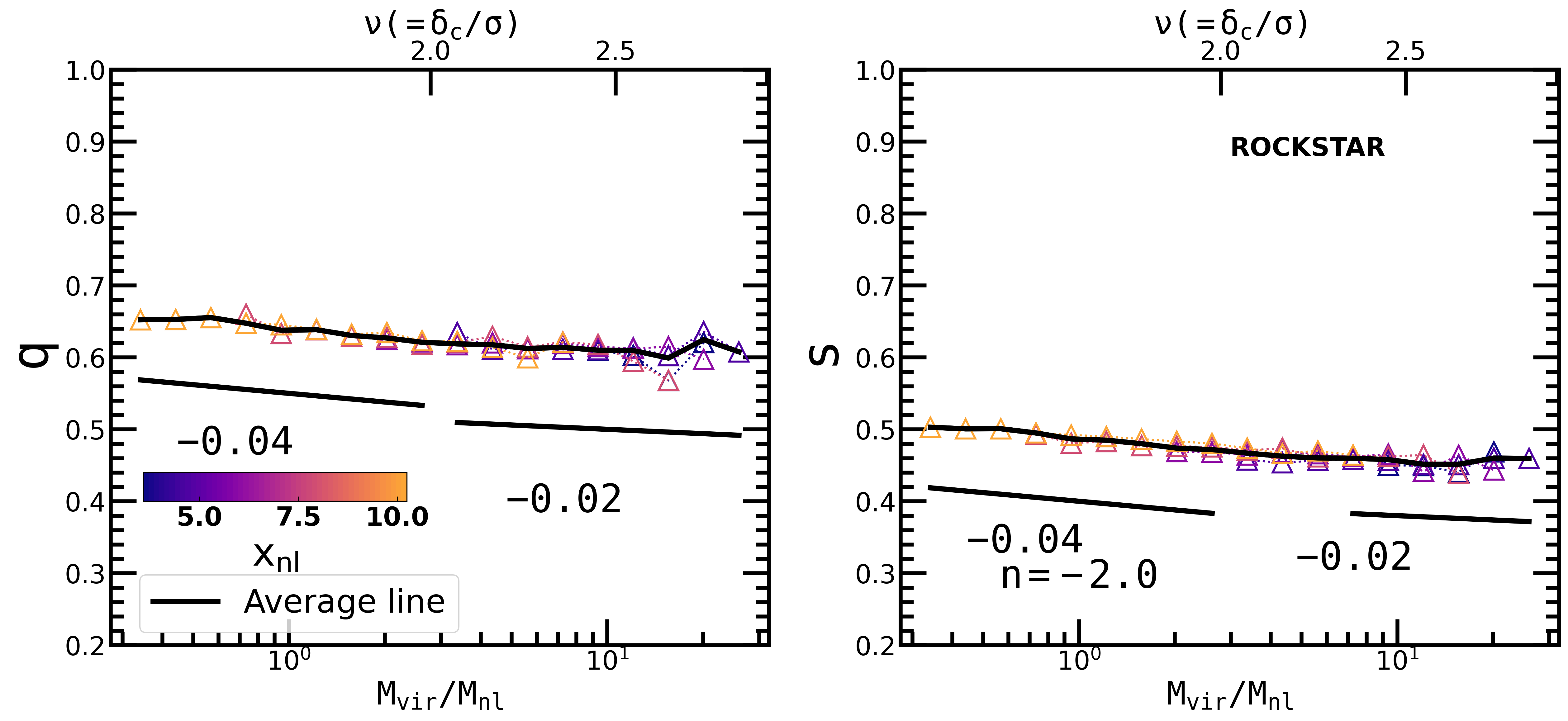}
    \includegraphics[width=\linewidth]{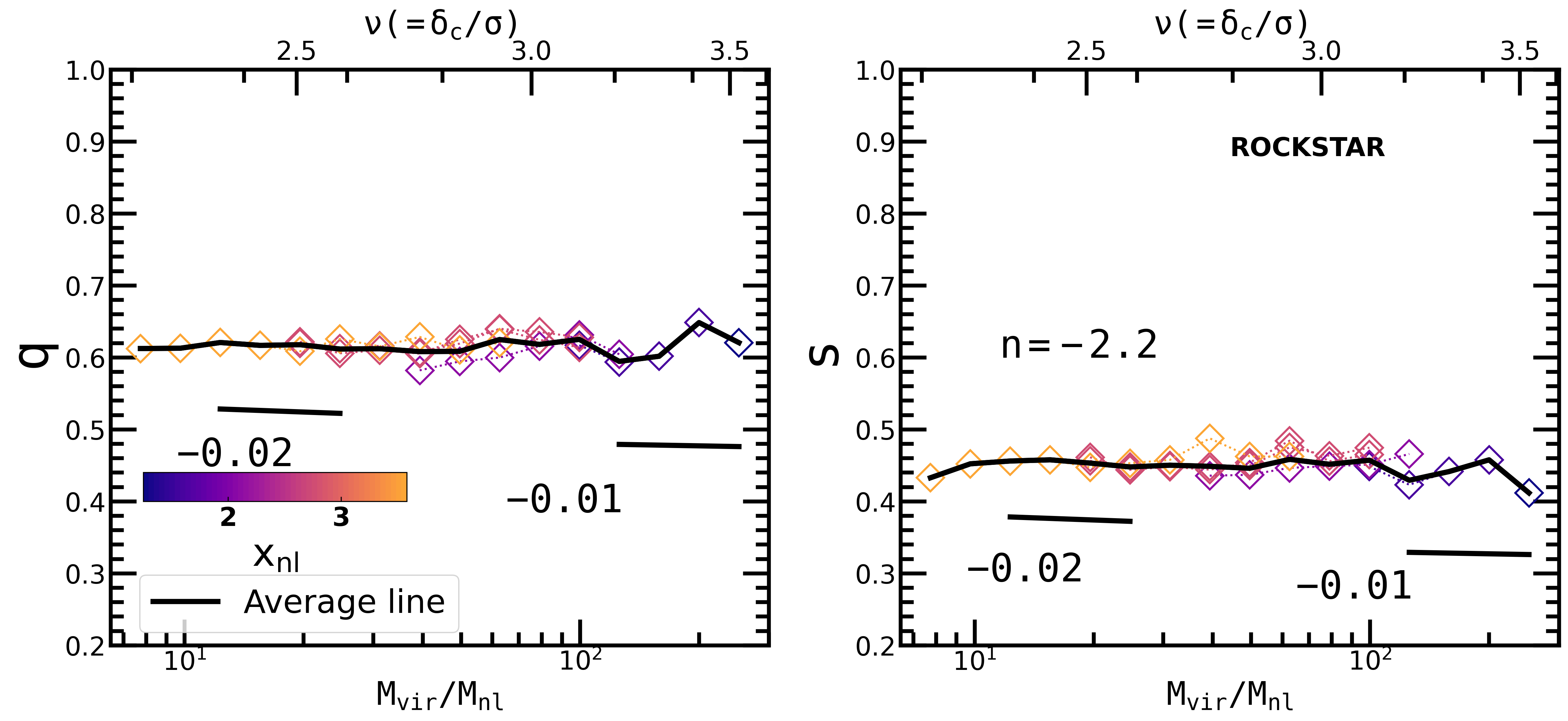}
    \endminipage
    \caption{The figure shows the median values of axis ratios as a function of $(M/\Mnl)$ (bottom x-axis) or $\nu$ (top x-axis)
      for all the scale-free models ranging from  $n=-1.0$ to $n=-2.2$ (top to bottom).
      The first two columns ($q$ and $s$) and the last two columns ($q$ and $s)$ are the results from \subfind and \rockstar respectively.
      The colour bar represents the results from different epochs, with darker (lighter) colours representing earlier (later) times with $\xnl$ being the time variable.
      The solid black curve and lines represent the average value of the data across all epochs and the local slope to guide the eye. }
    \label{fig_scalfree_qs_evolve_all}
    \end{figure*}

In figure~\ref{fig_scalfree_qs_evolve_all} we look at the evolution of the median values of the
axis ratios $q$ (first and third column) and $s$ (second and fourth) as a function of $(M/\Mnl)$ (bottom x-axis) or $\nu$ (top x-axis) for all the scale-free models ranging from
$n=-1.0$ to $n=-2.2$ (top to bottom). The first two columns are the results with \subfind and the next two are for \rockstar.
The colour bar encodes different epochs with darker (lighter) colours representing earlier (later) times with $\xnl$ as a time variable.
The solid black curve is the median value of the data across all epochs at fixed   $(M/\Mnl)$ or $\nu$, and the small line segment denotes the local slope to guide the eye.
The data is binned in $(M/\Mnl)$ and the same bins are used across all the epochs so that weighted averaging over epochs can be done conveniently.
For a fixed epoch, bins have a minimum of 30 halos to avoid large fluctuations. However large changes in the median $q$ and $s$ values are seen at the end points for some of the models.
Increasing the threshold removes these points but reduces the dynamic range. Including these points does not affect the analysis that will be done later.

\begin{figure*}[]
  \minipage{0.48\textwidth}
  \includegraphics[width=\linewidth]{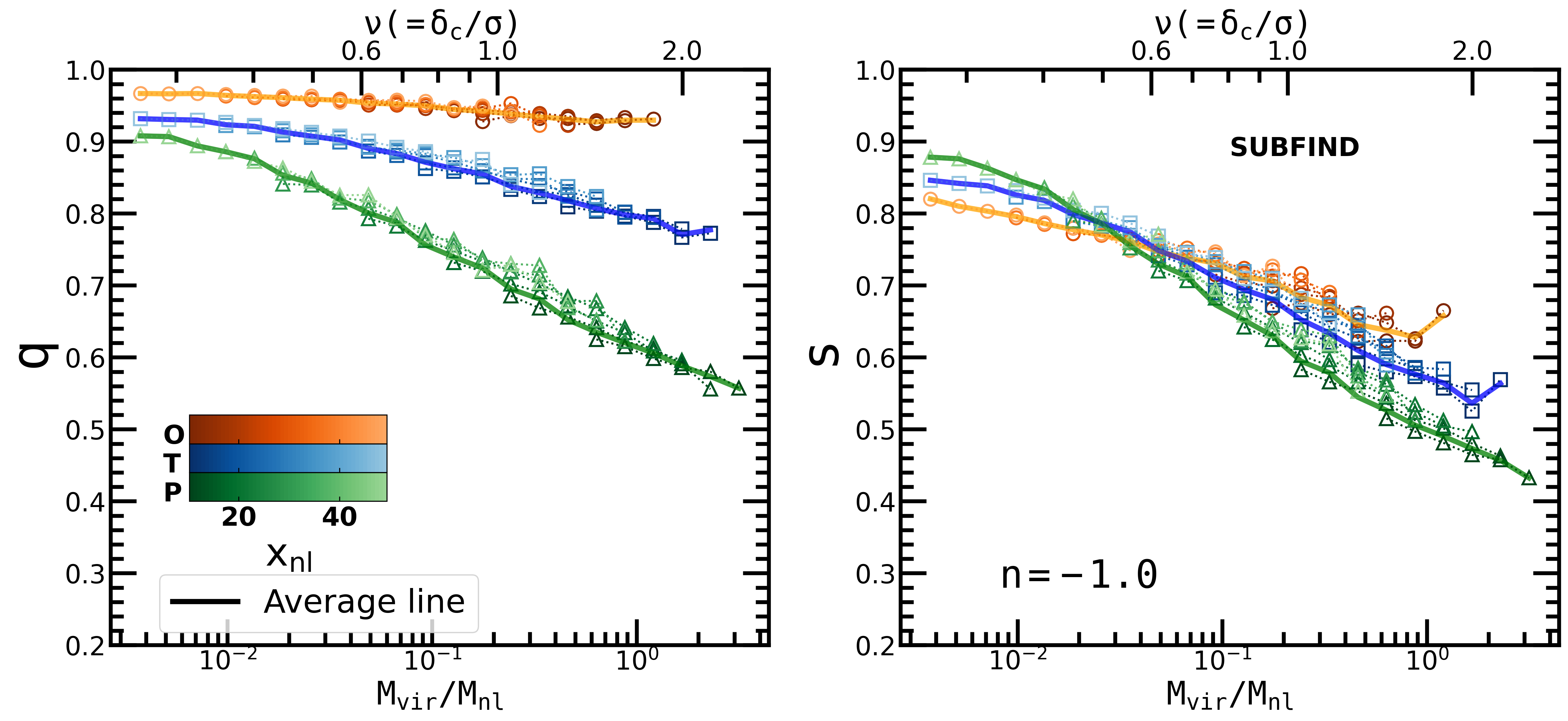}
  \includegraphics[width=\linewidth]{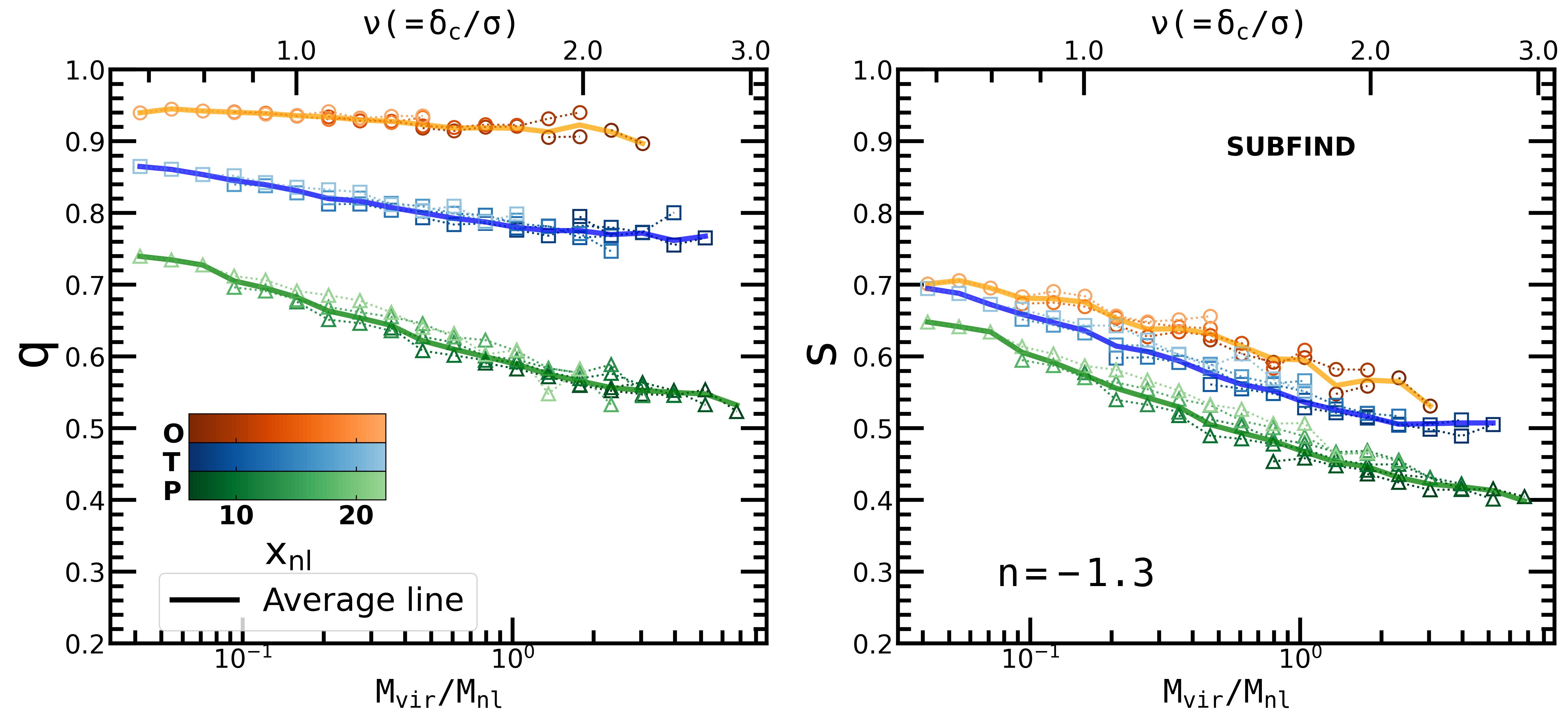}
  \includegraphics[width=\linewidth]{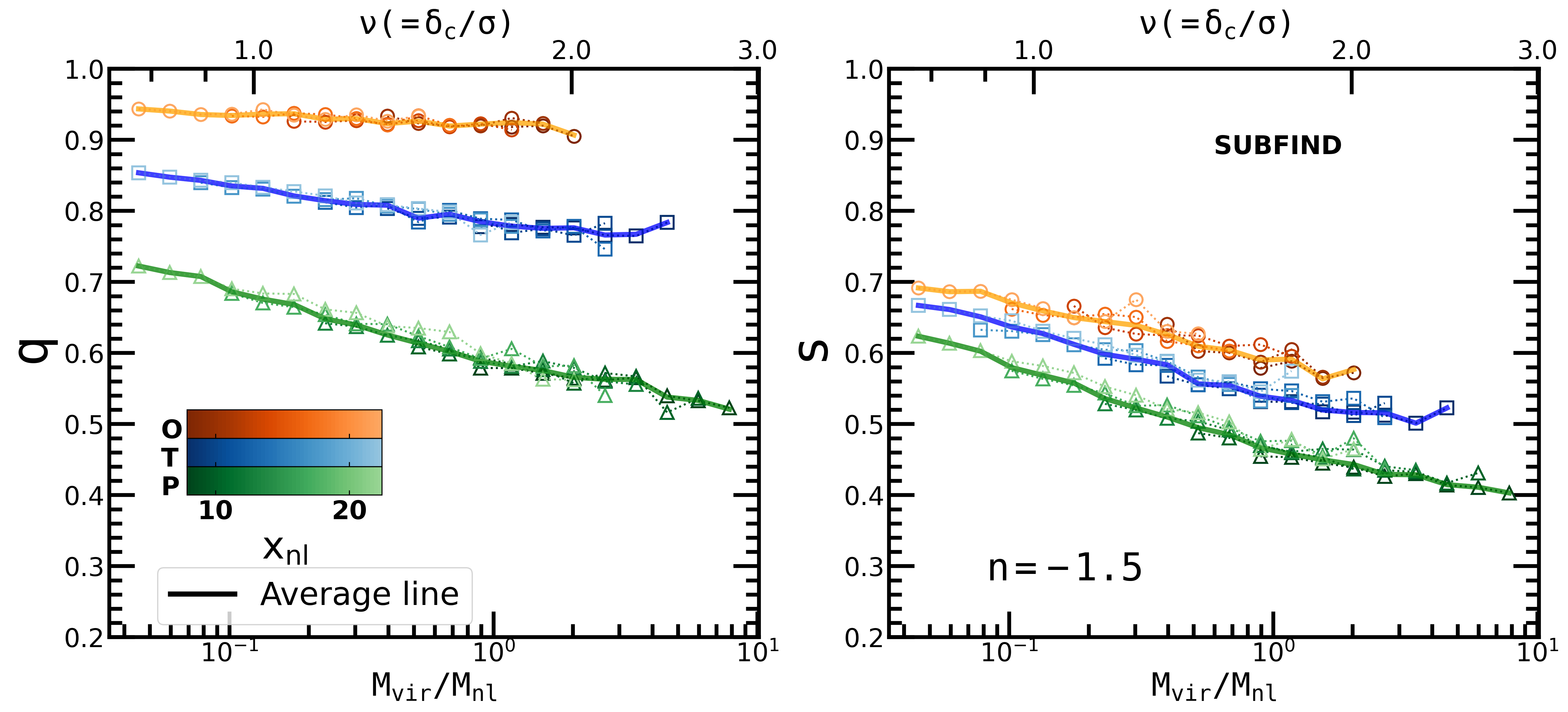}
  \includegraphics[width=\linewidth]{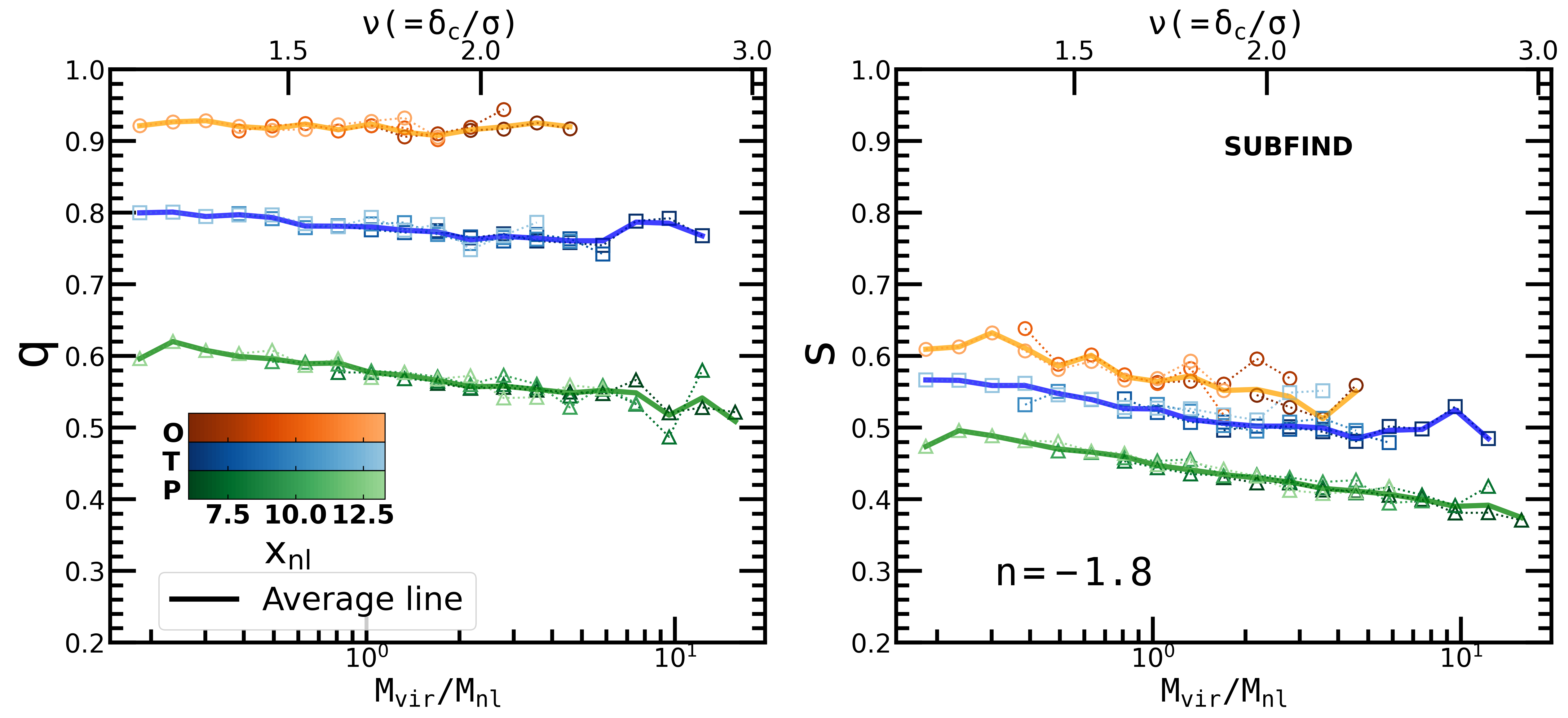}
  \includegraphics[width=\linewidth]{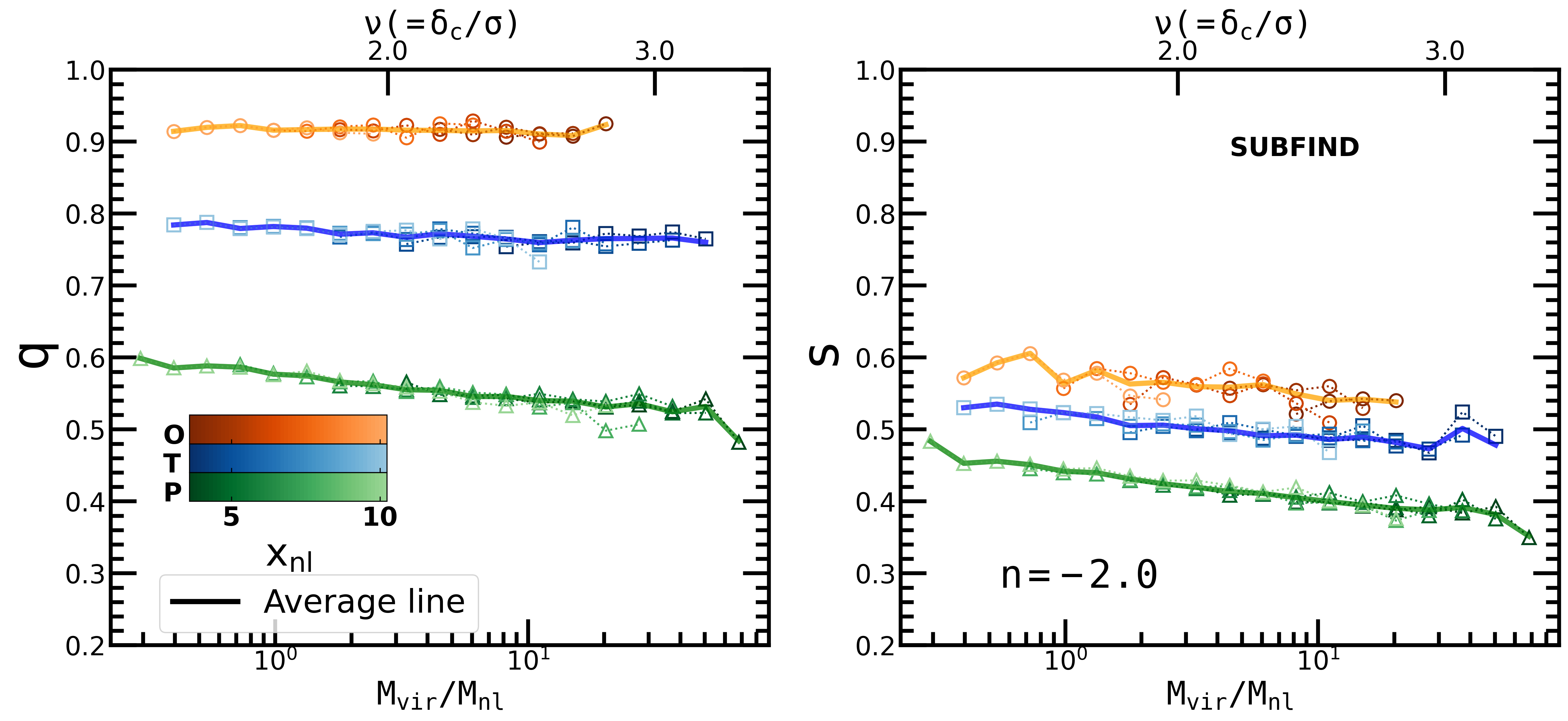}
  \includegraphics[width=\linewidth]{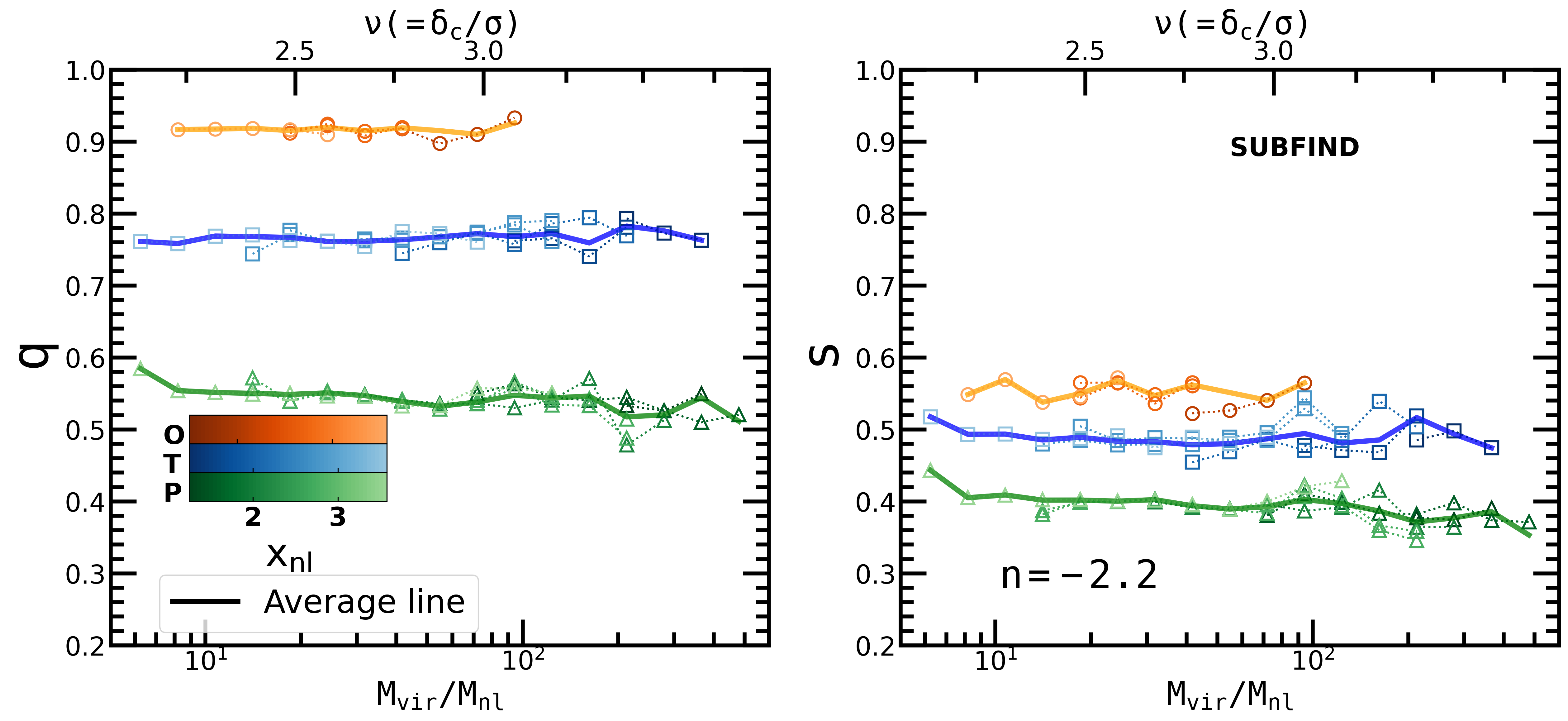}
  \endminipage\hfill
  \minipage{0.48\textwidth}
  \includegraphics[width=\linewidth]{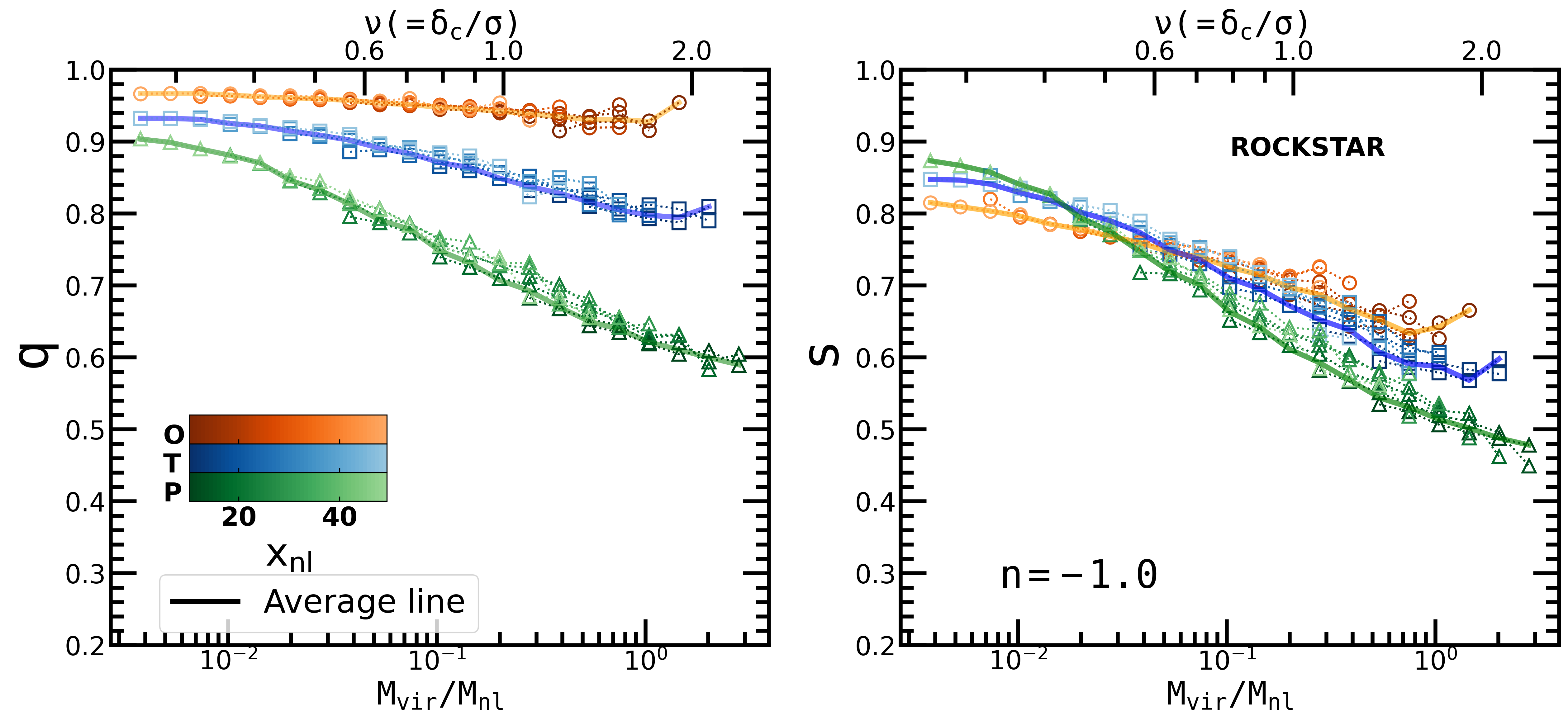} 
  \includegraphics[width=\linewidth]{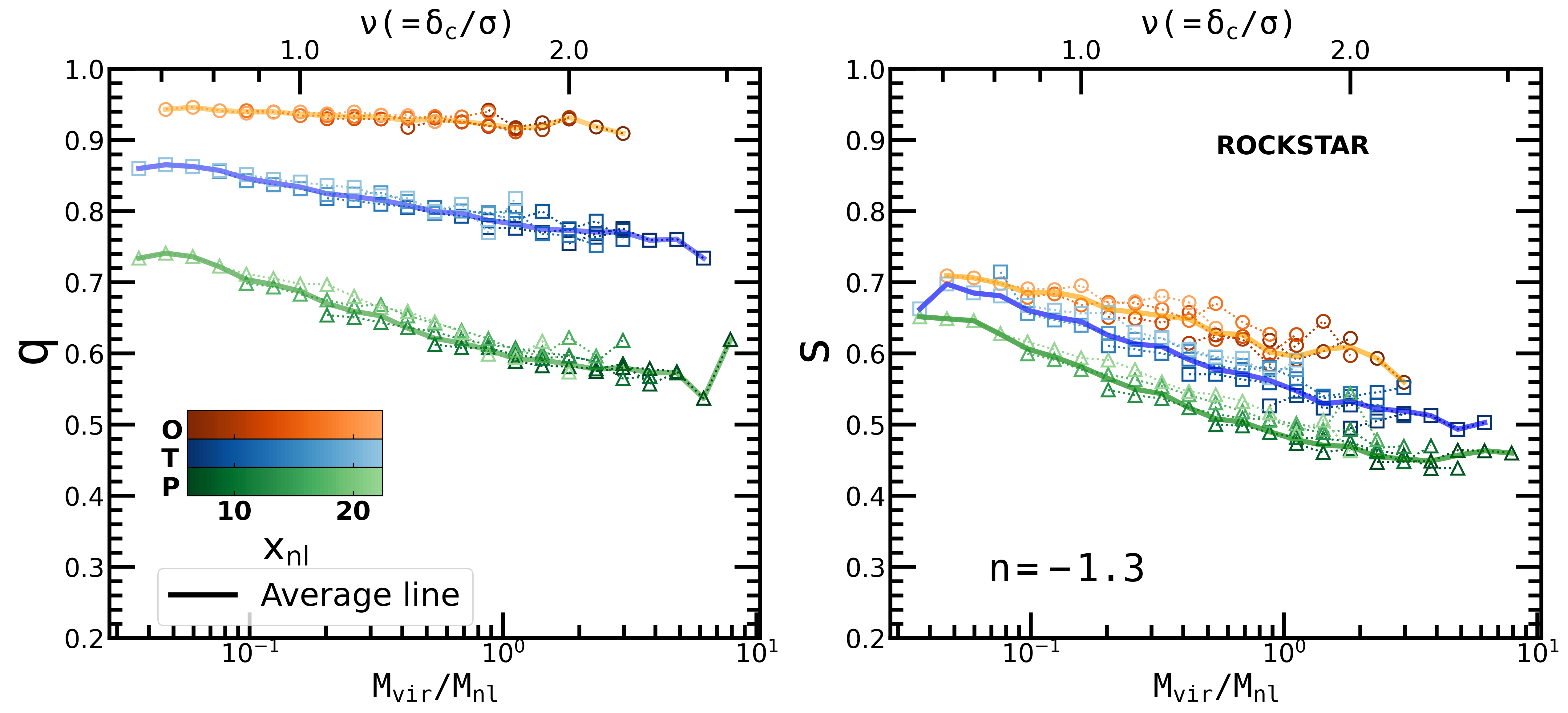}
  \includegraphics[width=\linewidth]{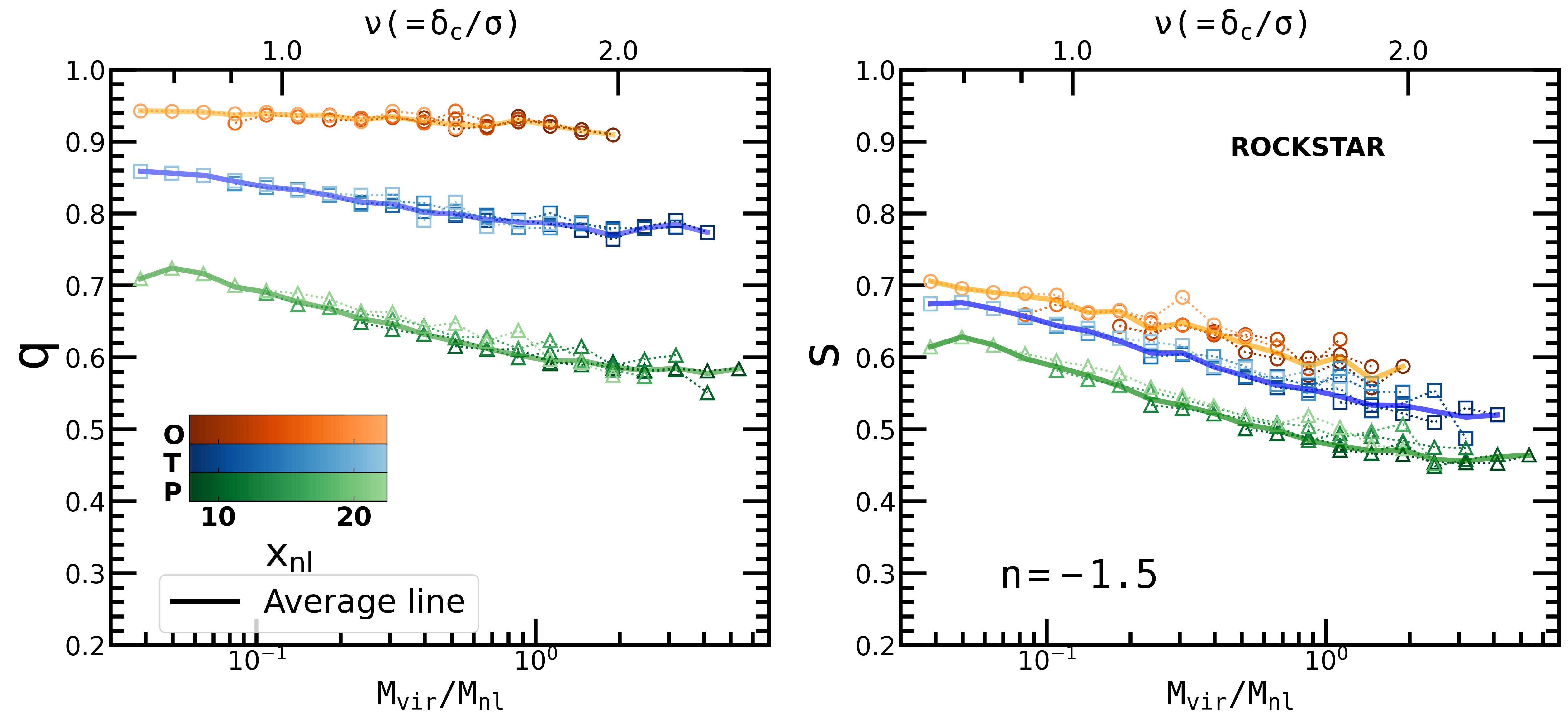}
  \includegraphics[width=\linewidth]{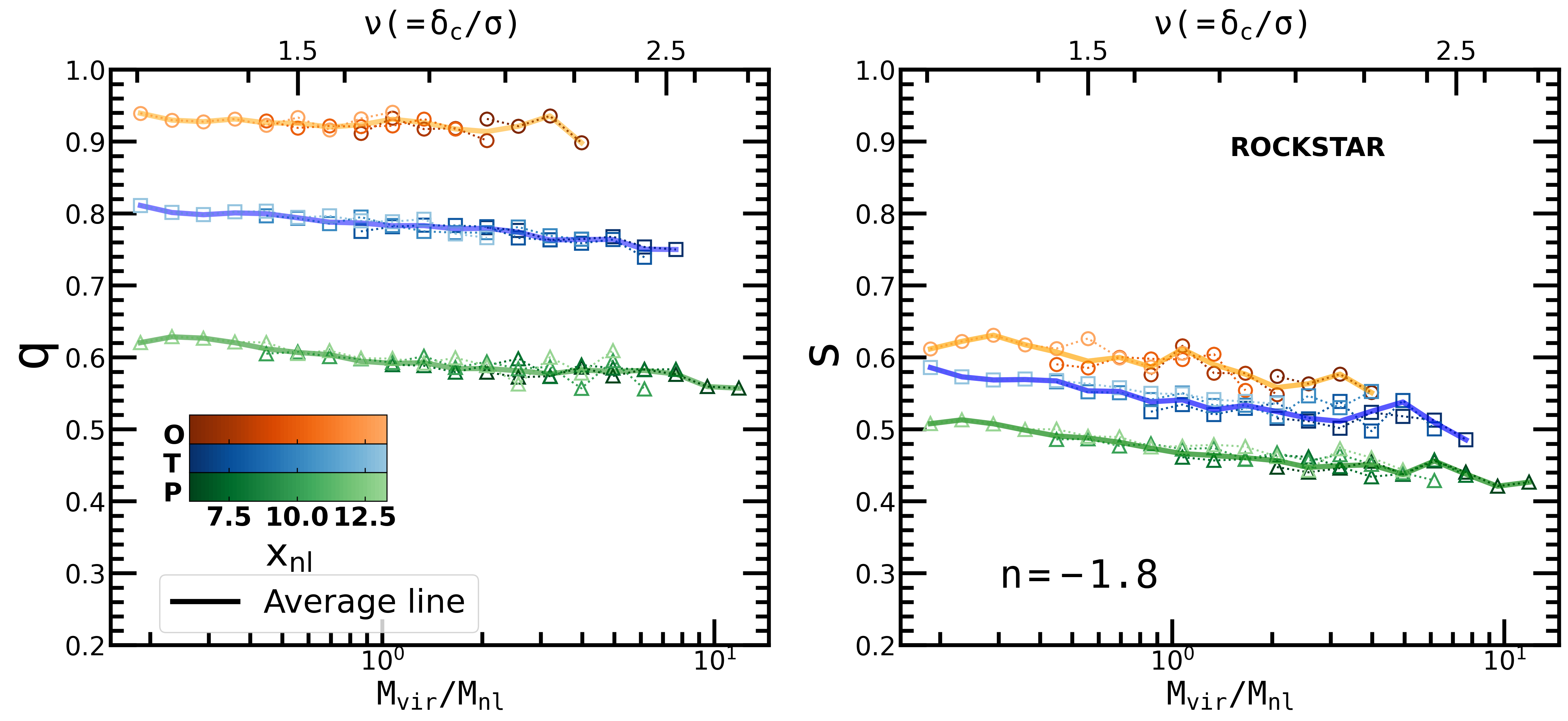}
  \includegraphics[width=\linewidth]{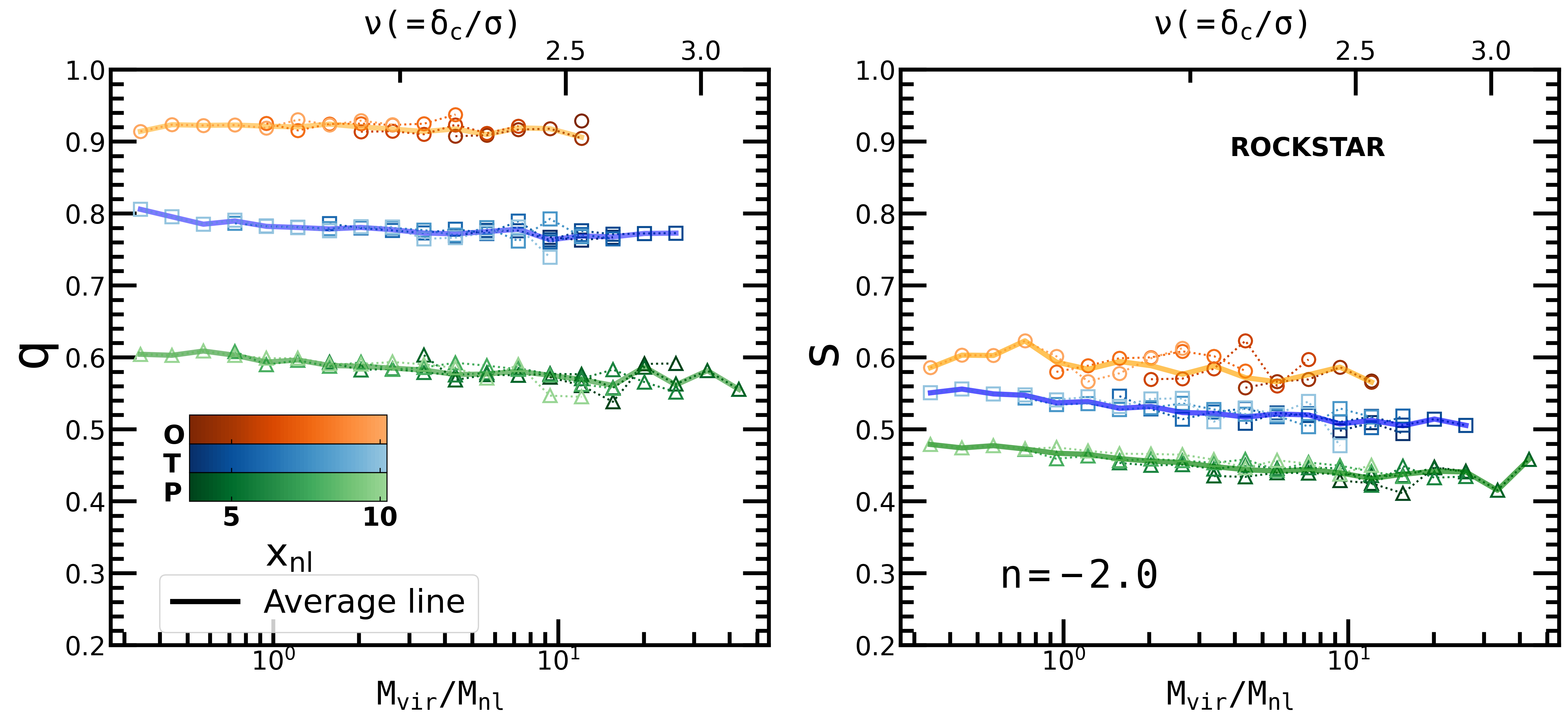}
  \includegraphics[width=\linewidth]{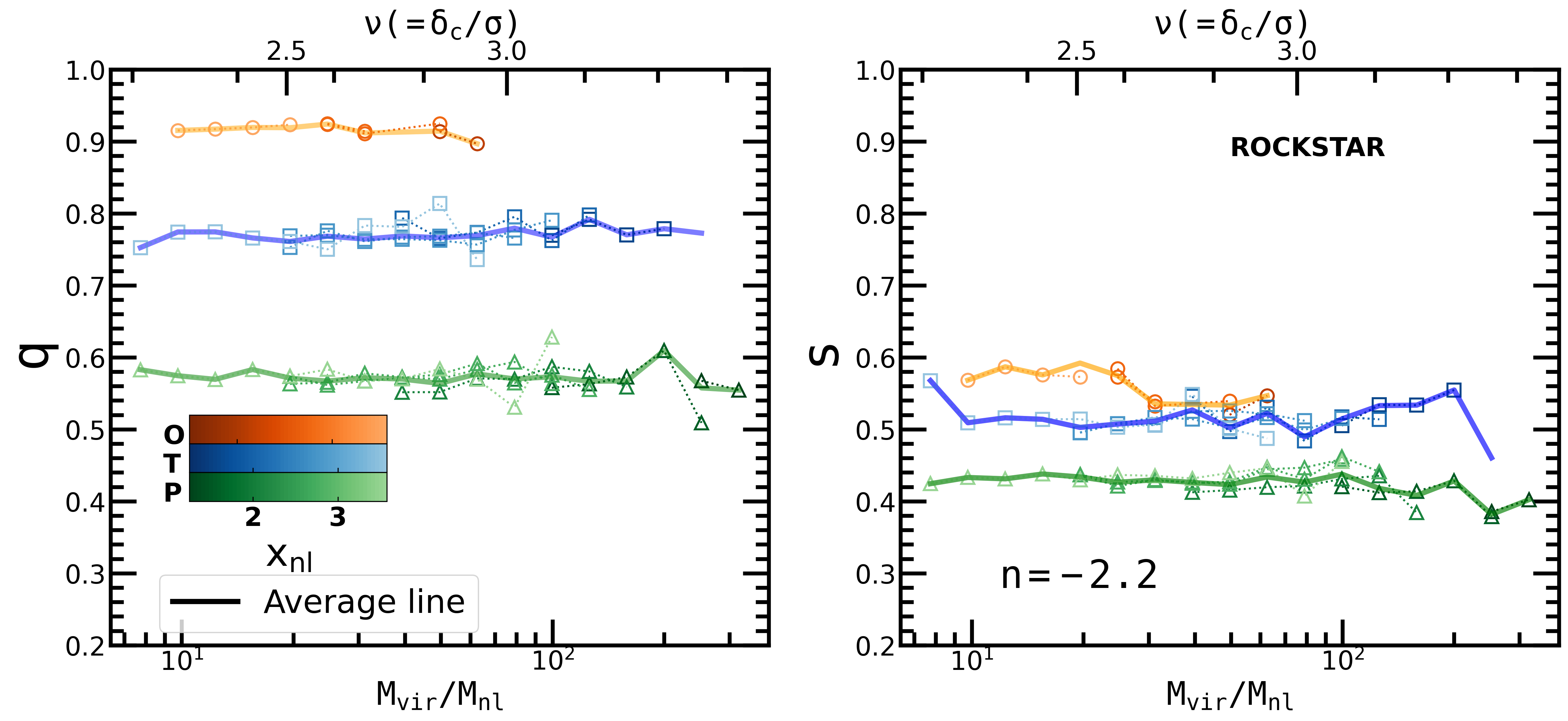}
  \endminipage
  \caption{The figure shows the median values of axis ratios as a function of $(M/\Mnl)$ (bottom x-axis) or $\nu$ (top x-axis)
      for all the scale-free models ranging from  $n=-1.0$ to $n=-2.2$ (top to bottom) for the oblate (O-orange), triaxial (T-blue) and prolate (P-green) halos.
      The first two columns ($q$ and $s$) and the last two columns ($q$ and $s)$ are the results from \subfind and \rockstar respectively.
      The colour bar represents the results from different epochs, with darker (lighter) colours representing earlier (later) times with $\xnl$ being the time variable.
      The solid black curve and lines represent the average value of the data across all epochs.}
    \label{fig_scalfree_qs_evolve_otp}
\end{figure*}

Across all models the results with \rockstar are comparable to \subfind. However the values of $q$ and $s$ in \rockstar are generally higher than those found
with \subfind. In figure~\ref{fig_scalfree_qs_evolve_all} we have not plotted the spread about the median values. We will discuss them when a fitting procedure is
employed in section~\ref{subsec_universality_qs_all}. A common feature seen across all models is the increase in the median $q$ and $s$ values with increasing time (or decreasing $(M/\Mnl)$ or $\nu$).
This suggests that on average, at fixed mass halos are more spherical at low redshift compared to their high redshift counterparts.
The evolution of the median $q$ and $s$ values follow a characteristic curve demonstrating self-similar behaviour across models moving from right to left with time.
The average line is relatively smooth except at the end points. Fluctuations of the data with respect to the average line is seen at higher masses values of  $(M/\Mnl)$.

The slope for $q(M/\Mnl)$ and $s(M/\Mnl)$ for the $n=-1.0$ model is shallow at $(M/\Mnl) \gtrsim 1$, gets steeper at $(M/\Mnl) \simeq 10^{-1}$ and then flattens again at $(M/\Mnl) \lesssim 10^{-2}$.
Before we compare this behaviour in the other models we look at the range of values for $(M/\Mnl)$. For $n=-1.0$ the range is   $ 4\times 10^{-2} < (M/\Mnl) < 5$. With decreasing
index $n$ the range of $(M/\Mnl)$ systematically shifts to larger values. For $n=-2.2$ the range $ 6\times 10^{-1} < (M/\Mnl) < 4\times 10^2$. 
The scale-free models are therefore probing different regimes of $(M/\Mnl)$.  For the $n=-1.3$ and $n=-1.5$ models we find a shallow slope at  $(M/\Mnl) \gtrsim 1$ which gets steeper at
$(M/\Mnl) \simeq 10^{-1}$ as in the $n=-1.0$ case. However the flattening seen in the $n=-1.0$ case for  $(M/\Mnl) \lesssim 10^{-2}$ is not seen here since these scales are not probed for the two models.
With decreasing index we see that the shallow slope seen for $(M/\Mnl) \gtrsim 1$ flattens even further at larger values and becomes zero for  $(M/\Mnl) \gtrsim 100$. This is seen both in $q$ and $s$. 

In figure~\ref{fig_scalfree_qs_evolve_otp} we look at self-similar behaviour of median values of axis ratios $q$ and $s$ for OTP halos which are colour coded as orange, blue and green respectively.
The columns and rows are ordered  in a manner similar to figure~\ref{fig_scalfree_qs_evolve_all}. In this case, at a fixed epoch, all bins have a minimum of 10 halos.  
The solid line curve is the median value of the data across all epochs at fixed $M/\Mnl$ or $\nu$. Similar to figure~\ref{fig_scalfree_qs_evolve_all}, the median line for $q$ and $s$
across all halo types is relatively smooth and follows a characteristic curve demonstrating self-similar evolution. Deviations from a smooth curve occur at the end-points where data is relatively sparse.
The median values of $q$ evolve along non-intersecting characteristic curves for each halo type. 
However the characteristic curves for the median values of $s$ intersect for the OTP at $M/\Mnl \approx 2\times 10^{-2}$ or $\nu \approx 0.5$ for the $n=-1.0$ model.
The results hold independent of the halo finding algorithm.

In summary, self-similar evolution of axis ratios  is seen across all scale-free models and separately for OTP halos.
Additionally we find a new trend when comparing models. Our results suggest a universal behaviour across models, which probe different regimes of $(M/\Mnl)$
which we will explore in the next section~\ref{subsec_universality_qs_all}.

\clearpage
\newpage

\subsection{Universality of Axis ratios}
\label{subsec_universality_qs_all}

\begin{figure}[h]
  \minipage{0.5\textwidth}
  \includegraphics[width=\linewidth]{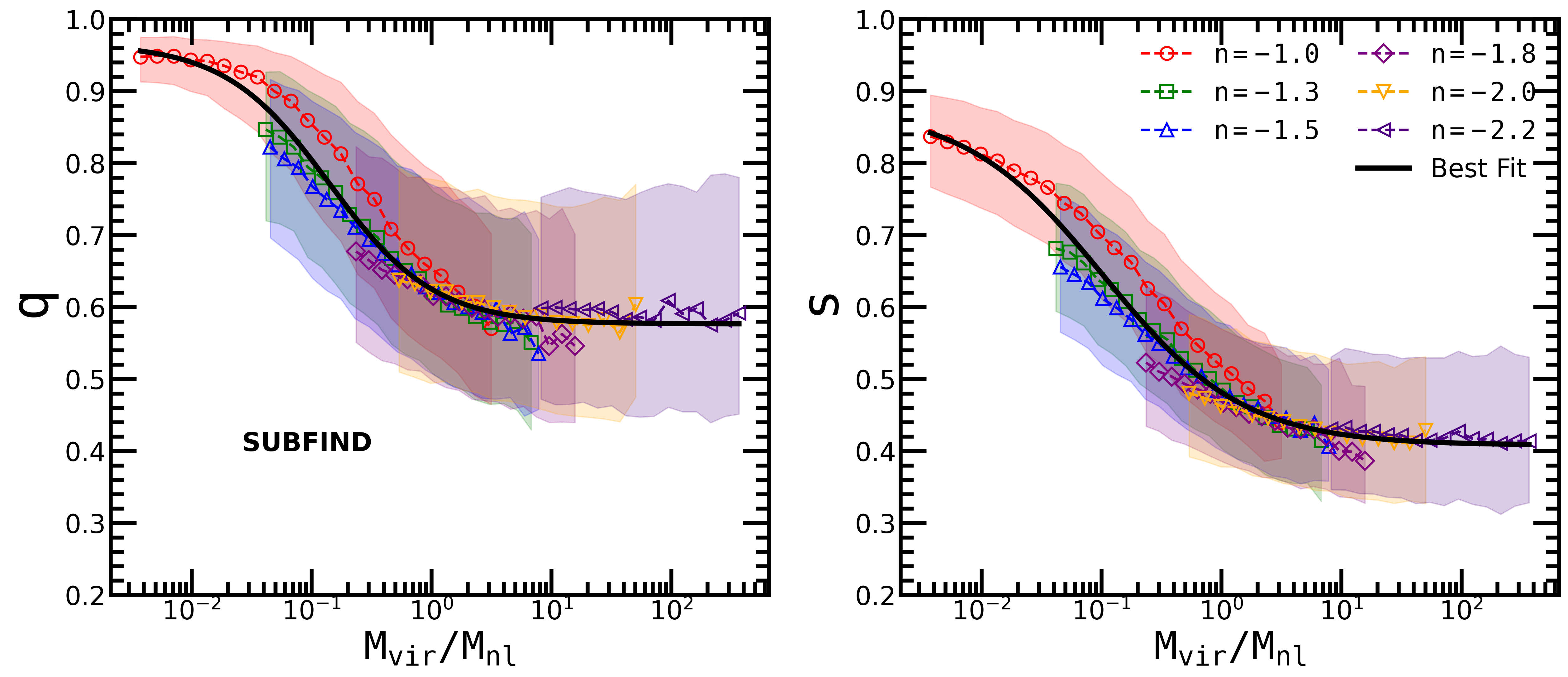}\\
  \includegraphics[width=\linewidth]{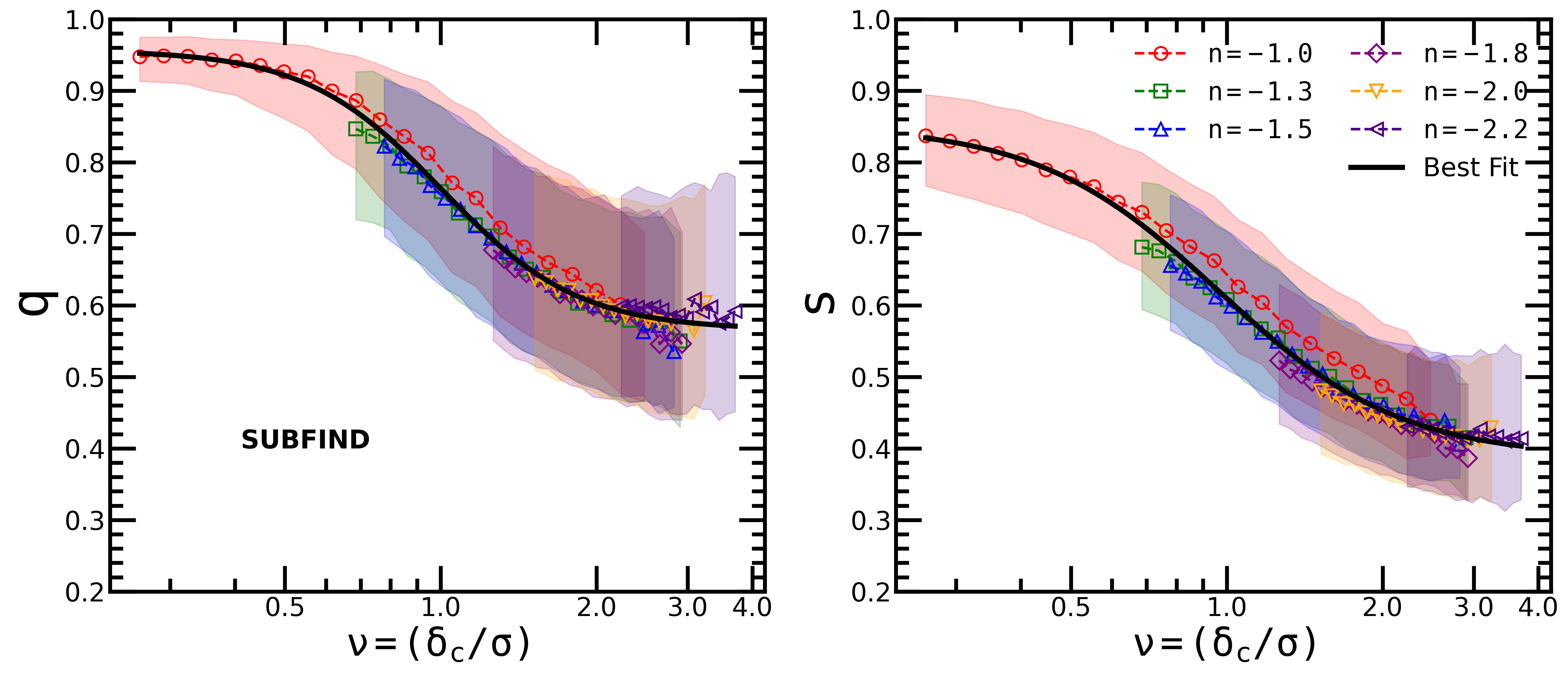}
  \endminipage
  \minipage{0.5\textwidth}
  \includegraphics[width=\linewidth]{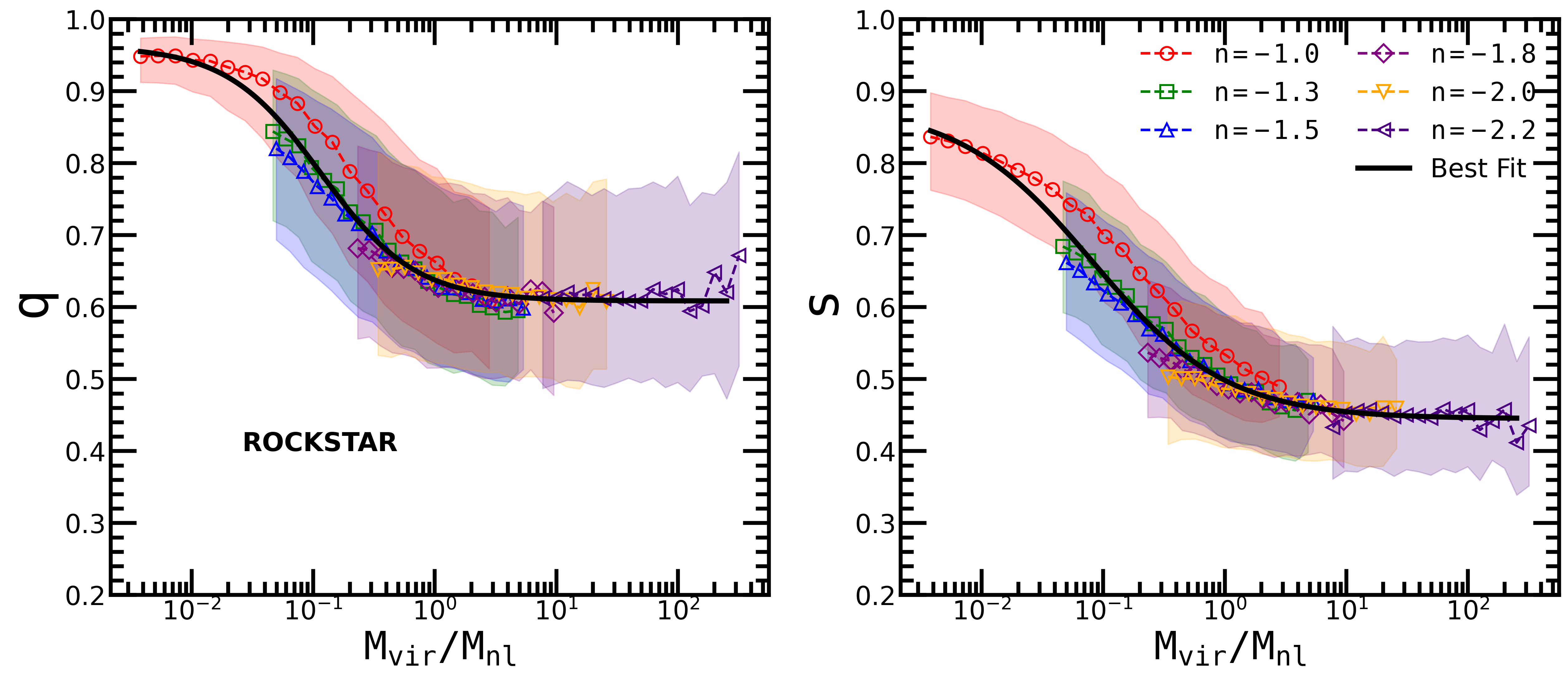}\\
  \includegraphics[width=\linewidth]{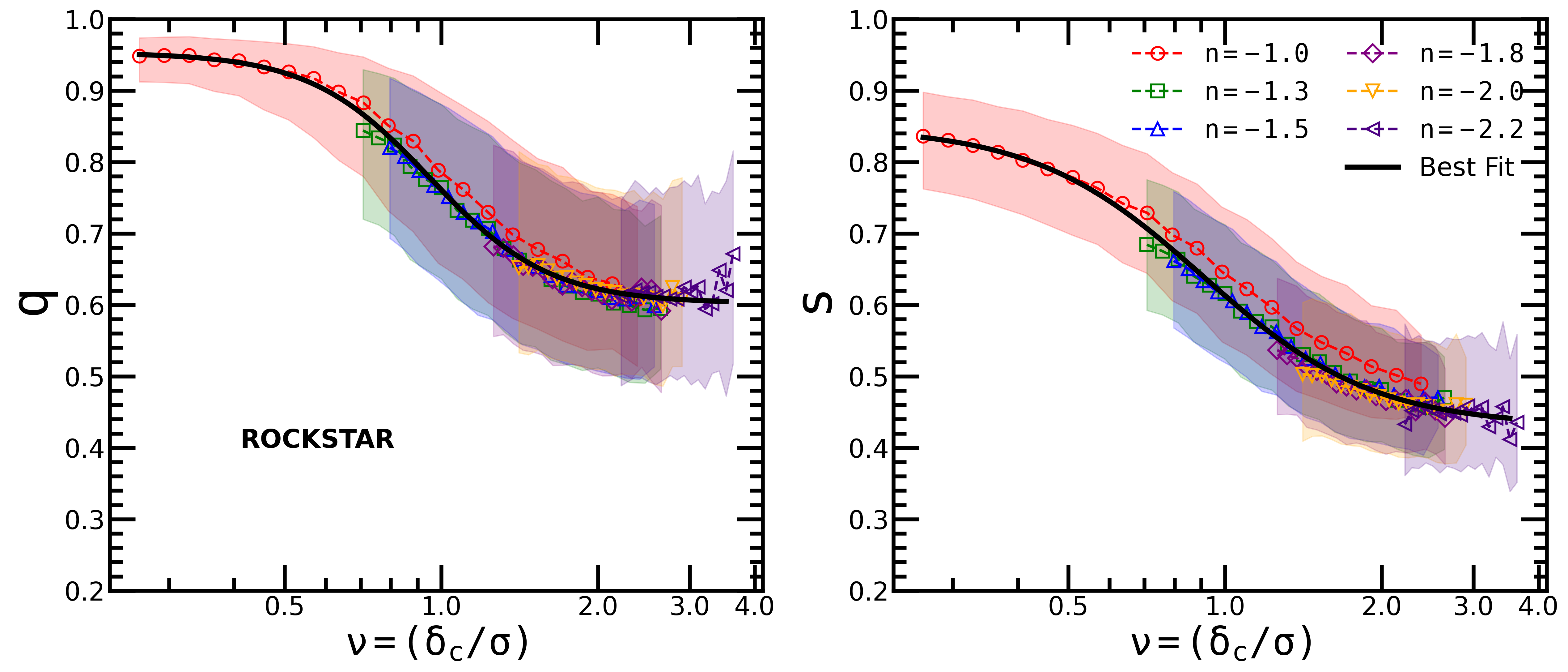}
  \endminipage
  \caption{The median values (points) of $q$ (column 1 and 3) and $s$ (column 2 and 4) in a bins of
    $(M/\Mnl)$ (top) and $\nu$ (bottom) bin. The circle, square, triangle, diamond, inverted triangle and rotated triangle are from the $n=-1.0, -1.3, -1.5, -1.8, -2.0, -2.2$ scale free runs
    respectively. These points are compiled from the solid curve (across models) in figure~\ref{fig_scalfree_qs_evolve_all}.
    The solid line is the parametric fit (equation~\ref{eq_tanh}) to the data. The best fit parameters are given in table~\ref{tab_powlaw_universal_all}.
    The results from \subfind and \rockstar are given in the first two and the next two columns
    respectively. The shaded region represent the $1\sigma$ percentile of the distribution.}
  \label{fig_scalefree_universal_all}
\end{figure}

\begin{figure}[h]
  \minipage{0.5\textwidth}
  \includegraphics[width=\linewidth]{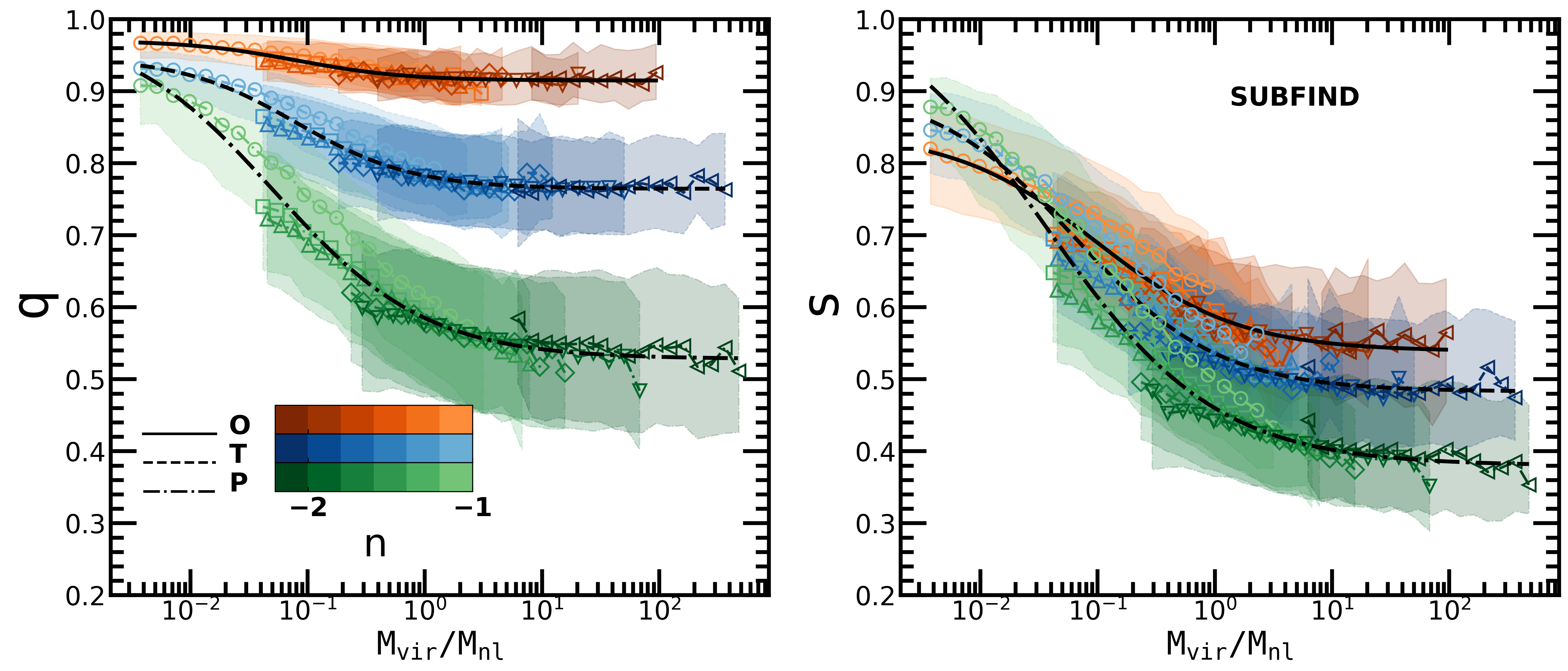}
  \includegraphics[width=\linewidth]{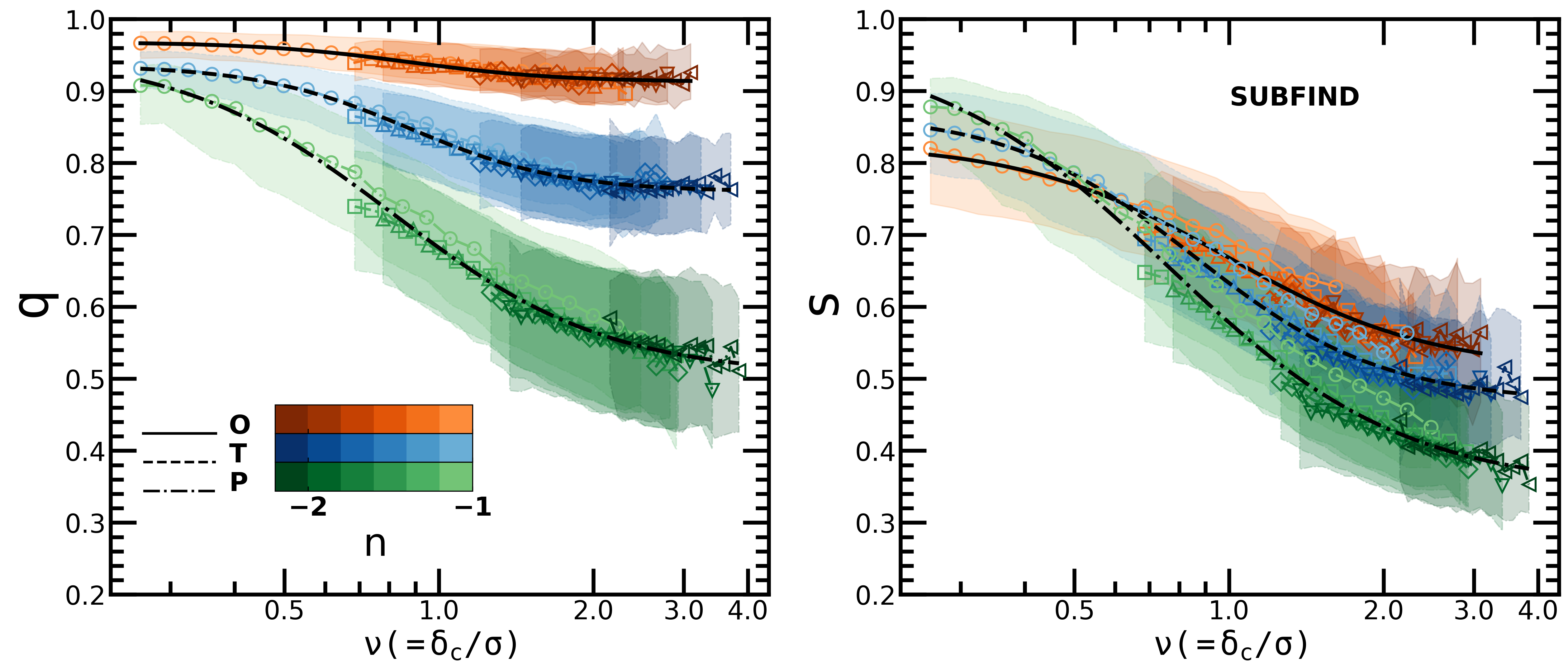}
  \endminipage
  \minipage{0.5\textwidth}
  \includegraphics[width=\linewidth]{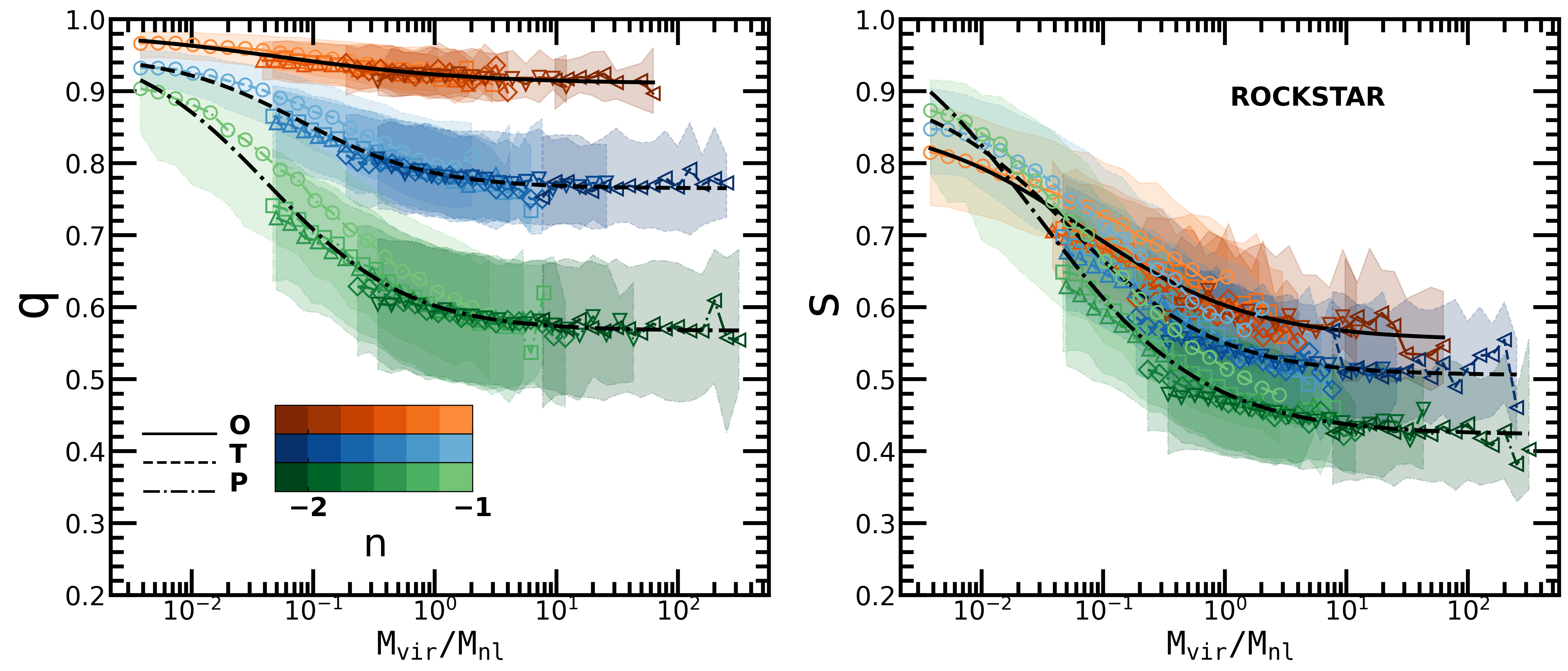}\\
  \includegraphics[width=\linewidth]{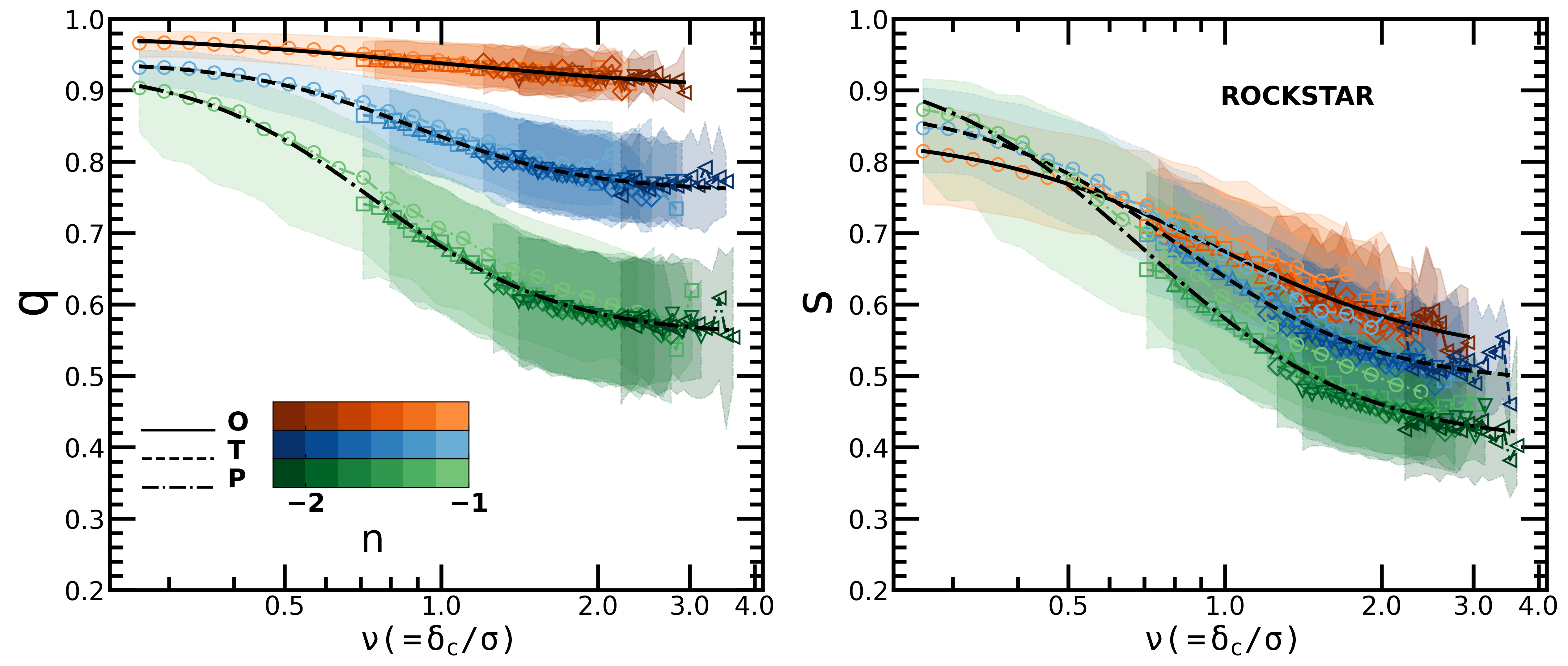}
  \endminipage  
  \caption{Same as in figure~\ref{fig_scalefree_universal_all}, but done separately for OTP (orange, blue, green) halos. Each data point is collected
    from the solid curve (across models) in figure~\ref{fig_scalfree_qs_evolve_otp}.
    The solid, dashed and dot-dashed lines represent the parametric fit (equation~\ref{eq_tanh}) for the OTP (orange, blue, green) halo data respectively.
    The best fit parameters of the universal curve for oblate, triaxial and prolate halos are given in
    tables~\ref{tab_powlaw_universal_oblate},\,\,\ref{tab_powlaw_universal_triaxial} and \ref{tab_powlaw_universal_prolate} respectively.
    The shaded region represents the $1\sigma$ percentile of the distribution.}
  \label{fig_scalefree_universal_otp}
\end{figure}

We concatenate all the data (median solid line) across scale-free models from figures~\ref{fig_scalfree_qs_evolve_all} and \ref{fig_scalfree_qs_evolve_otp} separately for the median $q$ and $s$ values,
to see if their evolution is described by a single universal curve. 
Our results are compiled in figures~\ref{fig_scalefree_universal_all} and \ref{fig_scalefree_universal_otp},
where the shaded region is determined
by the median (at fixed ($M/\Mnl)$ or $\nu$) 16 and 84 percentiles of the $P(q)$ and $P(s)$ distributions.
We will refer to these percentiles as $1\sigma$ percentiles.
We find the evolution of the distributions $P(q)$ and $P(s)$, as quantified by the median and the $1\sigma$ percentiles, across models to be consistent with a universal behaviour.
Since the data appears to have flat asymptotes and $\tanh$ is a good function for such cases, we fit the median $q$ and $s$ with the following function.
\begin{equation}
  \begin{split}
    y(x) = \alpha - \delta \tanh\left[\omega\left(\log_{10}(x)-\mu\right)\right]\\
  \end{split}
  \label{eq_tanh}
\end{equation}

\begin{table}
\centering
\renewcommand{\arraystretch}{1.2}
\begin{tabular}{|p{2.5cm}|c|c|c|c|c|}
\hline
  \textbf{All Halos} & $\alpha \pm \Delta\alpha$ & $\delta \pm \Delta\delta$ & $\omega \pm \Delta\omega$ & $\mu \pm \Delta \mu$ & rms \\
\hline
\textbf{\subfind} & & & & & $\times 10^{-2}$ \\
\hline
$q\left(\frac{M}{M_\mathrm{nl}}\right)$ & $0.771 \pm 0.006$  & $0.195 \pm 0.008$ & $1.149 \pm 0.082$ & $-0.848 \pm 0.042$ & $2.2$ \\
\hline
$s\left(\frac{M}{M_\mathrm{nl}}\right)$ & $0.643 \pm 0.010$  & $0.235 \pm 0.012$ & $0.856 \pm 0.064$ & $-0.981 \pm 0.064$ & $2.2$ \\
\hline
$q\left(\nu\right)$ & $0.761 \pm 0.003$  & $0.196\pm 0.005$ & $3.798 \pm 0.183$ & $0.003 \pm 0.006$ & $1.4$ \\
\hline
$s\left(\nu\right)$ & $0.620 \pm 0.004$  & $0.232 \pm 0.008$ & $2.847 \pm 0.163$ & $-0.015 \pm 0.010$ & $1.5$ \\
\hline
\textbf{\rockstar} & & & & & \\
\hline
$q\left(\frac{M}{M_\mathrm{nl}}\right)$ & $0.785 \pm 0.006$  & $0.177 \pm 0.007$ & $1.290 \pm 0.094$ & $-0.934 \pm 0.039$ & $2.1$ \\
\hline
$s\left(\frac{M}{M_\mathrm{nl}}\right)$ & $0.663 \pm 0.010$  & $0.218 \pm 0.013$ & $0.906 \pm 0.075$ & $-1.085 \pm 0.070$ & $2.3$ \\
\hline
$q\left(\nu\right)$ & $0.778 \pm 0.003$  & $0.175 \pm 0.004$ & $4.347 \pm 0.186$ & $-0.022 \pm 0.006$ & $1.2$ \\
\hline
$s\left(\nu\right)$ & $0.640 \pm 0.004$  & $0.210 \pm 0.007$ & $3.071 \pm 0.181$ & $-0.042 \pm 0.010$ & $1.4$ \\
\hline
\end{tabular}
\caption{Best fit parameters (and their uncertainties) of the universal curve (equation~\ref{eq_tanh})
  for median $q$ and $s$ as a function of $(M/M_\mathrm{nl})$ and $\nu$ using \subfind and \rockstar for all halos in the scale-free
  model. The last column is the rms fluctuation of the data with respect to the best fit curve. 
  The universal curves are plotted in figure~\ref{fig_scalefree_universal_all}.
  }
\label{tab_powlaw_universal_all}
\end{table}

\begin{table*}[b]
\centering
\renewcommand{\arraystretch}{1.2}
\begin{tabular}{|p{2.5cm}|c|c|c|c|c|}
\hline
    \textbf{Oblate} &$\alpha \pm \Delta\alpha$ & $\delta \pm \Delta\delta$ & $\omega \pm \Delta\omega$ & $\mu \pm \Delta \mu$ & rms \\
\hline
\textbf{\subfind} & & & & & $\times 10^{-2}$ \\
\hline
$q\left(\frac{M}{M_\mathrm{nl}}\right)$ & $0.943 \pm 0.002$ & $0.028 \pm 0.003$ & $1.068 \pm 0.168$ & $-1.090 \pm 0.102 $ & $0.6$ \\
\hline
$s\left(\frac{M}{M_\mathrm{nl}}\right)$ & $0.691 \pm 0.011 $ & $0.152 \pm 0.015 $ & $0.820 \pm 0.110 $ & $ -1.012 \pm 0.110$ & $2.2$\\
\hline
$q\left(\nu\right)$ & $0.940 \pm 0.001$ & $0.028 \pm 0.002$ & $3.378 \pm 0.523$ & $-0.059 \pm 0.023$ & $4.8$ \\
\hline
$s\left(\nu\right)$ & $0.669 \pm 0.007$  & $0.159 \pm 0.014$ & $2.503 \pm 0.351 $ & $-0.003 \pm 0.021 $ & $1.7$ \\
\hline
\textbf{\rockstar} & & & & & \\
\hline
$q\left(\frac{M}{M_\mathrm{nl}}\right)$ & $0.948 \pm 0.007 $ & $0.037 \pm 0.010 $ & $0.611 \pm 0.176 $ & $-1.278 \pm 0.361 $ & $0.7$ \\
\hline
$s\left(\frac{M}{M_\mathrm{nl}}\right)$ & $0.708 \pm 0.015 $ & $0.154 \pm 0.020 $ & $0.728 \pm 0.119 $ & $-1.148 \pm 0.158 $ & $0.2$ \\
\hline
$q\left(\nu\right)$  & $0.939 \pm 0.005 $ & $0.044 \pm 0.018 $ & $1.529 \pm 0.786 $ & $-0.013 \pm 0.084 $ & $0.6$ \\
\hline
$s\left(\nu\right)$ & $0.681 \pm 0.008$ & $0.159 \pm 0.020 $ & $2.217 \pm 0.394 $ & $-0.019 \pm 0.027 $ & $1.7$ \\
\hline
\end{tabular}
\caption{Best fit parameters (and their uncertainties) of the universal curve (equation~\ref{eq_tanh})
  for median $q$ and $s$ as a function of $(M/M_\mathrm{nl})$ and $\nu$ using \subfind and \rockstar for oblate halos in the scale-free
  model.
  The last column is the rms fluctuation of the data with respect to the best fit curve.
  The universal curves are plotted in figure~\ref{fig_scalefree_universal_otp}.
  }
\label{tab_powlaw_universal_oblate}
\end{table*}

\begin{table}[h]
\centering
\renewcommand{\arraystretch}{1.2}
\begin{tabular}{|p{2.5cm}|c|c|c|c|c|}
\hline
    \textbf{Triaxial} &$\alpha \pm \Delta\alpha$ & $\delta \pm \Delta\delta$ & $\omega \pm \Delta\omega$ & $\mu \pm \Delta \mu$ & rms \\
\hline
\textbf{\subfind} & & & & & $\times 10^{-2}$ \\
\hline
$q\left(\frac{M}{M_\mathrm{nl}}\right)$ & $0.856 \pm 0.004 $ & $0.092 \pm 0.005 $ & $1.004 \pm 0.083 $ & $-1.091 \pm 0.062 $ & $1.0$ \\
\hline
$s\left(\frac{M}{M_\mathrm{nl}}\right)$ & $0.696 \pm 0.013 $ & $0.213 \pm 0.016 $ & $0.819 \pm 0.075 $ & $1.188 \pm 0.093 $ & $2.2$ \\
\hline
$q\left(\nu\right)$ & $0.848 \pm 0.002 $ & $0.088 \pm 0.003 $ & $3.357 \pm 0.214 $ & $-0.057 \pm 0.010 $ & $0.7$ \\
\hline
$s\left(\nu\right)$ & $0.669 \pm 0.006$ & $0.202 \pm 0.010 $ & $2.730 \pm 0.210 $ & $-0.066 \pm 0.014 $ & $1.6$ \\
\hline
\textbf{\rockstar} & & & & & \\
\hline
$q\left(\frac{M}{M_\mathrm{nl}}\right)$ 0& $0.858 \pm 0.005 $ & $0.093 \pm 0.006 $ & $0.929 \pm 0.089 $ & $-1.010 \pm 0.117 $ & $1.1$ \\
\hline
$s\left(\frac{M}{M_\mathrm{nl}}\right)$ & $0.709 \pm 0.016 $ & $0.204 \pm 0.019 $ & $0.824 \pm 0.091 $ & $-1.265 \pm 0.117 $ & $2.4$ \\
\hline
$q\left(\nu\right)$ & $0.850 \pm 0.003 $ & $0.092 \pm 0.005  $ & $2.989 \pm 0.255$ & $-0.054 \pm 0.015 $ & $0.9$ \\
\hline
$s\left(\nu\right)$ & $0.685 \pm 0.008 $ & $0.197 \pm 0.013 $ & $2.608 \pm 0.260 $ & $-0.091 \pm 0.020 $ & $1.8$ \\
\hline
\end{tabular}
\caption{Best fit parameters (and their uncertainties) of the universal curve (equation~\ref{eq_tanh})
  for median $q$ and $s$ as a function of $(M/M_\mathrm{nl})$ and $\nu$ using \subfind and \rockstar for triaxial halos in the 
  scale-free model.
  The last column is the rms fluctuation of the data with respect to the best fit curve.        
  The universal curves are plotted in figure~\ref{fig_scalefree_universal_otp}.
}
\label{tab_powlaw_universal_triaxial}
\end{table}

\begin{table}[h]
\centering
\renewcommand{\arraystretch}{1.2}
\begin{tabular}{|p{2.5cm}|c|c|c|c|c|}
\hline
    \textbf{Prolate} &$\alpha \pm \Delta\alpha$ & $\delta \pm \Delta\delta$ & $\omega \pm \Delta\omega$ & $\mu \pm \Delta \mu$ & rms \\
\hline
\textbf{\subfind} & & & & & $\times 10^{-2}$ \\
\hline
$q\left(\frac{M}{M_\mathrm{nl}}\right)$ & $0.760 \pm 0.015 $ & $0.233 \pm 0.018 $ & $0.768 \pm 0.067 $ & $-1.282 \pm 0.104 $ & $2.1$ \\
\hline
$s\left(\frac{M}{M_\mathrm{nl}}\right)$ & $0.711 \pm 0.027 $ & $0.331 \pm 0.031 $ & $0.691 \pm 0.063 $ & $-1.436 \pm 0.136 $ & $2.6$ \\
\hline
$q\left(\nu\right)$ & $0.729 \pm 0.006 $ & $0.226 \pm 0.011 $ & $2.375 \pm 0.165 $ & $-0.089 \pm 0.015 $ & $1.3$ \\
\hline
$s\left(\nu\right)$ & $0.657 \pm 0.010 $ & $0.310 \pm 0.017 $ & $2.162 \pm 0.163 $ & $-0.120 \pm 0.020 $ & $1.7$ \\
\hline
\textbf{\rockstar} & & & & & \\
\hline
$q\left(\frac{M}{M_\mathrm{nl}}\right)$ & $0.769 \pm 0.014 $ & $0.202 \pm 0.015 $ & $0.867 \pm 0.075 $ & $-1.360 \pm 0.095 $ & $1.9$ \\
\hline
$s\left(\frac{M}{M_\mathrm{nl}}\right)$ & $0.727 \pm 0.028 $ & $0.305 \pm 0.031 $ & $0.727 \pm 0.069 $ & $-1.544 \pm 0.143 $ & $2.4$ \\
\hline
$q\left(\nu\right)$ & $0.745 \pm 0.006 $ & $0.189 \pm 0.008 $ & $2.797 \pm 0.186 $ & $-0.125 \pm 0.015 $ & $1.3$ \\
\hline
$s\left(\nu\right)$ & $0.676 \pm 0.010 $ & $0.271 \pm 0.015 $ & $2.393 \pm 0.181 $ & $0.153 \pm 0.020 $ & $1.7$ \\
\hline
\end{tabular}
\caption{Best fit parameters (and their uncertainties) of the universal curve (equation~\ref{eq_tanh})
  for median $q$ and $s$ as a function of $(M/M_\mathrm{nl})$ and $\nu$ using \subfind and \rockstar for prolate halos
  in the scale-free model.
  The last column is the rms fluctuation of the data with respect to the best fit curve.
  The universal curves are plotted in figure~\ref{fig_scalefree_universal_otp}.
}
\label{tab_powlaw_universal_prolate}
\end{table}

Here $y$ represents the median $q$ or $s$.
$x$ represents $(M/\Mnl)$ or $\nu$. $\alpha$ and $\mu$ describe the vertical and horizontal offsets respectively,
$\delta$ scales the $y$-range and $\omega$ represents the width of the derivative of the $\tanh$ function.
$\mu$ determines the transition scale where the slope flattens for $\log_{10}(x) < \mu$. 

The best fit curve for all halos is shown as the solid line in figure~\ref{fig_scalefree_universal_all}. The rows represent fits for the dependence of $q$ and $s$ for $(M/\Mnl)$ and $\nu$
respectively. The first two columns ($q$ and $s$) are for \subfind and the next two columns ($q$ and $s$) are for \rockstar.
We find the that the best fit curves (and data) are similar between \subfind and their corresponding \rockstar results. The amplitude for the results with \rockstar are however slightly larger for both $q$ and
$s$ compared to those from \subfind. The second thing we find is that, both for \rockstar and \subfind,  the fluctuation of the data with respect to the best fit curve
in terms  $(M/\Mnl)$ is larger compared to that with $\nu$. This suggests that $q(\nu)$ and $s(\nu)$ shows better universality when compared to $q(M/\Mnl)$  and $s(M/\Mnl)$
respectively. We quantify these observations in table~\ref{tab_powlaw_universal_all} where we summarise the best fit parameters ($\alpha, \delta, \omega, \mu$) and their uncertainties (columns 2-4)
of the universal curve (equation~\ref{eq_tanh}) for median $q$ and $s$ as function of $(M/M_\mathrm{nl})$ or $\nu$ using \subfind and \rockstar.

As observed visually, earlier, the \subfind and \rockstar values compare well with each other. The  amplitude with \rockstar is slightly higher compared to \subfind.
The relative uncertainties in the parameters for $\alpha$,  $\delta$ is small. The transition scale for both $q$ and $s$ occur at $\log_{10}(M/\Mnl) \approx -1$
and $\log_{10}(\nu) \approx 0$. Hence we cannot speak about relative uncertainties in $\mu$ when considering $\nu$ as a variable. 
The largest relative uncertainty of $\approx 10\%$ is found for the width parameter $\omega$. This arises due to a different behaviour at masses  below the transition
$\log_{10}(M/\Mnl) \lesssim -1$ and $\log_{10}(\nu) \lesssim 0$ for the $n=-1.0$ model compared to the $n=-1.3$ and $n=-1.5$ models. The differences are larger when $(M/\Mnl)$ 
is used as a variable as compared to $\nu$. We finally quantify the deviation of the data with respect to the best fit curve by computing the rms fluctuation
of the data with respect to the best fit curve. A smaller number represents better universal behaviour. Comparing between $(M/\Mnl)$ and $\nu$ as a variable, we find that
the rms fluctuations with the latter is about $\approx 26\%-32\%$ smaller  than those with $(M/\Mnl)$, again confirming our observations from the plots. The second take away is
that the rms fluctuations with \rockstar is $\approx 7\% - 18\%$ smaller compared to the results with \subfind. This suggests that using $\nu$ as a variable with \rockstar
provides the best universal behaviour.

We investigate if a universal evolution of shape parameters exists separately for  OTP halos in figure~\ref{fig_scalefree_universal_otp}.  
The panels are arranged in a similar manner to figure~\ref{fig_scalefree_universal_all}. Each data point is collected from the solid curve,
across models, in figure~\ref{fig_scalfree_qs_evolve_otp}.  The solid, dashed and dot-dashed lines represent the parametric fit (equation~\ref{eq_tanh})
for the OTP (orange, blue, green) halo data respectively. The shaded region represents the $1\sigma$ percentile of the distribution. Darker shades
are for more negative spectral indices.
We find that the $1\sigma$ percentile spread and the rms fluctuation (table~\ref{tab_powlaw_universal_oblate},\ ,\ref{tab_powlaw_universal_triaxial}
and \ref{tab_powlaw_universal_prolate}) of the data with respect to the best fit curve
increases systematically from oblate to triaxial to prolate halos. The rms fluctuation of the data with respect to the universal
curve for all halos (table~\ref{tab_powlaw_universal_all}) is
\,i) larger than that of the oblate halos,
\,\,ii) comparable to triaxial halos and
\,\,iii) smaller than those of the
prolate halos. They are however of the same order of magnitude.
The universal curve of OTP halos for median $q$ are non-intersecting and increase with decreasing mass (or $\nu$).
The universal curve of OTP halos for median $s$ intersect at $\nu \approx 0.5$.
This is occurring, as mentioned earlier, mainly due to the $n=-1.0$ model which probes this regime.

In summary, we find that the distribution of shape parameters, as characterised by the median and $1\sigma$ percentiles, in scale-free models
are self-similar as postulated. Additionally the evolution can be described universally, across models and redshifts, with a single parametric curve (equation~\ref{eq_tanh}.
OTP halos evolve separately along different universal curves which implies that all halos, which is a weighted sum of OTP halos, should evolve along
a universal curve. This suggests that we should see something similar for the $\Lambda$CDM model.  The $\Lambda$CDM model is not scale-free as discussed earlier
since the cosmological constant and radiation are added to the total energy budget. Secondly the power spectrum in the $\Lambda$CDM has a non-trivial shape.
We therefore expect that the universality that is seen in scale-free models, may be degraded somewhat, in the $\Lambda$CDM context when compared
to the results of the scale-free runs. We explore this in the next section.

\section{Results II: the $\Lambda$CDM model}
\label{sec:result2}

\begin{figure}[b]
\centering
    \includegraphics[width=1.0\linewidth]{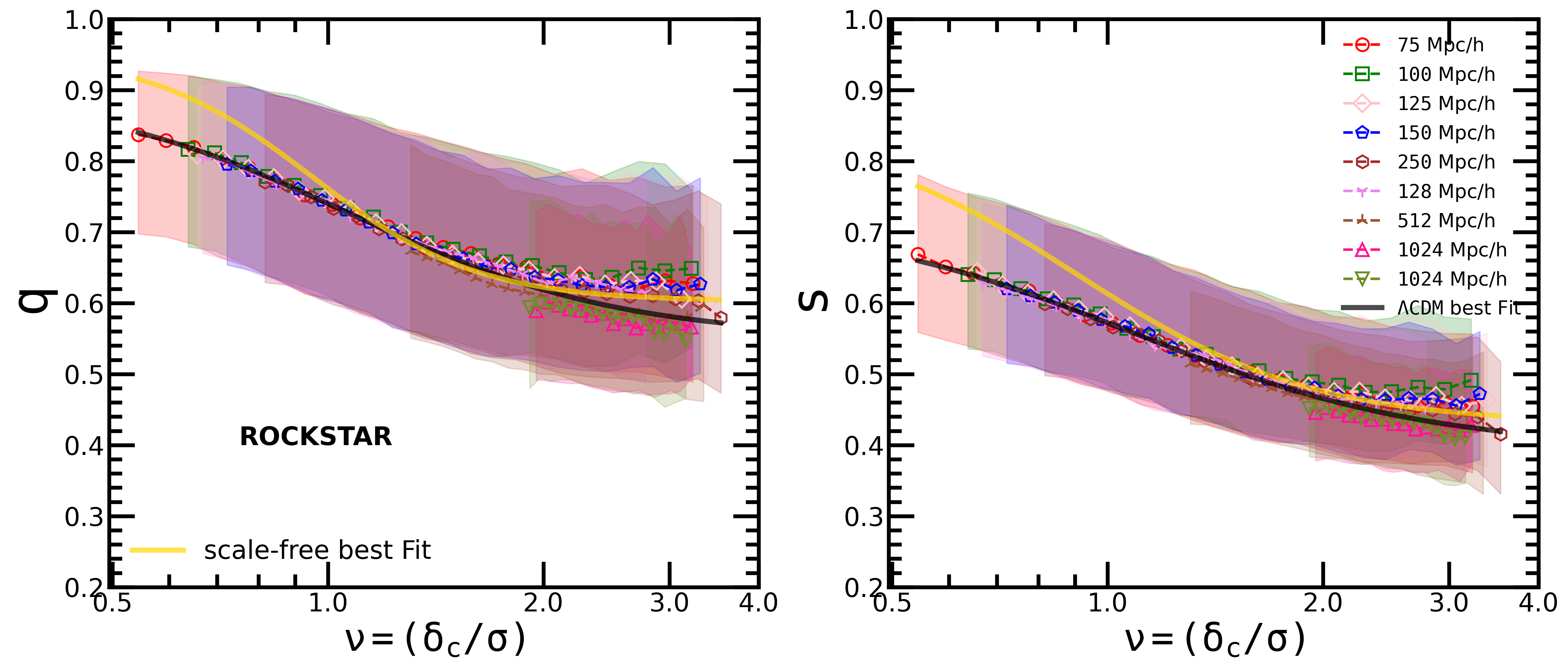}
    \caption{The figure shows the evolution of median $q$ (left) and $s$ (right) values as a function of $\nu$ using \rockstar for all halos in the $\lcdm$ model.
      The different points are from different boxes across corresponding epochs (see table~\ref{tab_lcdm}). The shaded region
      represents the $1\sigma$ percentile of the distribution. 
      The solid black line represents best fit $\tanh$ parametric curve, equation \ref{eq_tanh}, to all the points. 
      The reader is referred to table~\ref{tab_lcdm_universal} for the parameters of the best fit curve. 
      The solid yellow line, is the best fit universal scale-free curve (last two columns of second row of 
      figure~\ref{fig_scalefree_universal_all} and last two rows of table~\ref{tab_powlaw_universal_all}).} 
    \label{fig_universal_lcdm}
\end{figure}

In this section we explore the evolution of the distribution of shape parameters of dark matter halos $P(q)$ and $P(s)$ for the $\Lambda$CDM model.
As discussed in the previous section, we find that better self-similar and universal behaviour of the evolution
of shape parameters when using \rockstar with $\nu$ as a mass variable. In what follows we will restrict our analysis
to this combination for halos with a minimum of 3000 particles. Although our resolution analysis was based on the particular
scale-free run $(n=-1.0)$, we find that this particle threshold applies to the $\Lambda$CDM model as well.

\subsection{Universal Evolution of Halo Shapes }
In figure~\ref{fig_universal_lcdm} we collate results from the various boxes of the $\lcdm$ model (see table~\ref{tab_lcdm}) to obtain
the evolution of $P(q)$ (left) and $P(s)$(right) as a function of $\nu$. The different markers represent median values from different boxes
and the shaded region is the $1\sigma$ percentile of the distribution.
Following the procedure outlined in the scale-free case, for a given box, we bin masses in  predetermined $\nu$ bins
such that there are a minimum of 30 objects in every bin. This is done at every redshift and the median value at every bin represents a point in the figure.
The solid black line is the parametric  fit (equation~\ref{eq_tanh}) to the points. The minimisation is done by Poisson weighting with the total number
of objects in a given $\nu$ bin. 

The smallest value of $\nu$ is determined by the smallest mass halos at $z=0$ in every simulation. The low and intermediate $\nu$ values have good statistics
across simulations and are not affected by finite volume effects. 
A minimum of 30 halos  per bin ensures that
for the smaller boxes the high-$\nu$ data come from relatively rare well resolved
high-redshift halos which are not captured in the larger, low-resolution boxes.   
On the other hand, massive cluster sized objects at $z=0$ contribute to the high-$\nu$ end from the large volume boxes with good statistics.
We find that there is an average  upward shift for the median $q$ and $s$ values with decreasing box size for $\nu \gtrsim 2$. The spread
in the median $q$ and $s$ at $\nu = 3.0$ is about 0.1. However, the shift is not completely systematic with the size of the box. 
 For example, the largest
deviation for $\nu \geq 2.5$ comes from  the $\Lbox = 100 {\rm Mpc/h}$ run rather than the $\Lbox = 75 {\rm Mpc/h}$ run whose results
are consistent with the $\Lbox = 150 {\rm Mpc/h}$ run. On closer inspection, we find
that the relatively poorer statistics of the smaller boxes at the large $\nu$ end, 
coming from high-redshift objects, are responsible for this shift.
Indeed, the best fit line at large $\nu$ is closer to the data from $\Lbox = 1024 {\rm Mpc/h}$ due to its better statistics of cluster sized halos at $z=0$.

The best fit parameters of the universal curve, equation~\ref{eq_tanh}, and uncertainties are listed in table~\ref{tab_lcdm_universal}. The last column
is the rms of data with respect to the best fit curve. The quality of the fit is similar to those obtained for the scale-free case (last two rows of table~\ref{tab_powlaw_universal_all}).

In figure~\ref{fig_universal_lcdm} we also plot the universal curve from the scale-free runs 
(see the last two columns of the second row of figure~\ref{fig_scalefree_universal_all} and the last two rows of table~\ref{tab_powlaw_universal_all}) 
as solid yellow line. Universal curves  for both scale-free and 
$\lcdm$ runs agree well with each other for $\nu \gtrsim 1.5$ corresponding to $n_{\rm eff}\sim -1.8$(cluster scale). 
For lower values of $\nu$, the change to larger values of $q$ and $s$ is shallower for the $\lcdm$ model
compared to scale-free runs. This results in a smaller of $q$ and $s$ at $\nu = 0.5$ for the $\lcdm$ model compared 
to the scale-free runs. 

\begin{figure}[h]
    \centering
    \includegraphics[width=\linewidth]{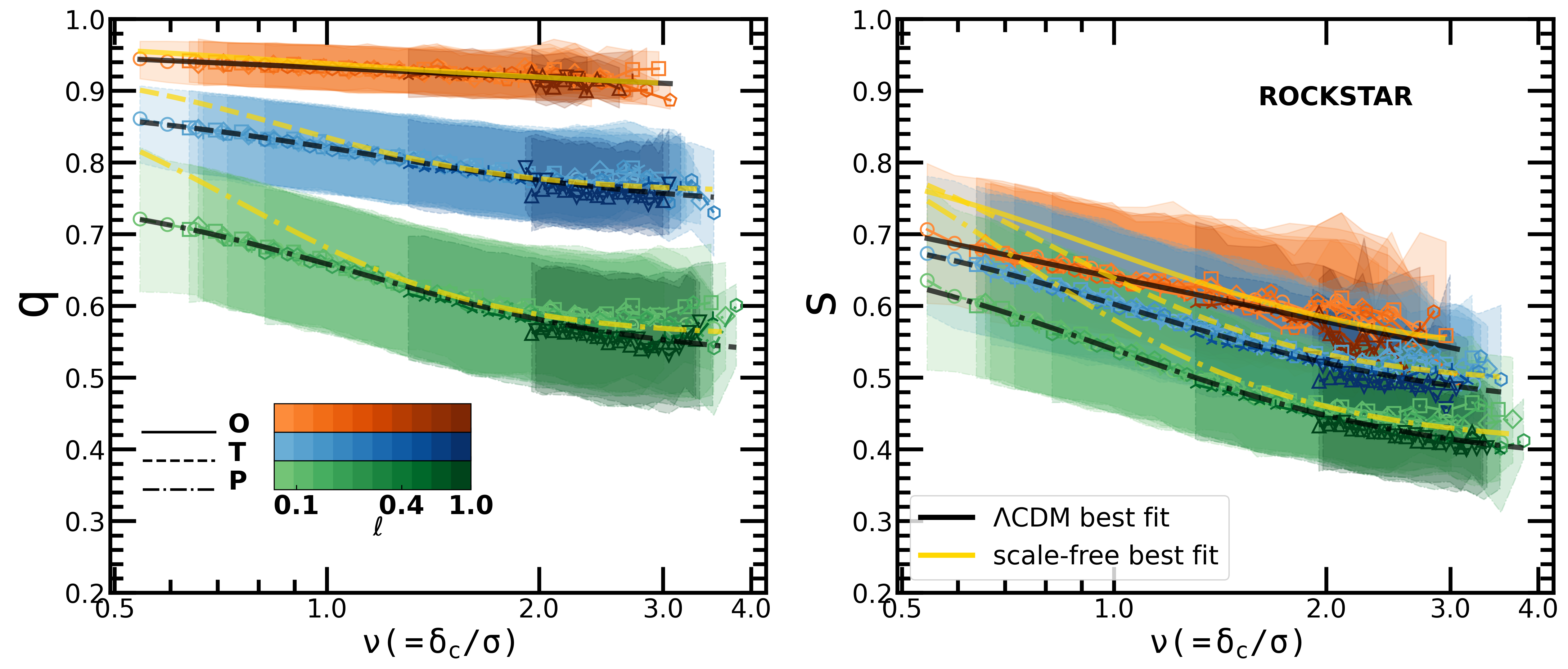}
     \caption{The figure shows the evolution of median $q$ (left) and $s$ (right) values as a function of $\nu$ using \rockstar separately
       for OTP (Orange, Blue, Green) halos in the $\lcdm$ model. 
       Darker shades represent lower resolution (in terms of mean interparticle separation $l$) simulations.
       The different points are from different boxes across corresponding epochs (see table~\ref{tab_lcdm}). The point styles are the same as
       in figure~\ref{fig_universal_lcdm}. The shaded region
       represents the $1\sigma$ percentile of the distribution. The black dashed and dot-dashed line represents the parametric fit, equation \ref{eq_tanh}, to all the points
       for triaxial and prolate halos respectively. For oblate halos (solid line), a linear fit, equation
       \ref{eq_linear}, does the job adequately with fewer parameters. 
       We refer the reader to table~\ref{tab_lcdm_universal} and table~\ref{tab_shape_fit_params} for the parameters
       of the best fit curves. 
       The yellow line, is the best fit universal scale-free curve (last two columns of second row of 
       figure~\ref{fig_scalefree_universal_otp} and last two rows of 
       tables~\ref{tab_powlaw_universal_oblate}, \ref{tab_powlaw_universal_triaxial}, \ref{tab_powlaw_universal_prolate}).
    }
    \label{fig_universal_lcdm_otp}
\end{figure}

In figure~\ref{fig_universal_lcdm_otp} we repeat the analysis separately for OTP halos for the $\lcdm$ model. 
As seen in figure~\ref{fig_universal_lcdm}, compared to the scale-free runs the 
change in slope for median $q(\nu)$ and $s(\nu)$ values are smaller in the $\lcdm$. 
We therefore explore both the $\tanh$ curve (equation~\ref{eq_tanh}) and a linear curve, 
equation~\ref{eq_linear} below, as potential universal curves:
\begin{equation}
    \begin{split}
        y = A - B \log_{10}(\nu), 
    \end{split}
    \label{eq_linear}
\end{equation}
where $y$ represents the median $q$ or $s$ value for OTP halos.

The range of $q$ and $s$ is smaller for oblate halos as compared to the triaxial and prolate halos, we therefore
plot the linear universal curve for oblate halos (black sold line) and the $\tanh$ universal 
curve for the triaxial and prolate halos (black dashed and dot-dashed curves respectively).
We also plot the $\tanh$ universal curves from the scale-free runs as corresponding
yellow curves for comparison (see the last two
columns of the second row of figure~\ref{fig_scalefree_universal_otp} and the last two rows of 
tables~\ref{tab_powlaw_universal_oblate}, \ref{tab_powlaw_universal_triaxial} and \ref{tab_powlaw_universal_prolate}.
The shaded region is the $1\sigma$ percentile of the $q$ and $s$ distributions. 
Darker shades represent smaller mean interparticle separations, or higher resolution simulations. The point 
styles are the same as in figure~\ref{fig_universal_lcdm}. 

We find that the scale-free and $\lcdm$ universal  curves agree with each other for $\nu \gtrsim 1.5$
for all halo types. This translates to a similar deviation between both curves when considering all halos in 
figure~\ref{fig_universal_lcdm}. For all halo types we find that the scale-free model predicts larger values 
of median $q$ and $s$ values compared to the $\lcdm$ model for smaller values of $\nu \lesssim 1.5$.
The scatter of the data with respect to the universal curve (black lines) is smaller for lower values of $\nu \lesssim 2$
as compared to larger values, $\nu \gtrsim 2$. This is because of relatively lower counts at the large-$\nu$ end.

The best fit parameters for the $\tanh$ and the linear curves are given in 
table~\ref{tab_lcdm_universal} and table~\ref{tab_shape_fit_params} respectively. 
Both models describe the data very well when we consider OTP halos separately.
This is seen by the small but comparable rms fluctuation of the data with respect to the universal curves (last column of 
tables~\ref{tab_lcdm_universal} and table~\ref{tab_shape_fit_params}). 
However, these fluctuations are nearly an order of magnitude larger when compared to all halos. This is again due to sparser data at the large-$\nu$ end.

\begin{table}[h]
\centering
\renewcommand{\arraystretch}{1.}
\begin{tabular}{|p{1.9cm}|c|c|c|c|c|}
\hline
\textbf{\footnotesize{ROCKSTAR}} & $\alpha \pm \Delta\alpha$ & $\delta \pm \Delta\delta$ & $\omega \pm \Delta\omega$ & $\mu \pm \Delta \mu$ & rms\\
\textbf{$\Lambda$CDM} & & & & & $\times 10^{-2}$\\
\hline
\multicolumn{6}{|c|}{\textbf{All}} \\
\hline
$q\left(\nu\right)$ & $0.726 \pm 0.007 $ & $0.176 \pm 0.015 $ & $2.627 \pm 0.257 $ & $0.030 \pm 0.016 $ & $2.0$  \\
\hline
$s\left(\nu\right)$ & $0.558 \pm 0.006 $ & $0.164 \pm 0.014 $ & $2.401 \pm 0.232 $ & $0.036 \pm 0.016 $ & $1.4$ \\
\hline
\multicolumn{6}{|c|}{\textbf{Oblate}} \\
\hline
$q\left(\nu\right)$ & $0.950 \pm 0.022 $ & $0.020 \pm 0.053 $ & $1.843 \pm {5.105} $ & $0.010 \pm 0.613 $ & $5.0$ \\
\hline
$s\left(\nu\right)$ & $0.751 \pm 0.014 $ & $0.085 \pm 0.040 $ & $1.833 \pm 0.930 $ & $0.043 \pm 0.092 $ & $17.0$ \\
\hline
\multicolumn{6}{|c|}{\textbf{Triaxial}} \\
\hline
$q\left(\nu\right)$ & $0.811 \pm 0.009$ & $-0.077 \pm 0.029 $ & $-2.090 \pm 0.912 $ & $0.065 \pm 0.058 $ & $11.6$ \\
\hline
$s\left(\nu\right)$ & $0.593 \pm 0.008 $ & $0.143 \pm 0.022 $ & $2.077 \pm 0.358 $ & $0.033 \pm 0.030 $ & $17.0$\\
\hline
\multicolumn{6}{|c|}{\textbf{Prolate}} \\
\hline
$q\left(\nu\right)$ & $0.646 \pm 0.007$ & $0.123 \pm 0.018 $ & $2.273 \pm 0.382$ & $0.046 \pm 0.027 $ & $20.5$ \\
\hline
$s\left(\nu\right)$ & $0.540 \pm 0.009$ & $0.164 \pm 0.019 $ & $2.118 \pm 0.261 $ & $-0.002 \pm 0.028 $ & $16.4$\\
\hline
\end{tabular}
\caption{Best fit parameters (and their uncertainties) of the $\tanh$ universal curve (equation~\ref{eq_tanh})
  for median $q$ and $s$ as a function of $\nu$  for the $\lcdm$ model, using the \rockstar algorithm.
  The last column is the rms fluctuation of the data with respect to the best fit curve. We refer the reader to figure~\ref{fig_universal_lcdm} (all halos)
  and figure~\ref{fig_universal_lcdm_otp} (OTP halos).}
\label{tab_lcdm_universal}
\end{table}

\begin{table}[h]
    \centering
    \renewcommand{\arraystretch}{1.}
    \begin{tabular}{|p{2.5cm}|c|c|c|}
        \hline
        \textbf{\rockstar} & $A \pm \Delta A$ & $B \pm \Delta B$ & rms  \\
        \textbf{$\Lambda$CDM} & & & $\times 10^{-2}$\\
        \hline
        \multicolumn{4}{|c|}{\textbf{Oblate}} \\
        \hline
        $q(\nu)$ & $0.931 \pm 0.001$ & $0.045 \pm 0.006$ & $6.4$ \\
        \hline
        $s(\nu)$ & $0.640 \pm 0.001$ & $0.205 \pm 0.004$ & $18.5$ \\
        \hline
        \multicolumn{4}{|c|}{\textbf{Triaxial}} \\ 
        \hline
        $q(\nu)$ & $0.820 \pm 0.001$ & $0.148 \pm 0.005$ & $13.5$ \\
        \hline
        $s(\nu)$ & $0.602 \pm 0.001$ & $0.269 \pm 0.003$ & $22.8$ \\
        \hline
        \multicolumn{4}{|c|}{\textbf{Prolate}} \\ 
        \hline
        $q(\nu)$ & $0.658 \pm 0.001$ & $0.269 \pm 0.003$ & $13.5$ \\
        \hline
        $s(\nu)$ & $0.658 \pm 0.001$ & $0.303 \pm 0.003$ & $22.8$ \\  
        \hline
    \end{tabular}
    \caption{Best fit parameters (and their uncertainties) of the linear universal curve (equation~\ref{eq_linear})
      for median $q$ and $s$ as a function of $\nu$ for the $\lcdm$ model for OTP halos, using the \rockstar algorithm .
      The last column is the rms fluctuation of the data with respect to the best fit curve. We refer the reader to figure~\ref{fig_universal_lcdm_otp} 
      where the linear fit has been plotted only for the oblate halos.}
    \label{tab_shape_fit_params}
\end{table}

\clearpage
\newpage

\section{Discussion and Summary}
\label{sec:discussion_summary}
In this work, we present a detailed analysis of the evolution of the distribution of halo shapes $P(q)$, $P(s)$.
We do this with the help of a large suite of $N-$body simulations. We study two classes of models: scale-free models with different indices of the initial power spectrum, 
$n= -2.2, -2.0, -1.8, -1.5, -1.3, -1.0$, and the $\lcdm$ model. 
After carefully accounting for finite volume effects, 
the dynamic range in masses in the rescaled variable, the peak height $\nu$, which 
captures mass, redshift evolution and cosmology, spans the range [$0.3$, $4.0$] when all scale-free models are combined and [$0.5$, $3.5$] when all $\lcdm$ boxes are combined. 
We argued that scale-free models should 
exhibit self-similar evolution of statistical quantities. Indeed, for each of the scale-free models,
we find that the median values of $q$ and $s$
evolve along a characteristic smooth curve, a universal curve, when plotted in terms of $\nu$.
The rms deviation of the data with respect to the universal curve is small.
Surprisingly, a single universal curve for median $q(\nu)$ and $s(\nu)$ describes the data concatenated from all scale-free models.
There are hints of deviations from a single universal curve for the $n=-1.0$ model for $\nu \gtrsim 0.5$ 
(see figure~\ref{fig_scalefree_universal_all}), but these deviations are nevertheless small. 

The $\lcdm$ model also exhibits a tight universal evolution for the median $q(\nu)$ and $s(\nu)$. Deviations from universality 
occur at large $\nu$, i.e. $\nu \gtrsim 2.0$ and is due to cosmic variance. 
Based on the argument of using the effective spectral index $n_{\rm eff}$ to map the $\lcdm$ model to scale-free models
we expect that the median $q(\nu)$ and $s(\nu)$ for $\lcdm$  to be consistent with the scale free models for $n \lesssim -1.8$.
This is indeed the case, as can be seen, when comparing figures~\ref{fig_scalefree_universal_all}  
and \ref{fig_universal_lcdm}. For smaller values of $\nu$ the scale-free and $\lcdm$ models differ from each other but evolve 
nevertheless along separate universal curves. 
The $\lcdm$ model has a more complex evolution of mode-coupling compared to the scale-free 
case and this may be the reason for these differences. We also alluded to hints of a different evolution 
of the median $q(\nu)$ and $s(\nu)$ for the $n=-1.0$ model compared to the rest which may also partially be responsible 
for the differences between the $\lcdm$ and scale-free models, since the small $\nu$ values are probed by the $n=-1.0$ model only.

Our results are not specific to any subhalo finding algorithms 
and hold for both \subfind and \rockstar. 
We next classify halos as oblate, triaxial or prolate, based on their 
triaxiality values (equation~\ref{eq_triaxiality_def}). 
We find, yet again, that the median $q(\nu)$ and $s(\nu)$ are described by a universal curve for each of the halo types.
This holds for both scale-free  (figure~\ref{fig_scalefree_universal_otp}) and $\lcdm$ 
(figure~\ref{fig_universal_lcdm_otp}) models. 

\begin{figure}[h]
    \centering
    \includegraphics[width=0.8\linewidth]{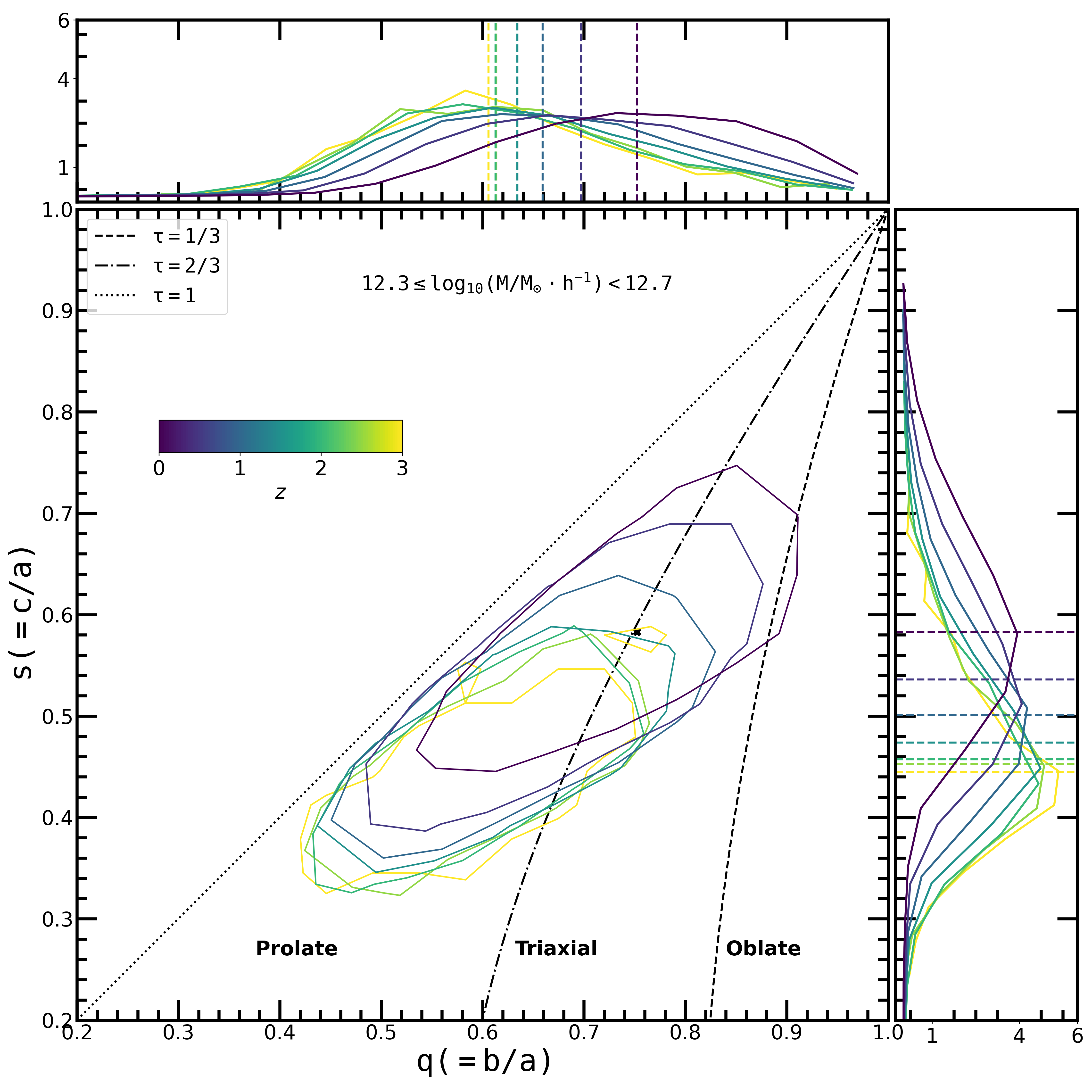}
    \caption{\emph{Main panel}: The figure shows the distribution axis ratios of halos $P(q,s)$, in the mass range
    ($12.3 \leq \log_{10}M/M_{\odot}h^{-1}<12.7$) from the $\Lbox = 75 
    \textrm{Mpc/h}$ run. The contours represent $68\%$ of the distribution. 
    The dashed, dot-dashed and dotted black lines mark the boundary separating oblate, triaxial and prolate halos. 
    These boundaries have been determined using equation~\ref{eq_triaxiality_def}. The colours represent different redshifts.
    The top and right panels show the marginalized distributions  of $P(q)$ and $P(s)$ respectively.
    for the same redshifts mentioned above. The vertical dashed lines in these panels, represent the median values of $q$ and $s$. }
    \label{fig_zevolv_qs-plane}
\end{figure}

Median $q(\nu)$ and $s(\nu)$ increase with decreasing $\nu$. This means that, at fixed mass, 
low  redshift halos are more spherical compared to their high-redshift counterparts. 
This is best seen by looking at how the distribution of shapes, $P(q,s)$ in the $q-s$ plane evolves with 
redshift.
In figure~\ref{fig_zevolv_qs-plane} we plot the 68\% contour of the distribution $P(q,s)$
in the mass range ($12.3 \le \log_{10}(M/M_{\odot}h^{-1}) < 12.7$) from the
$\Lbox = 250\,\mathrm{Mpc}/h$ run from $z=3$ to $z=0$ (different colours).
The dashed, dot-dashed and dotted black lines mark the boundary 
separating oblate, triaxial and prolate halos. 
These boundaries have been determined using equation~\ref{eq_triaxiality_def}.
Halos approach the spherical limit when $(q,s) \rightarrow (1,1)$ with an average rate of change 
${\rm d}q/{\rm d}s > 0$. This limit can be approached separately 
from oblate, triaxial or prolate halos. For this particular mass range, we find that $P(q,s)$ is peaked in the 
prolate population and primarily spread over the prolate and oblate populations at $z=3$. With decreasing 
redshift the peak and the distribution move in the direction of increasing $q$ and $s$, or the spherical limit. This in turn increases the triaxial and oblate 
populations at the cost of the prolate population. 
Although this is done at a specific mass range, we find a similar qualitative behaviour for halos 
at other mass scales. 
In summary low redshift counterparts of high redshift halos (fixed mass) are, on average, more spherical. 
At fixed redshift, lower mass halos are more spherical than higher mass halos. 
Given the relative increase in oblate and triaxial halos and a corresponding decrease in prolate halos, 
we look at the mass function of halos by halo shape and then use it to compute their relative fractions.

\begin{figure}[h]
    \centering
    \includegraphics[width=\linewidth]{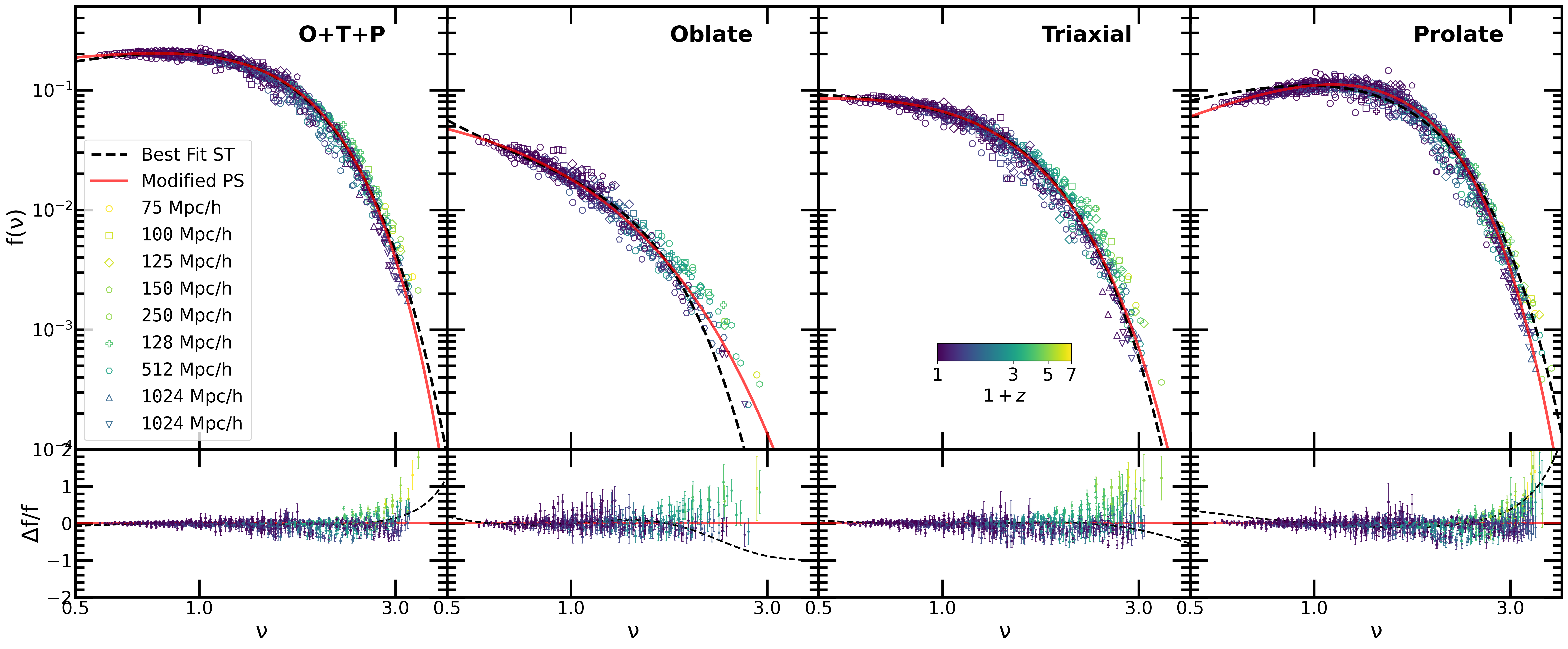}
    \caption{The mass function of central halos for the full catalogue (first panel) and separately 
    for OTP halos (panels 2-4) in the $\lcdm$ model. 
    The point styles are the same as in figures~\ref{fig_universal_lcdm} and \ref{fig_universal_lcdm_otp} and denote different simulation boxes. The points are colour coded according to redshift. 
    The dashed black line and solid red lines 
    in each panel represent the best-fit ST 
    (equation~\ref{eq_massfn_st}) and modified PS (equation~\ref{eq_massfn_ps_modified}) mass functions respectively. 
    The corresponding best fit parameters are given in tables~\ref{tab_massfn_psm} and \ref{tab_massfn_st}.
    The bottom row are the residuals with respect to the modified PS mass function fit. Poisson errors are shown only in the lower row.}
    \label{fig_massfn_all-otp}
\end{figure}

In figure~\ref{fig_massfn_all-otp} we plot the mass function of central halos in its universal form \cite{1994MNRAS.271..676L}
$f(\nu)$ where 
\begin{equation}
    \frac{{\rm d}n}{{\rm d}\ln M} = \frac{\bar{\rho}}{M}\frac{{d}\ln \sigma(M)^{-1}}{{\rm d}\ln M} f(\nu),
    \label{eq_mf_definition}
\end{equation}
here $\bar{\rho}$ is the mean matter density. $f(\nu)$ is computed from the halo abundance, which is on the left hand side of the equation. The runs are from the $\lcdm$ 
model and the data and point styles 
are the same as in figures~\ref{fig_universal_lcdm} and \ref{fig_universal_lcdm_otp}. 
We use Poisson errors to do a $\chi^2$ fit for the mass function. These errors are plotted in the lower panel of figure~\ref{fig_massfn_all-otp}.  
We explore two functional forms of the mass function.
i. A modified Press-Schechter mass function which generalizes the functional form 
of the Press-Schechter \citep{1974ApJ...187..425P} mass function with four parameters 
$\left\{A_{\rm ps,m},l,m,r\right\}$ 
\beq
f_{\rm ps,m}(\nu) = A_{\rm ps,m} \nu^l \exp{\left[-\frac{m\nu^r}{2}\right]},
\label{eq_massfn_ps_modified}
\eeq
ii. The Sheth-Tormen mass function \citep{1999MNRAS.308..119S} with three parameters
$\left\{A_{\rm st},Q,p\right\}$. 

\beq
f_{\rm st}(\nu) = A_{\rm st} \nu \left[\frac{2Q}{\pi}\right]^{1/2} \left[ 1 + \left(Q\nu^2\right)^{-p}\right]
\exp{\left[-\frac{Q\nu^2}{2}\right]}.
\label{eq_massfn_st}
\eeq
The amplitude $A_{\rm st}$ is treated as a separate parameter, since we are looking at central halos only.
The modified Press-Schechter and the Sheth-Tormen mass functions are plotted as the solid red and dashed black lines in figure~\ref{fig_massfn_all-otp}. 
We present the mass function for all halos (panel 1) and separately for oblate, triaxial and prolate halos (panels 2-4). We note that data is not independent 
in all the four panels since the sum of OTP halos (panels 2-4) is the full catalogue (panel 1).
The lower row is the residual with respect to 
the best-fit modified Press-Schechter mass function. In tables~\ref{tab_massfn_psm} and \ref{tab_massfn_st}
we report the best fit parameters of modified Press-Schechter and Sheth-Tormen mass functions respectively.
For oblate halos we obtained the parameter $l \simeq 0$   in the fitting procedure for the 
modified Press-Schechter mass function. This in turn gave poor fits and estimates for other parameters, 
like the amplitude $A_{\rm ps,m}$. Fixing $l=0$, resolves this issue. Similarly 
we fixed $p=0$ for prolate halos when fitting the Sheth-Tormen mass function.
The last two columns of tables~\ref{tab_massfn_psm} and \ref{tab_massfn_st}, 
are the reduced $\chi^2$ and the rms of the data with respect to the best fit model.

The modified Press-Schechter and Sheth-Tormen mass functions do a reasonable job in describing the 
data in all cases. The scatter is larger at large-$\nu$, due to fewer halos at this end.  
In the case of all halos and triaxial halos, both mass functions are comparable in their 
quality of fits. The quality of fit is however poor. The $\chi^2_{\rm red}$ for all halos is 
$\sim 4$ and $\sim 2.7$ for triaxial halos, 
for both the Sheth-Tormen and modified Press-Schechter mass functions.
The best fit values $\{A_{\rm st}, p, Q\}$ = $\{0.208,0.166,1.183\}$ for the Sheth-Tormen mass function for all 
halos, is very different from the standard values of $\{A_{\rm st}(p), p, Q\} = \{0.322, 0.3, 0.707\}$ 
\citep{1999MNRAS.308..119S}.
This is because we are considering central halos only, which reduces the mass compared to the virial mass
of the full halo. Our mass functions are therefore shifted left with respect to the standard mass functions 
in the literature. 
We have checked that using the virial mass of the full halo and fitting for the 
Sheth-Tormen mass function we obtain $\{A_{\rm st}, p, Q, \chi^2_{\rm red}\} = \{0.358, 0.335, 0.745, 2.80\}$ for halos resolved 100 particles. The parameters change to 
$\{A_{\rm st}, p, Q, \chi^2_{\rm red}\} = \{0.360, 0.317, 0.750, 1.92\}$ for halos resolved with 1000 particles 
and $\{A_{\rm st}, p, Q, \chi^2_{\rm red}\} = \{0.360, 0.167, 0.789, 2.39\}$ for halos resolved with 3000 particles. The value of $(p,Q)$ change significantly 
as we go from a 1000 to a 3000 particle cut.  
The quality of fit in the mass function up to a 1000 particle cut is broadly consistent with 
the standard values of the Sheth-Tormen mass function and \cite{2023MNRAS.521.5960G}.
It improves dramatically with a drop in  $\chi^2_{\rm red}$ by a factor of $\sim 2$ (compared to 
fitting it with a mass definition for central halos) 
when considering the virial mass of the entire halo. 
This suggests that the halo mass function for central
halos has more scatter  compared to the full halo or is less universal than the halo mass function of 
the full halo and may require additional parameters to describe the a more universal mass function 
(see for example \cite{2025arXiv251116730F}). The four parameter modified Press-Schechter parametrization 
\ref{eq_massfn_ps_modified} for the virial mass definition results in a better fit compared to the three parameter Sheth-Tormen mass function \ref{eq_massfn_st}. For a 100-particle 
cut we obtain $\{A_{\rm ps,m}, l,m,r,\chi^2_{\rm red}\}$ = \{0.452,0.419,0.480,2.268,1.91\}. The best fit 
parameters for a 1000 and 3000 particle resolution are 
$\{A_{\rm ps,m}, l,m,r,\chi^2_{\rm red}\} = \{0.470,0.496,0.546,2.190,1.84\}$  and 
$\{A_{\rm ps,m}, l,m,r,\chi^2_{\rm red}\} = \{0.707,1.302,1.408,1.657,2.10\}$ respectively.

\begin{table*}
\centering
\renewcommand{\arraystretch}{1.}
\begin{tabular}{|l|c|c|c|c|c|c|}
\hline
 \textbf{Halo} & $A_{\rm ps,m} \pm $ & $l \pm $ & $m \pm  $ & $r  \pm  $ & $\chi^2 _{\rm red}$ & RMS   \\
\textbf{Type} & $\Delta A_{\rm ps,m}$ & $\Delta l$ & $\Delta m$ & $\Delta r$ &  & ($\times 10^{-2}$)\\

\hline
\textbf{O+T+P} & $0.276 \pm 0.003$ & $0.464 \pm 0.022$ & $0.697 \pm 0.023$ & $2.382 \pm 0.025$ & $3.98$ & $1.0$ \\
\hline
\textbf{Oblate} & $0.096 \pm 0.006$ & $0 $ & $3.348 \pm 0.120$ & $1.243 \pm 0.035$ & $1.26$ & $0.2$   \\
\hline
\textbf{Triaxial} & $0.145 \pm 0.009$ & $0.455 \pm 0.067$ & $1.569 \pm 0.116$ & $1.826 \pm 0.050$ & $2.63$ & $0.4$  \\
\hline
\textbf{Prolate} & $0.193 \pm 0.004$ & $1.496 \pm 0.035$ & $1.140 \pm 0.042$ & $2.104 \pm 0.025$ & $3.11$ & $0.8$  \\
\hline
\end{tabular}
\caption{Best-fit parameters and uncertainties of the modified PS mass function (equation~\ref{eq_massfn_ps_modified}) 
for all halos and separately for OTP halos. 
These best fit curves are plotted as the dashed-red line in figure~\ref{fig_massfn_all-otp}.
For oblate halos the parameter $l$ 
has been set to zero, since the best-fit value was close to zero.}
\label{tab_massfn_psm}
\end{table*}

For oblate halos, the fit for the modified Press-Schechter and Sheth-Tormen 
mass functions results in $\chi^2_{\rm red} = 1.26$ and  $\chi^2_{\rm red} = 1.84$ respectively.
A look at the residuals in figure~\ref{fig_massfn_all-otp} suggests that the relative increase 
$\chi^2_{\rm red}$  in the case of the Sheth-Tormen mass function 
comes from deviations at the large-$\nu$ end. 
We also find that for prolate halos, the  modified Press-Schechter parametrization does a better job 
($\chi^2_{\rm red} = 1.84$) in describing data as compared to Sheth-Tormen ($\chi^2_{\rm red} = 2.71$),
but here the deviations come from both the large and small 
$\nu$ ends of the mass function.

\begin{figure*} 
    \centering 
    \includegraphics[width=\linewidth]{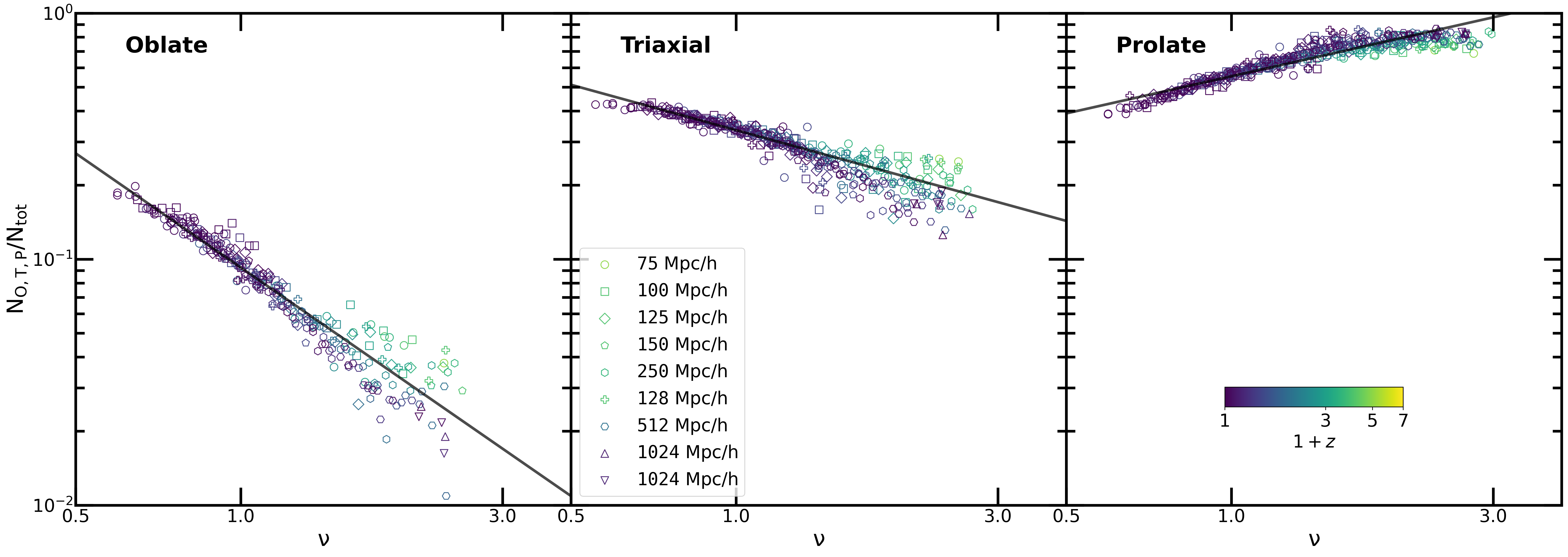} 
    \caption{The figure shows the oblate(left), triaxial(middle), prolate(right) 
    number fractions as a function of $\nu$ in the $\lcdm$ model. 
    The points (and colours) are the same as figure~\ref{fig_massfn_all-otp}. 
    The solid line is a linear, $\log_{10}(N_{\rm O,T,P}/N_{\rm tot}) = \log_{10}(F_{\rm O,T,P}) = u\log_{10}
(\nu) + v$, fit to the data. 
     Here $N_{\rm O,T,P}$ refers to the number of Oblate, Triaxial or Prolate halos and 
     $N_{\rm tot} = N_{\rm O} + N_{\rm T} + N_{\rm P}$.} 
     \label{fig_fraction_otp}
 \end{figure*}

In figure~\ref{fig_fraction_otp} we look at the evolution of fractional abundances $F_{\rm O,T,P}$
by halo type as a function of $\nu$. The fractional abundance of oblate, triaxial or prolate (left to right) halos is $F_{\rm O,T,P}(\nu) = {N_{\rm O,T,P}}(\nu)/{N_{\rm tot}}(\nu)$.
Here $N_{\rm O,T,P}(\nu)$ refers to the number density of oblate,   triaxial or prolate halos and 
$N_{\rm tot}(\nu) = N_{\rm O}(\nu) + N_{\rm T}(\nu) + N_{\rm P}(\nu)$.  
The data points and colours are the same as in figure~\ref{fig_massfn_all-otp}. The solid line is a linear fit to the logarithm of the data, given by:
\beq
\log_{10} F_{\rm O,T,P} = u \log_{10}(\nu) + v
\eeq
which is equivalent to 
\beq
F_{\rm O,T,P} = \left(\frac{\nu}{\nu_0}\right)^\kappa
\eeq
with $\kappa = u$ and $\nu_0 = 10^{-(v/u)}$.  We note that only two of the panels are independent (since the fractions should add to 1) as in figure~\ref{fig_massfn_all-otp} 
but the fits have been done assuming all three to be independent. Indeed 
we see greater than 10\% deviation from the expected $N_{\rm tot}$ for $\nu \leq 0.59$ and $\nu \geq 2.43$.

For oblate and triaxial halos the fraction increases with decreasing $\nu$.
This means that the fraction of oblate and triaxial halos at all mass scales 
increases with decreasing redshift.
Conversely the fraction of prolate halos decrease with decreasing redshift at all mass scales. 
The best fit parameters 
$(u,v)$ are  $(-1.545,-1.034)$, $(-0.619,-0.475)$ and $(0.517,-0.253)$ for oblate, triaxial and prolate halos 
respectively. These numbers translate to best fit values for 
$(\kappa, \nu_0)$ as $(-1.545,0.214)$ $(-0.619,0.170)$ $(0.517,3.086)$ for oblate, triaxial and prolate 
halos respectively. 

Using the best fit lines we find that for $\log_{10} (\Mhalo/(\msun/h)) = 12.0$ at $z=1$ the fraction of oblate, triaxial and prolate halos are roughly 6\%, 28\% and 64\%. 
These numbers change to 13\%, 39\% and 49\% for the same mass scale at $z=0$. For $\log_{10} (\Mhalo/(\msun/h)) = 13.0$ at $z=1$, the fraction of oblate, triaxial and 
prolate halos are roughly 2\%, 22\%, 77\% and 
7\%, 30\% and 60\% at $z=0$. The fraction of triaxial halos dominate over prolate halos for $\nu \leq 0.631$
which translates to a mass scale of $\log_{10} (\Mhalo/(\msun/h)) = 11.3$.\\

\begin{table*}
\centering
\begin{tabular}{|l|c|c|c|c|c|}
\hline
\textbf{Halo Type} & $A_{\rm st} \pm \Delta A_{\rm st}$ & $p \pm \Delta p$ & $Q \pm \Delta Q$ & $\chi^2_{\rm red}$ & RMS ($\times 10^{-2}$) \\


\hline
\textbf{O+T+P} & $0.208 \pm 0.0003 $ &  $0.166 \pm 0.008$ & $1.183 \pm 0.003$ & $4.23$ & $1.0$ \\
\hline
\textbf{Oblate} & $0.032 \pm 0.0004$ & $1.548 \pm 0.052$ & $1.969 \pm 0.018$ & $1.84$ & $0.2$ \\
\hline
\textbf{Triaxial} & $0.077 \pm 0.0002$ & $0.667 \pm 0.013$ & $1.351 \pm 0.006$ & $2.71$ & $0.5$  \\
\hline
\textbf{Prolate} & $0.225 \pm 0.0004$ & $0$ & $1.077 \pm 0.002$ & $7.03$ & $1.0$  \\
\hline
\end{tabular}
\caption{Best-fit parameters and uncertainties of the ST mass function (equation~\ref{eq_massfn_st}) 
    for all halos and separately for OTP halos. 
    These best fit curves are plotted as the solid black line in figure~\ref{fig_massfn_all-otp}
    For prolate halos the parameter $p$, has been set to zero, since the best-fit value was close to zero.}     
\label{tab_massfn_st}
\end{table*}

In this study, we examine the evolution of the distribution of shapes of dark matter halos using the shape tensor weighted 
by elliptical distance($r_{\rm ell}$). We analyse scale-free simulations with spectral indices ranging from $n=-1.0$ to $n=-2.2$ to test the self-similarity ansatz in the context 
of the distribution of of axis-ratios. We extend the analysis to a suite of $\lcdm$ simulations, with the same cosmology, spanning a broad range of halo masses from galactic scales to cluster 
scales. We also explore the evolution of the distribution of halo shapes across different morphological types and study how this evolution is connected to the 
abundance of distinct halo populations in the universe. Our results are summarized as follows: \\
\begin{itemize}
     \item In the scale free simulations listed in table~\ref{tab_scalefree} we find that the median axis ratios($q=\frac{b}{a};s=\frac{c}{a}$) exhibit self-similar evolution while expressed in terms of scaled mass 
     $(M/\Mnl)$ or equivalently peak height($\nu$) (figure~\ref{fig_scalfree_qs_evolve_all}). 
     Halos at low $\nu$ are more spherical compared to those at high $\nu$. 
     For spectral indices $n=-1.0$, $-1.3$, and $-1.5$, the slopes of median $q(\nu)$ and $s(\nu)$ change from shallower to steeper values with increasing $\nu$. 
     For more negative spectral indices, the dependence of median $q$ and $s$ on $\nu$ weakens, and becomes asymptotically flat for $n=-2.2$. 
     
     \item The median axis ratios from all scale-free simulations vary monotonically with both, $(M/\Mnl)$ and $\nu$ and  are well-described by a 
     single universal curve (figure~\ref{fig_scalefree_universal_all}) 
     fitted with the hyperbolic tangent function (equation~\ref{eq_tanh}). 
     
     \item We further classify halos morphologically, based on their triaxiality parameter,  as oblate, triaxial or prolate. 
     We find that each morphological class also exhibits self-similar evolution for their median axis ratios (figure~\ref{fig_scalfree_qs_evolve_otp}). 
     A single universal curve[\ref{eq_tanh}], albeit with different parameters for each morphological type,
     describes the evolution of median axis ratios (\ref{fig_scalefree_universal_otp}). 

     \item For scale-free simulations we employ two halo finders: \subfind and \rockstar to identify bound structures from simulation. 
     Although both algorithms provide consistent results for the median axis ratios, in terms of universal behaviour, 
     we find that the median $q,s$ values for \rockstar to be slightly higher than those identified with \subfind. 
     For the $\lcdm$ simulations we present results only with \rockstar.

     \item For scale-free models, we explore the universality of the evolution of median axis ratios as a function of $M/\Mnl$ and $\nu$ and find that the latter 
     shows better universal behaviour. 
     Our results for the $\lcdm$ simulations are present only for the evolution with $\nu$.

     \item We find that the median axis ratios exhibit universal behaviour for the $\lcdm$ (table~\ref{tab_lcdm}) 
     simulations consistent with \cite{2014MNRAS.443.3208D}. Unlike the combined scale-free runs, the increase in $q$ and $s$ values in the $\lcdm$ model is weaker with decreasing $\nu$. Both the $\tanh$ (equation~\ref{eq_tanh})
     and a linear curve (equation~\ref{eq_linear}) describe the universal behaviour consistent with \cite{2014MNRAS.443.3208D}.  
     At $z=0$, galactic scale halos (low-$\nu$) are more spherical than cluster sized halos (high-$\nu$), consistent with \citep{2006MNRAS.367.1781A}. However, analytical studies based on triaxial collapse \cite{2011MNRAS.416..248R,2018MNRAS.475.3553N} predict the opposite trend likely due to their simple assumptions, such as treating them as isolated objects and neglecting environmental effects.
     We find that the universal curve for the scale-free model is consistent with that of the $\lcdm$ model for  $\nu(\gtrsim 1.5)$, which corresponds to $n_{\rm eff} \lesssim -1.8$. This scale is 
     consistent with the range of validity of mapping scale-free to $\lcdm$ models [Figure~\ref{fig_universal_lcdm}].

     \item As in the scale-free models, we find separate universal of evolution of median axis halos based on their morphological type [Figure~\ref{fig_universal_lcdm_otp}].

     \item We show that two dimensional distribution $P(q,s)$  for halos in the mass 
     range[$12.3 \le \log_{10}(M/M_{\odot}h^{-1}) < 12.7$] moves in the direction of increase $q$ and $s$ i.e. $\frac{dq}{ds}>0$ (figure~\ref{fig_zevolv_qs-plane}). 
     The distribution moves in the direction of prolate $\rightarrow$ triaxial $\rightarrow$ oblate regimes with decreasing redshift.
     
     \item We find that the mass function of central halos is described by an approximate universal curve, distinct for each halo morphological class (figure~\ref{fig_massfn_all-otp}. 
     We explore two forms of the mass function: a four-parameter modified Press-Schechter function (equation~\ref{eq_massfn_ps_modified}) and  a three-parameter Sheth-Tormen function(equation~\ref{eq_massfn_st}). 
     We find that the modified Press-Schechter function does a better job compared to the Sheth-Tormen function in fitting the mass function for oblate and prolate halos. For triaxial 
     and all halos both mass functions describe the data equally well (tables~\ref{tab_massfn_psm} \& \ref{tab_massfn_st}).

     \item We examine the fractional evolution of each distinct halo population over cosmic time. 
     We find that the oblate and triaxial halo fractions increase with decreasing redshift across all mass scales whereas the prolate fractions decrease (figure~\ref{fig_fraction_otp}).
     We find that the universe is mostly populated with triaxial and prolate halos with a small fraction($\sim 18\%$) of oblate halos($\Mvir\geq10^{11}\msun/h$) today.
     
\end{itemize}

We note that we are not tracking here the growth of halos over cosmic time i.e. mass assembly histories($M(z)$) and therefore our analysis does not directly account for the roles of mergers\cite{2025ApJ...979..223W}, smooth accretion, or environmental effects in shaping halo structure. It would be interesting to see how the halo shape evolves in response to these processes and what determines the change of halo shapes \textemdash\ whether it is dominated by merger activity, anisotropic accretion, the tidal field. 
We leave the investigation of these questions to a future work. 

\section{Acknowledgements}
AN and SG acknowledge useful discussions with Ravi Sheth at the Cosmology Summer School held in June 2024 at ICTP, Italy. AN thanks  Ranit Behera for helpful discussion on the \rockstar halo finder. 
This work acknowledges the use of HPC facilities and storage servers in NISER which have been funded by the Department of Atomic Energy and the Department of Science and Technology, Government of India. NK acknowledges support from the IUCAA Associateship programme.

The data analysis and visualization has been carried out for this work using NumPy \cite{2020Natur.585..357H}, SciPy\cite{2020NatMe..17..261V}, Matplotlib \cite{2007CSE.....9...90H}, \pylians \citep{Pylians} and \colossus \citep{2018ApJS..239...35D}. The simulations were run with 
\gadgetfour \citep{2021MNRAS.506.2871S} and \mpgadget \citep{yu_feng_2018_1451799} and the halo catalogs were 
generated with \rockstar \citep{2013ApJ...762..109B} and \subfind \citep{2001MNRAS.328..726S}.

\bibliography{references}

\end{document}